\documentclass[rmp,aps,preprint,showpacs,preprintnumbers,amsmath,amssymb,showkeys]{revtex4-1}

\usepackage{graphicx}
\usepackage{dcolumn}
\usepackage{bm}
\usepackage{epsfig}
\usepackage{graphicx}
\usepackage{xcolor}
\usepackage{stmaryrd} 
\usepackage{amsmath}
\usepackage{braket}
\usepackage[version=3]{mhchem} 

\newcommand{\D}{\mathrm{d}}

\begin{document}
\preprint{Front. Appl. Math. Stat., 21 May 2021;
{\bf doi.org/10.3389/fams.2021.650716}}

\title{Monitoring and Forecasting COVID-19: \\
Statistical Heuristic Regression, \\
Susceptible-Infected-Removed model and, \\
Spatial Stochastics}

\author{Pedro L. de Andres}
\affiliation{ICMM, Consejo Superior de Investigaciones Cientificas, E-28049 Madrid, Spain}

\author{Lucia de Andres-Bragado}
\affiliation{Department of Biology, University of Fribourg, CH-1700 Fribourg, Switzerland}
\author{Linard D. Hoessly}
\affiliation{Department of Mathematics, University of Copenhagen,  DK-2100 Copenhagen, Denmark}

\date{\today}     

\begin{abstract}
The COVID-19 pandemic has had worldwide devastating effects on human lives, highlighting the need for tools to predict its development. The dynamics of such public-health threats can often be efficiently analysed through simple models that help to make quantitative timely policy decisions. We benchmark a minimal version of a Susceptible-Infected-Removed model for infectious diseases (SIR) coupled with a simple least-squares Statistical Heuristic Regression (SHR) based on a lognormal distribution. We derive the three free parameters for both models in several cases and test them against the amount of data needed to bring accuracy in predictions. The SHR model is $\approx \pm 2\%$ accurate about 20 days past the second inflexion point in the daily curve of cases, while the SIR model reaches a similar accuracy a fortnight before. All the analysed cases assert the utility of SHR and SIR approximants as a valuable tool to forecast the disease's evolution. Finally, we have studied simulated stochastic individual-based SIR dynamics, which yields a detailed spatial and temporal view of the disease that cannot be given by SIR or SHR methods.
\end{abstract}


\maketitle

The consequences of a pandemic like COVID-19 caused by the virus SARS-CoV-2 cannot be overstated~\cite{Nature}. 
Accurate mathematical tools allowing to monitor and forecast the evolution of the contagious disease are useful to guide social, economic and public health decisions made by governments. Nevertheless, despite the availability of powerful mathematical models~\cite{anderson2020}, initial forecasting by some organizations underestimated the evolution of the epidemics, hampering the immediate taking of necessary actions~\cite{TheEconomist,LeMonde}.

This study aims to take advantage of available worldwide data on COVID-19 \cite{owidcoronavirus} \cite{johnshopkins} to benchmark and assign error bars to minimal models, like the susceptible-infected-recovered (SIR) with different sophistication levels~\cite{kermak1927,weiss2013,HeChenJiang,YangWang,khan2020,ANNAS2020}, a straightforward least-squares best-fit (LS) Statistical Heuristic Regression (SHR) based on a lognormal distribution~\cite{lam1988}, or basic Monte-Carlo simulation~\cite{girona2020,gang2020}.
It is well-known that finding a global minimum of non-linear least-squares problems for
$p$ free parameters requires, at worst, a brute force search in p-dimensional parameter space. If each parameter can take $m$ values inside a given interval, it is a non-polynomial task that scales like $m^{p}$ and becomes non-practical for large or moderate values of $p$. 
Correspondingly, there are no general polynomial bounds on the time complexity given in the number of samples and the search space dimension. 
These models are gauged against two variables measured daily: (i) the number of deaths, and (ii) the number of new infections. Such indicators both possess advantages and disadvantages. Deaths are usually counted using a consistent methodology\footnote{However, some countries have varied these criteria during the course of the epidemic. Moreover, different countries can use different rules to define these variables. Those inconsistencies can be compensated and do not significantly affect the conclusions of our analysis. However, a comparison between countries is only an approximated one.} 
and, undeniably, it is an observable proportional to the spread of the disease, but the tally of deaths carry a delay of about one month on the actual dynamics of the disease. On the other hand, the number of infections is timely, but incorporates more uncertainties since it depends on details not related to the disease, e.g. on the number of tests performed. We show the simultaneous monitoring of both observables supplemented with relatively simple mathematical approaches can be used to follow and forecast the evolution of the disease with enough accuracy to help decision-making processes and we discuss the associated error bars.

Other efforts to modeling the pandemics include sensitivity and meta-analysis to estimate averaged values for the reproduction number, incubation time, infection rate and fatality rate~\cite{HeYiZhu}, 
wavelet-coupled random vector functional link networks \cite{HazarikaGupta},
machine learning~\cite{Singh2020}, and Advanced Autoregressive Integrated Moving Average Model~\cite{Singh2020B}. 
Approaches via learning algorithms are usually compared via corresponding tests~\cite{Demsar2006a}, where we recall the significant differences to statistics\cite{Breimann01}.

The paper is organised as follows. In Section 2 (results and discussion), we introduce the Statistical Heuristic Regression model (SHR, section 2.1), the Susceptible-Infected-Removed model (SIR, section 2.2), and the Spatial Stochastic Individual-Based model (MC, section 2.3). After each topic, we analyse the corresponding application to different countries or regions, most notably Spain and Germany. Finally, in the section of Conclusions, we summarise and review our approaches and further discuss the application of the cases analysed and future venues.

\section{Results and Discussion}

\subsection{Statistical Heuristic Regression (SHR)}

Epidemics can efficiently be modelled as a geometric process related to independent random events~\cite{lam1988}. This method yields a regression curve that describes the temporal variation of a contagious disease for the number of deaths, infections or some other relevant observable variable. Such a statistical heuristic approach results in a lognormal function, 

\begin{equation}
c_{c_M,\mu,\sigma}(t)= c_M \frac{e^{-\frac{\left( \ln{(t-t_0)}-\mu \right)^2}{2 \sigma^2}}}{\sqrt{2\pi} \sigma (t-t_0)} ~~~~~~t-t_0>0 
\label{eqn:PDF}
\end{equation}
\noindent
which is the probability distribution function of a random variable whose logarithm, $u=\ln{(t)}$, is normally distributed around its mean value $\mu$ with a dispersion $\sigma$~\cite{lloyd1994}.
The beginning of the propagation is determined by
$t_0$
and the value of the single maximum $c_M=c(t_M)$ happens at $t_M=t_0+\mu-\sigma^2$.

Starting from the model $\frac{\D c(t)}{\D t} = \alpha(t) c(t)$ and imposing general requirements on $\alpha(t)$ (which follow from the observed
behaviour of the number of daily cases) also leads to the same expression~\cite{entropy_lognormal}. 
Using entropy-related arguments, these authors have estimated that
$\sigma \approx 0.4$, which compares well with the
averaged values for ten different western countries for deaths and infected,
$0.6\pm 0.2$ and $0.5 \pm 0.2$ respectively, cf. Tables~\ref{tbl:SHRd}
and~\ref{tbl:SHRi}. Finally, we notice that
such lognormal distribution derived on~\cite{lam1988} was proved useful to model the SARS outbreak in 2003~\cite{chan2006}. 

The corresponding accumulated cases are,
\begin{equation}
C_{c_M,\mu,\sigma}(t)= \int_{0}^{t} c(u)~ \D u =
\frac{c_M}{2} \text{Erfc} \left( -\frac{\ln{(t-t_0)}-\mu}{\sqrt{2} \sigma} \right) ~~~~~~ t-t_0>0 \label{eqn:CDF}
\end{equation}
\noindent
Given arbitrary precision, $C(t)$ and $c(t)$ carry the same information about the 
set of three parameters, $\mathcal{F}=\{c_M,\mu,\sigma\}$,
since $c(t)$ is simply the temporal derivative of $C(t)$.
However, in practical terms, $c(t)$ is
heavily affected by noise in the collection of the data series $\{c_i\}$, and least-squares fits to the functions $C(t)$ and $c(t)$ are expected to determine slightly different values for $\mathcal{F}$. 
Therefore, we chose to report values related to
$C(t)$, which are less affected by noise.
Still, we notice that the information contained in $c(t)$ is equally
valuable and sometimes simpler to obtain, in particular, the position and value of its inflexion points and single maximum.

Next, we aim to prove that the ansatz in Equations 1 and 2 reproduces the behaviour of COVID-19 in ten different western countries using actual data up to the time of submission (revised version, March 2021). We observe a first wave that is relatively homogeneous among all countries (if properly normalized) and that could be considered strongly mitigated around May-June everywhere (see Section~\ref{sct:averaged}). Other waves have later developed, which are more heterogeneous because they reflect each country's different responses to the epidemics. A superposition of individual elementary peaks has been used to model these ulterior waves. 
Even if SHR merely amounts to a precise fit of the data, we observe that it carries significant advantages over the mere manipulation of the data series, $\{ c_i \}$, as: (i) it can be extrapolated
to the near future (extrapolations should be treated with great care, but an informed
extrapolation about the behavior in the future is always better than
a wild guess) and, (ii) it reduces long lists of numbers to an
analytical expression which only depends on three parameters. Such an analytical function
can be then easily manipulated to get integrals, derivatives of any order, or to search
for extrema/inflexion points, etc.


\subsubsection{Spain.} 
\begin{figure}[!t]
\includegraphics[width=0.49\columnwidth]{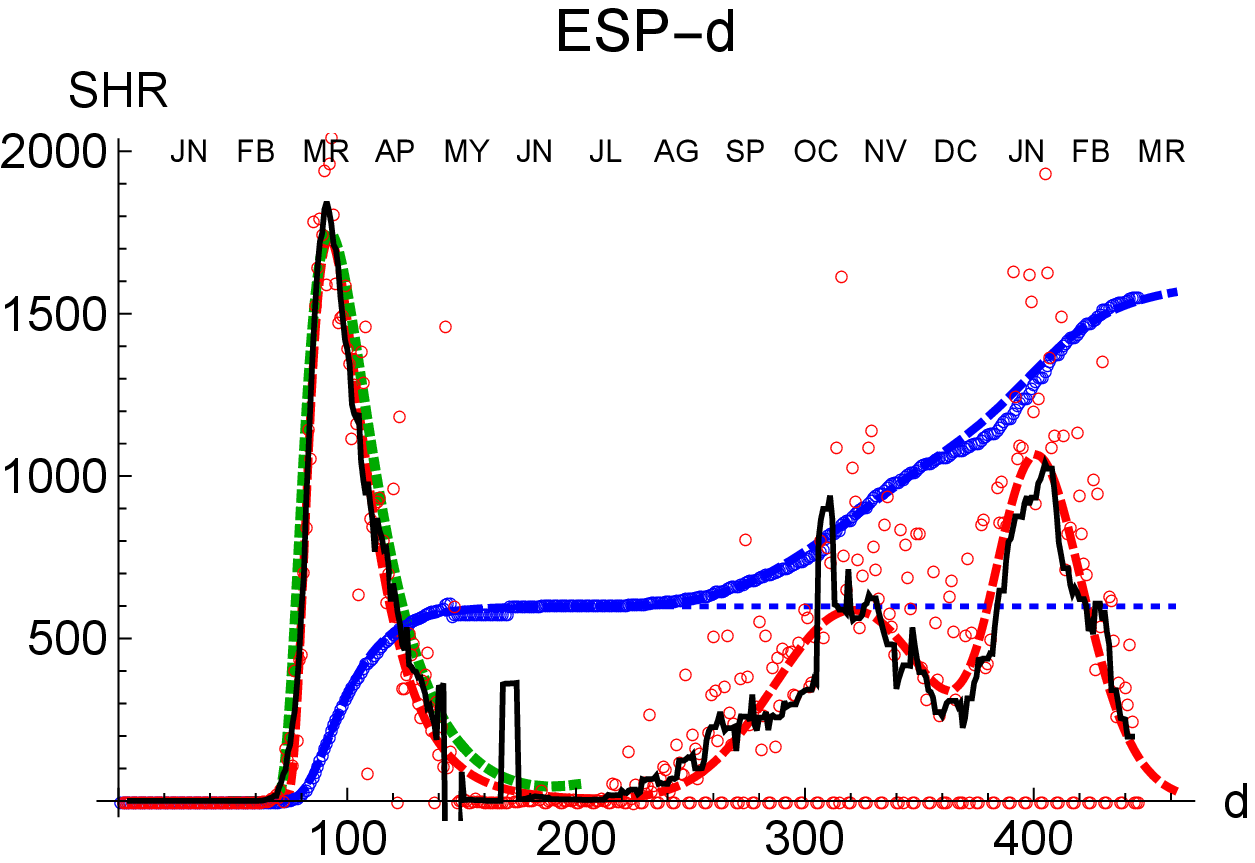}
\includegraphics[width=0.49\columnwidth]{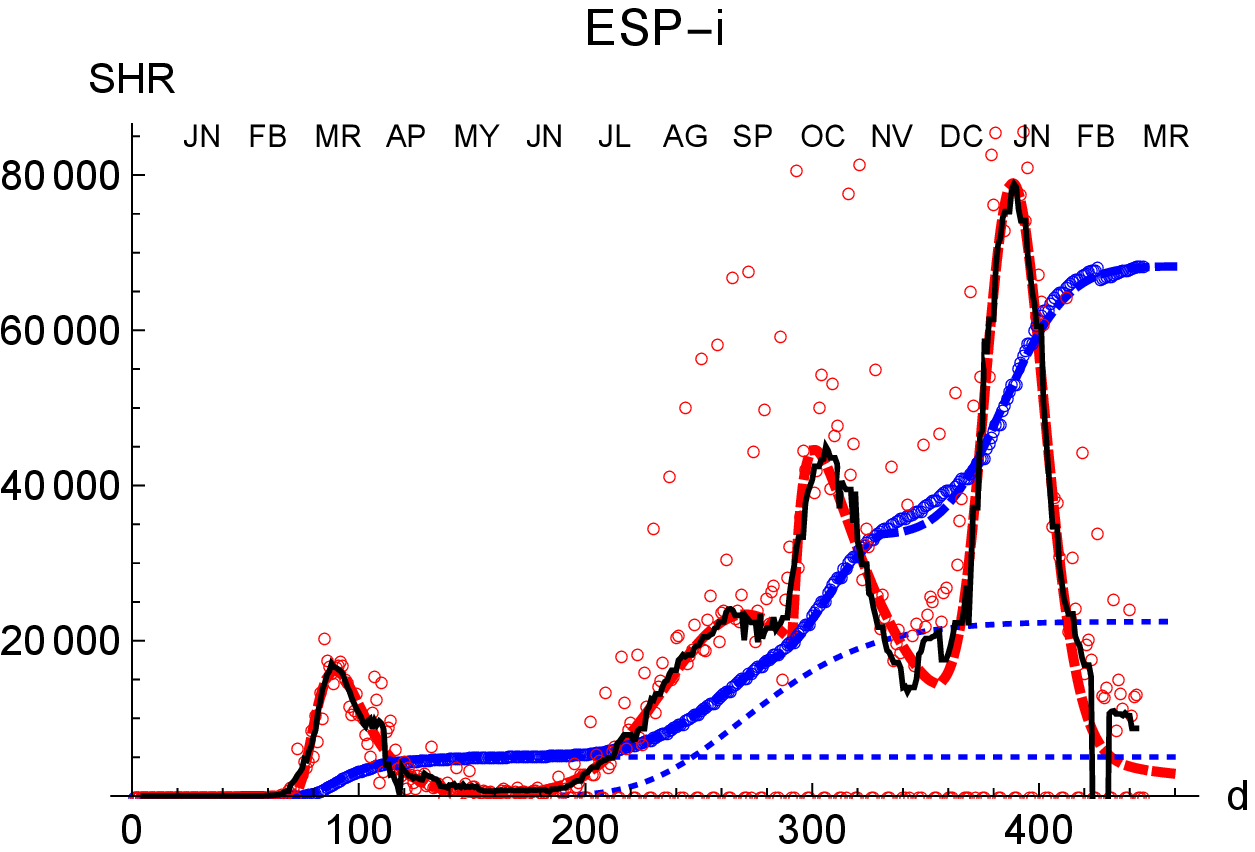}
\smallskip \caption{
{\bf SHR/Spain.}
Left/Right panels: deaths/infections related to COVID-19. 
Data (circles) are taken from~\cite{owidcoronavirus,johnshopkins}.
Dashed curves fit the data using Eqs~\ref{eqn:PDF} and~\ref{eqn:CDF}.
Blue: total accumulated cases per million inhabitant.
Red: daily cases per one hundred million inhabitants (the factor $\times 100$ 
is introduced for the sake of better visibility
on the scale of total cases only).
The black thin line is a 7-day moving average of data.
The green dashed line is the averaged representative curve discussed
in section~\ref{sct:averaged}.
Red and blue thin dotted lines give the
contributions of individual waves. 
The inset (left) gives the percentage between deaths and infections
from March to September. 
The inset (right) enlarges region II where a 
nearly constant number of infections takes place
(red: least-squares fit to data and constant mean value. Black: 7-day moving average).
Changes in data collection methodology
took place on April 19th, April 25th and November 4th, producing discontinuities on the data.
}
\label{fgr:shrESP}
\end{figure}

Spain is a country where the disease was particularly virulent
in its first wave,
spreading with remarkable strength.
The SHR model agrees well with the data for both deaths and infections (Fig.~\ref{fgr:shrESP} and Tables~\ref{tbl:SHRd} and~\ref{tbl:SHRi}~; a common colour code is applied in Figs. 1, 2, 3, 4, 8, 9, 10 and 11 to facilitate comparisons: red is used for daily cases (empty dots for actual data and dashed/dotted lines for models), black for 7-days moving averages of daily cases, and blue is used for accumulated cases (again, empty dots for actual data and dashed/dotted lines for models). In Fig. 11 we have used red vs magenta for daily cases and blue vs cyan for accumulated cases to allow easy comparison between models. Other details are given in the figure captions.
Together, these two variables provide a better idea of the epidemic's course by identifying two critical items: the impact on the population via infections and, the impact on the health system via deaths. 
Three simple features defining the epidemics that will be rationalized later in the context of the SIR model are: (i) the exponential behavior near the origin, (ii) the position and value of the single maximum in the daily number of cases and, (iii) 
an asymmetric decay towards the future w.r.t the past. 
The ratio between total infections and deaths has evolved
from about 1\% in March to a maximum of 12\% in August, but it has significantly decayed for the
second wave to about 4\% at the end of September (inset in left-hand side in Fig.~\ref{fgr:shrESP}).

Three regions are identified in the plots, both for
deaths and infections. The first region (I, wave 1) finishes approximately on the 1st of May, 2020 ($d=152$) and it clearly shows the three aforementioned features marking its association with an infectious disease.
The second region (II, inset in right-hand side in Fig.~\ref{fgr:shrESP}), 
goes approximately between the 150th and 200th day, 
and its hallmark is to sustain a fairly constant level of daily infections, $c(t)=<c_i>$, which reflects
in a linear increase of the accumulated number of cases, $C(t)$. 
Region II approximately terminates near the end of the general lockdown in Spain, on the
21st of June ($d=173$). 

Neither the SHR nor the SIR models can
account for a sustained period of constant infections, 
although they can accommodate this regime via the slowly decaying queue of the distribution where derivative of the function is very low. In contrast, such behavior can be naturally described via Monte-Carlo simulation. 
Likewise, while MC can describe several waves by producing more than one local maxima due to spatial inhomogeneous dynamics,
SHR and SIR can only describe such a scenario via a linear combination of 
individual waves, each one governed with its own parameters. 

Finally, in a third region (III, wave 2) the collective transmission displays again a similar behavior to the region I, marking the evolution of an out-of-control disease.
The superposition of these multiple regimes, plus other waves if needed, describes the overall function well. 
We notice that the accuracy in the fit for any wave is not expected to be 
reasonably stable
until at least the corresponding maximum is well developed (Section~\ref{sct:accuracySHR}).
However such incertitude, the model predicts that in Spain the number of infections due to the second wave should be reaching its maximum in December 2020, at most in January 2021.\footnote{Assuming fixed conditions, in particular disregarding the possible effect of Christmas festivities.} 
In addition,
the model predicts that the strength of the second wave
is approximately weaker than the first one by a factor two, as measured by the
number of accumulated certified deaths from SARS-Cov-2.
Although these predictions may be affected by large error bars since the maximum in the second wave is not yet well developped, those values offer sound guidance about the course of the disease. We have used this model to extrapolate the shape of the curve by a fortnight after the last day of the corresponding available data;
\footnote{On the 4th of November, $1385$ previously unaccounted deaths were added in Spain, producing a discontinuity in the curve. To make possible a prediction based on an extrapolation, we have not included those extra deaths.}
the resemblance to the ulterior course of the disease will be seen in the next weeks. 

The accompanying number of registered infections yield a picture of the 
likely evolution of deaths in the following days, even if the variation in
the absolute numbers from the first to the second wave is dominated by the change in the number of tests performed.
Given the large dispersion of raw data due to difficulties to collect them it is clear the necessity to perform moving averages and the advantages of working with least-square approximants that can be extrapolated a few days ahead, a statement that is true for the behaviour of other countries.
While deaths only show two waves so far, infections identify at least four local maxima that can be correlated with different events, like the end of the summer vacations or the occurrence of several bank holidays in Spain where the population has been moving and mixing in great numbers.

\subsubsection{Germany.} 
Compared with other countries with large populations, Germany has managed the pandemics quite well, as it is observed by comparing the number of cases
per inhabitant. Moreover, its evolution has been recorded with consistency both for deaths and infections. Therefore, it is an appropriate benchmark for any model. 

\begin{figure}[!t]
\includegraphics[width=0.49\columnwidth]{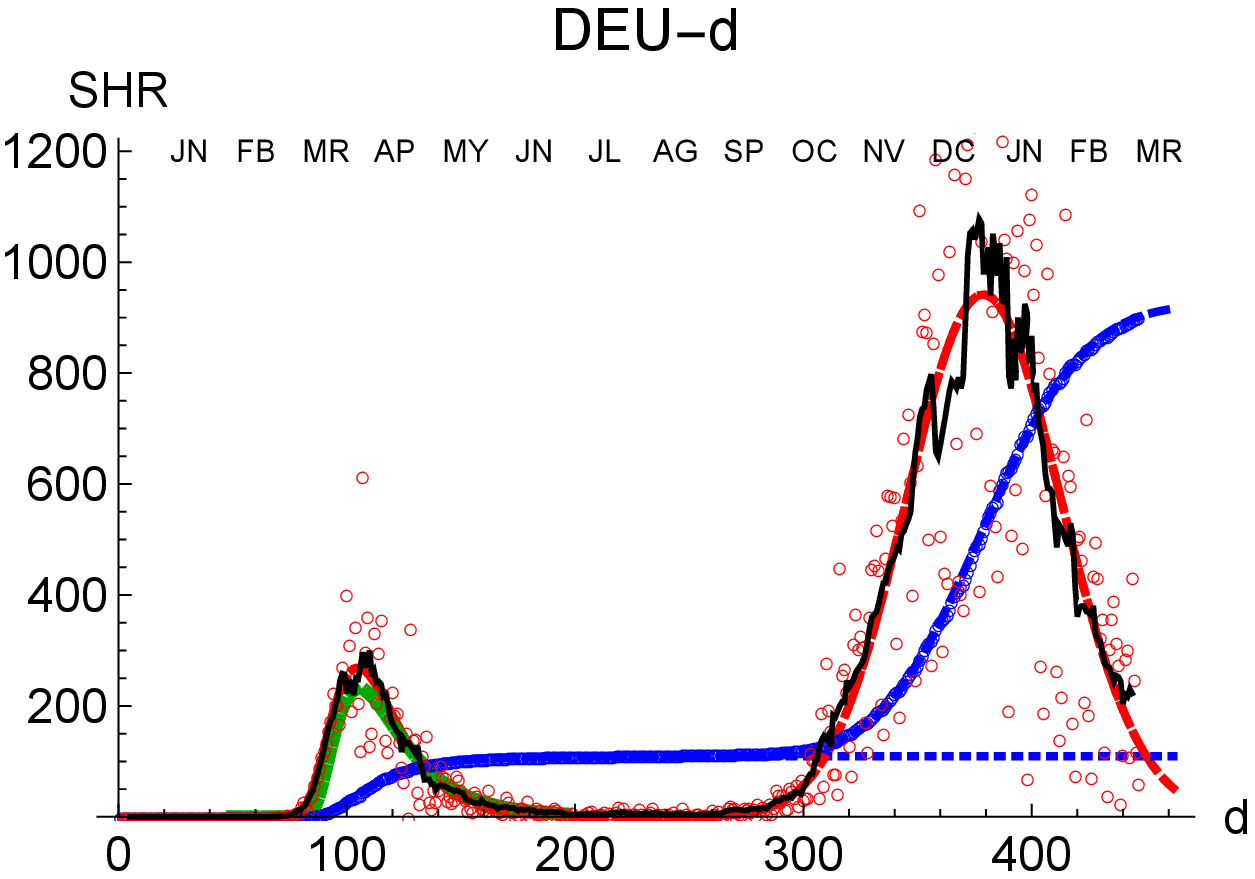}
\includegraphics[width=0.49\columnwidth]{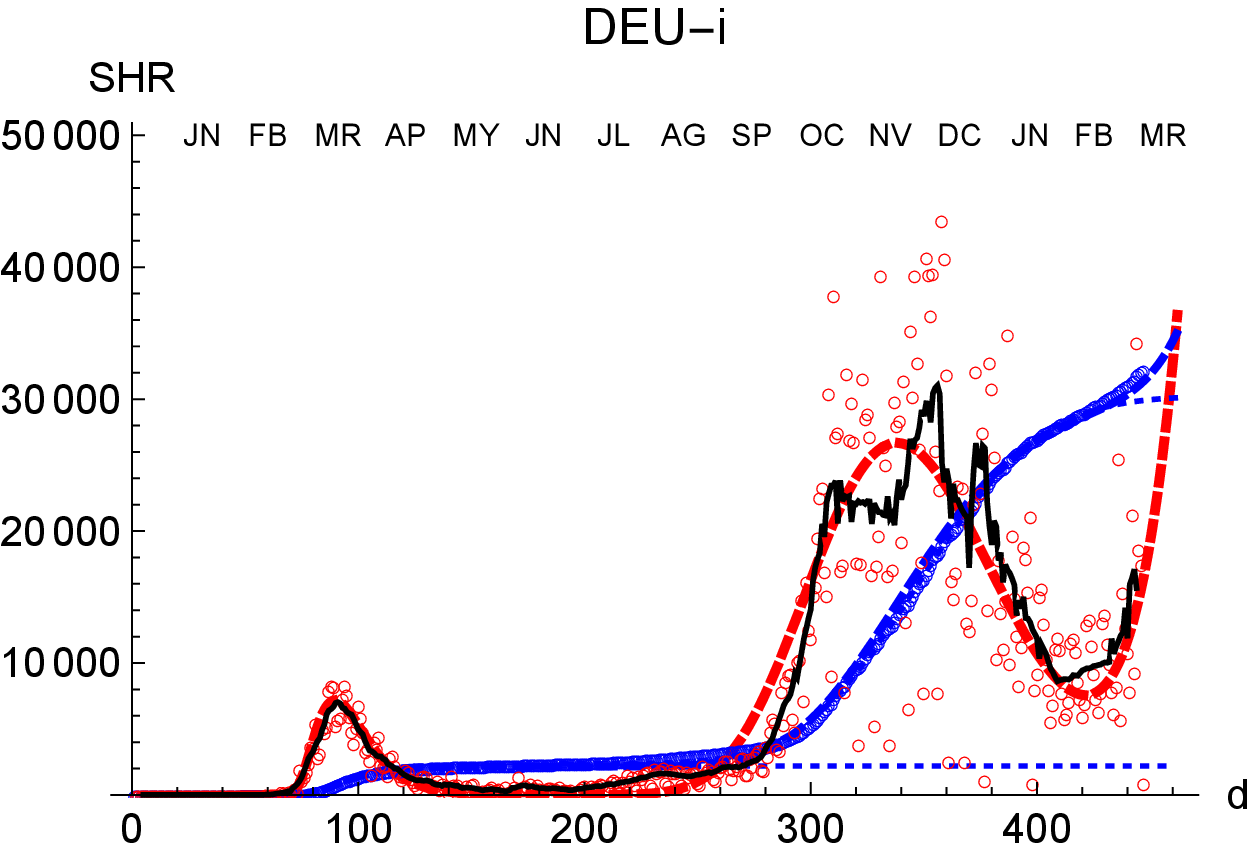}
\smallskip \caption{
{\bf SHR/Germany.}
Symbols as in Fig.~\ref{fgr:shrESP}. 
}
\label{fgr:shrDEU}
\end{figure}

Similarly to Spain, the SHR model can be used to accurately represent the disease evolution using only three parameters per wave (Fig.~\ref{fgr:shrDEU}). 
Curiously enough, best-fit values for $\mu$ and $\sigma$ are quite similar to Spain
(Tables~\ref{tbl:SHRd} and~\ref{tbl:SHRi}), indicating that, independently of the absolute strength, there are common underlying features in both cases.
Therefore, it is interesting to explore the ability
of a {\it single normalized averaged} curve to represent such contrasting cases as Spain and Germany, using $C(\infty) \propto c_M$ as the single only free parameter. 
Such a curve is represented in Figs.~\ref{fgr:shrESP} and~\ref{fgr:shrDEU} 
by the green dashed line 
having $\mu=3.53$ and $\sigma=0.56$
(Section~\ref{sct:averaged}), and it is clear
that despite having such a limited freedom 
for fitting
(since it only depends on one parameter), it provides a very reasonable approximation to the data.
In contrast to Spain, the ratio between total infections and deaths in Germany evolved from about 1\% in April to 4\% in September, which is about three times lower than for Spain (inset in Fig.~\ref{fgr:shrDEU}).


\subsubsection{Other countries.}
\begin{figure}[!t]
\includegraphics[width=0.46\columnwidth]{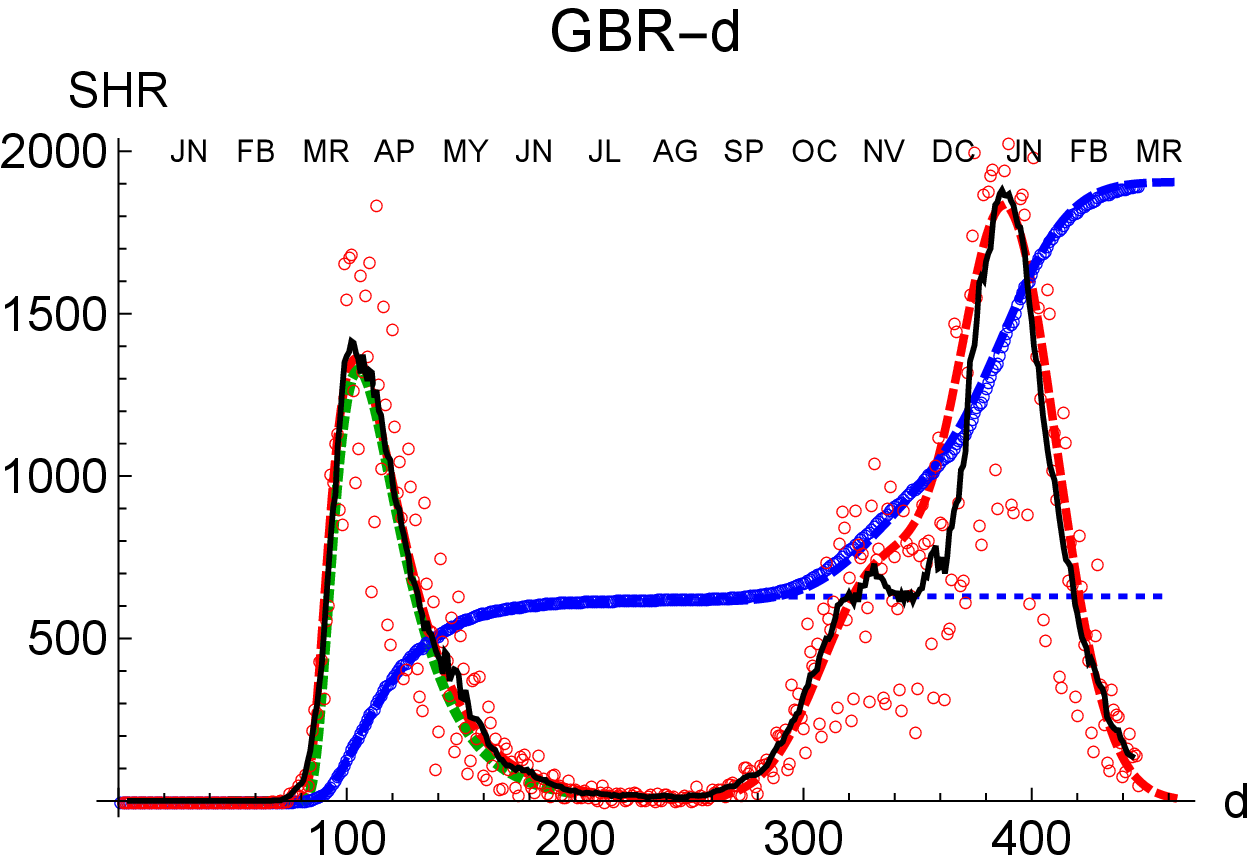}
\includegraphics[width=0.46\columnwidth]{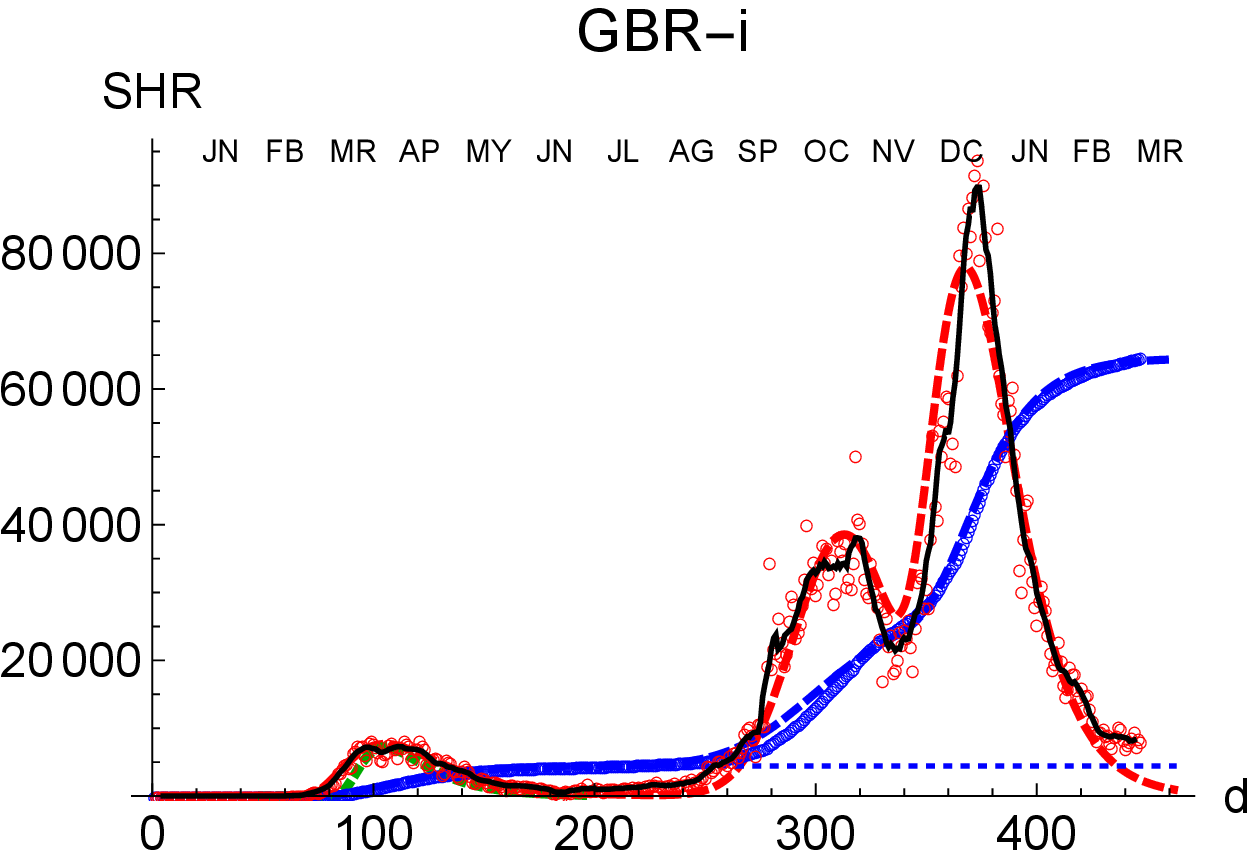} \\
\includegraphics[width=0.46\columnwidth]{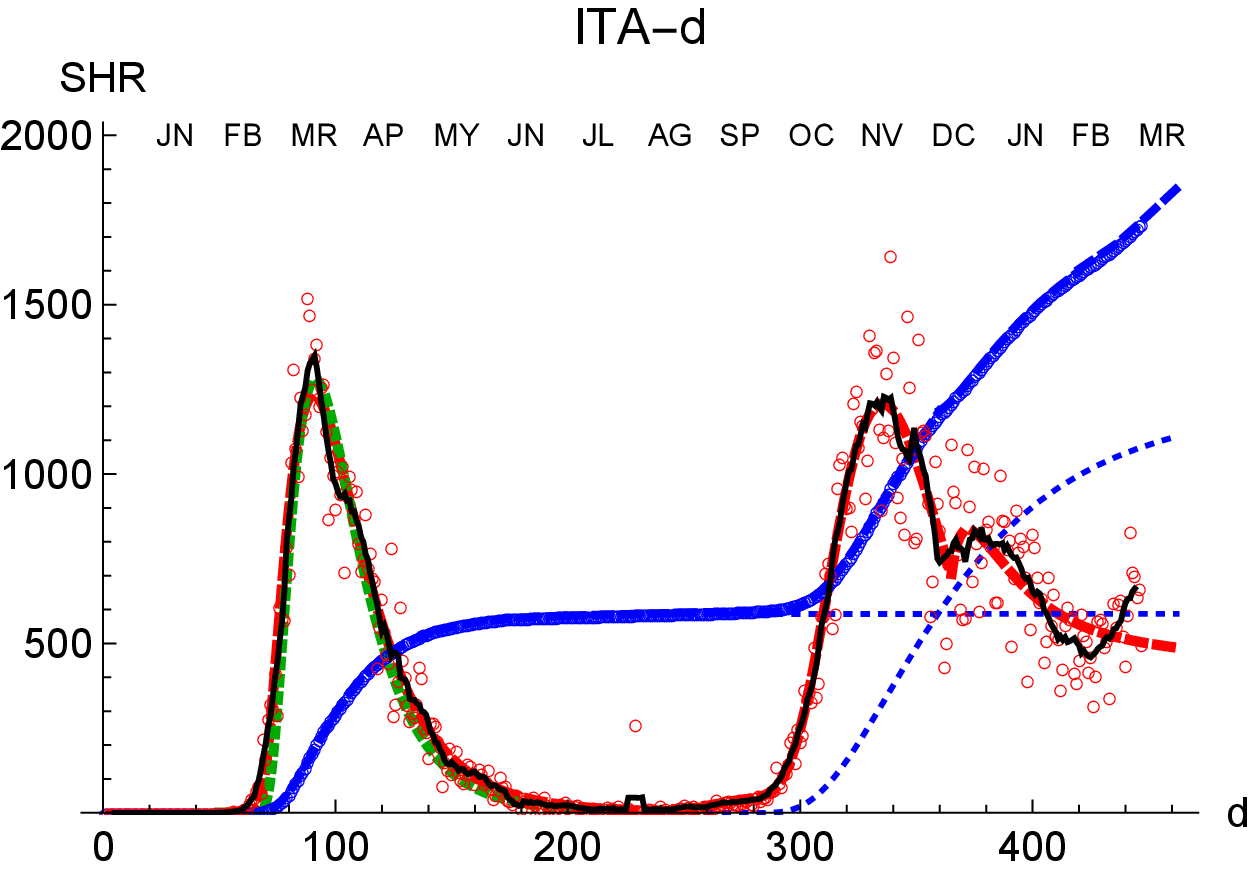}
\includegraphics[width=0.46\columnwidth]{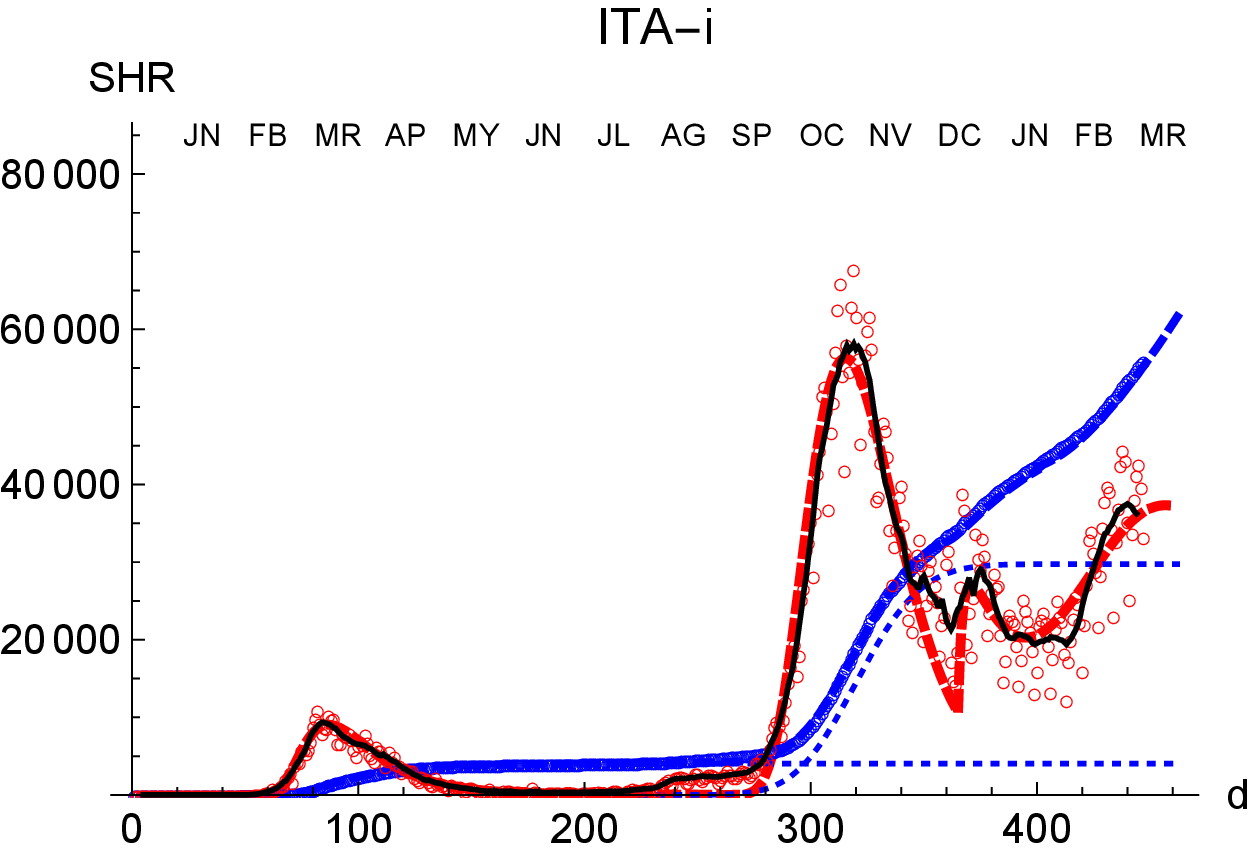} \\
\includegraphics[width=0.46\columnwidth]{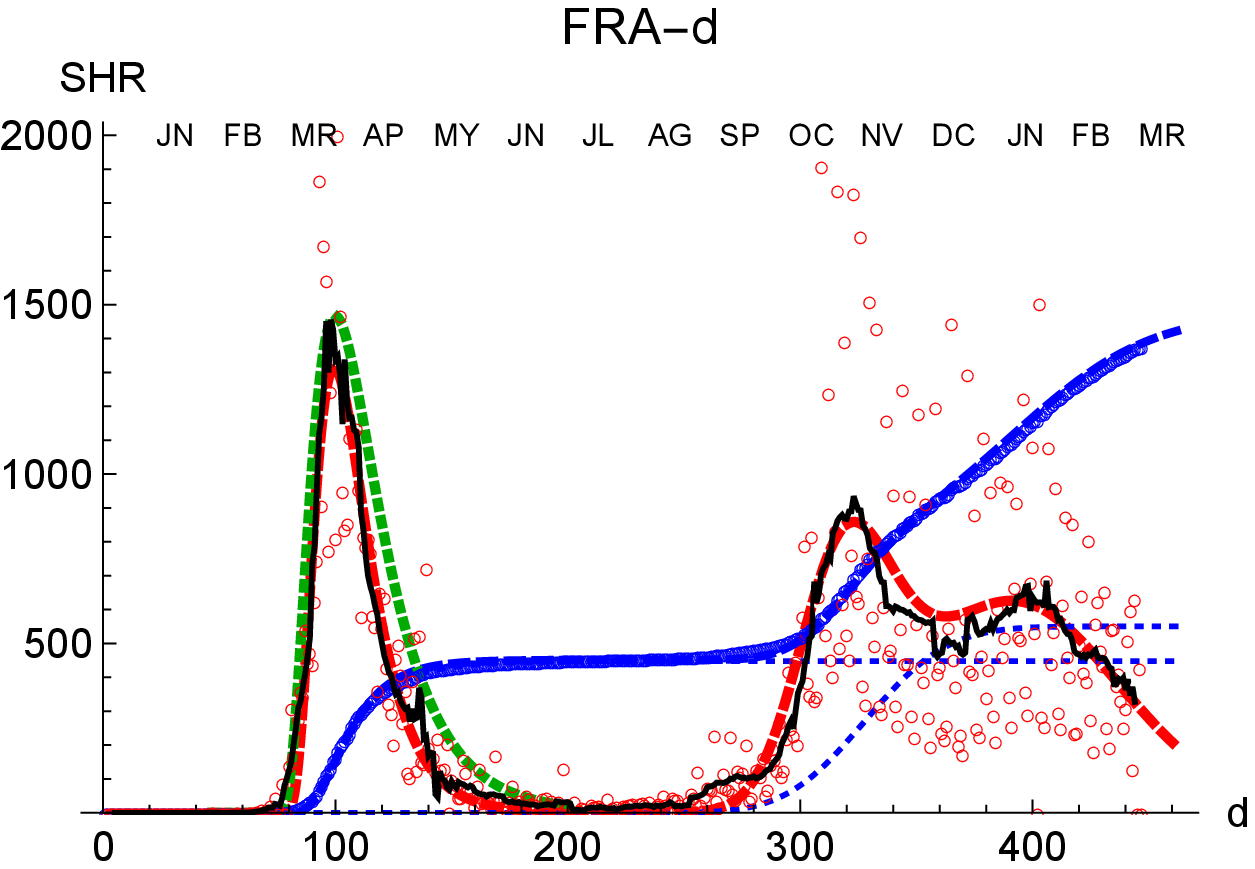}
\includegraphics[width=0.46\columnwidth]{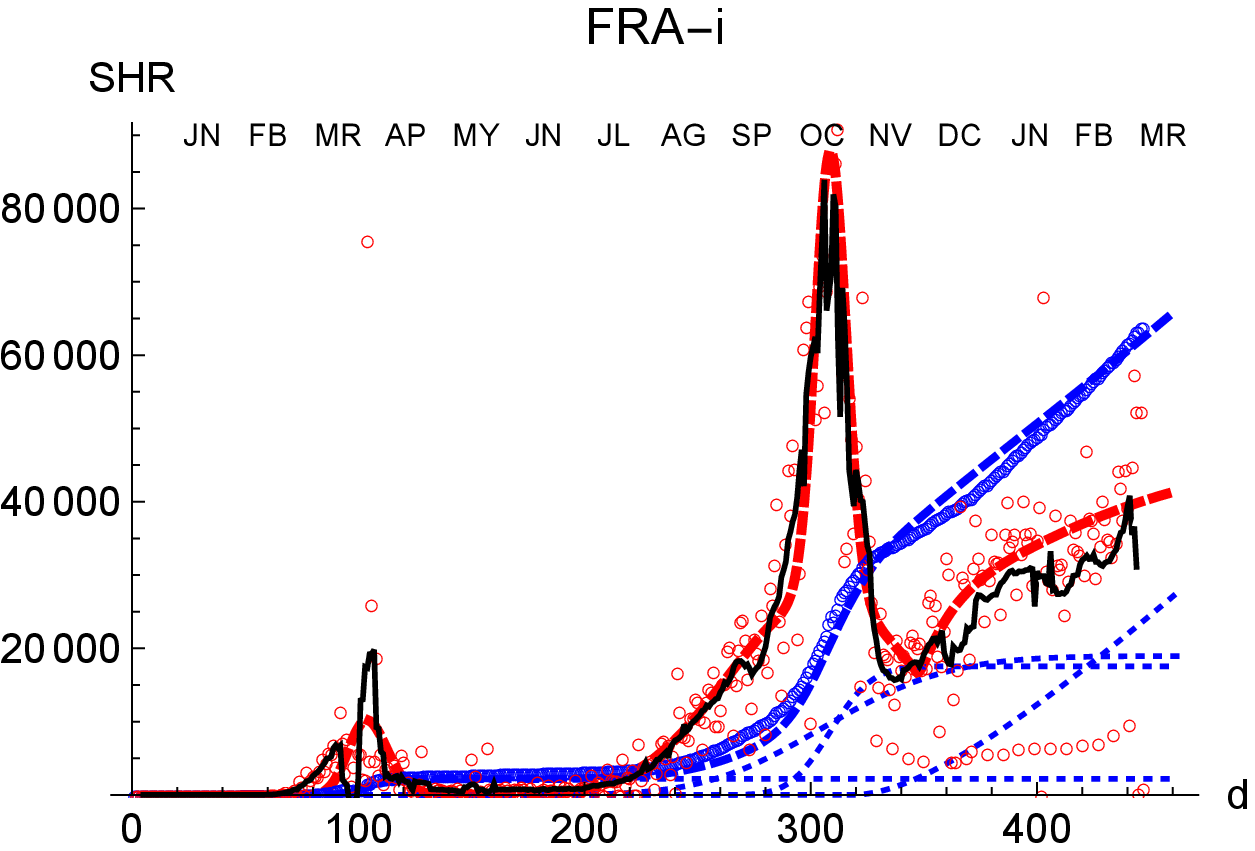} \\
\includegraphics[width=0.46\columnwidth]{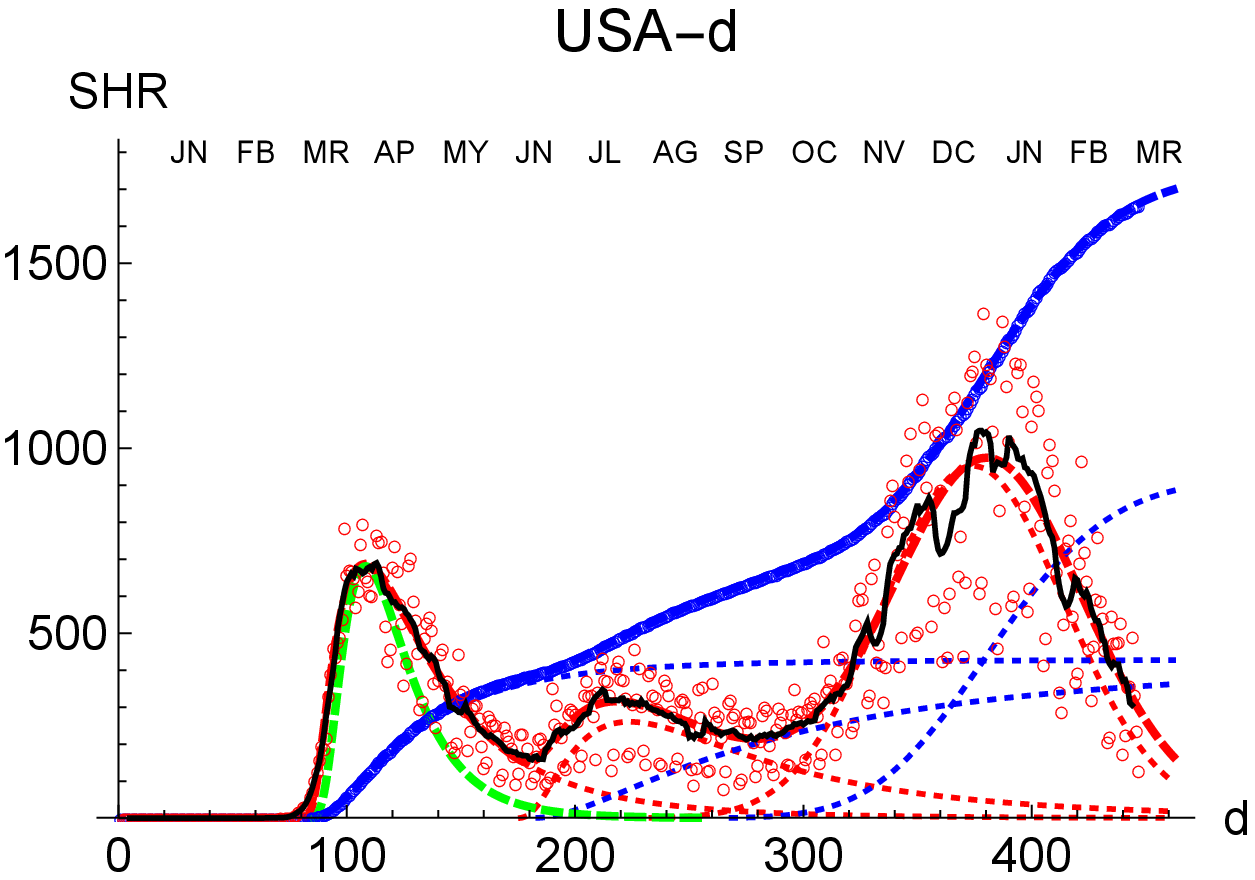}
\includegraphics[width=0.46\columnwidth]{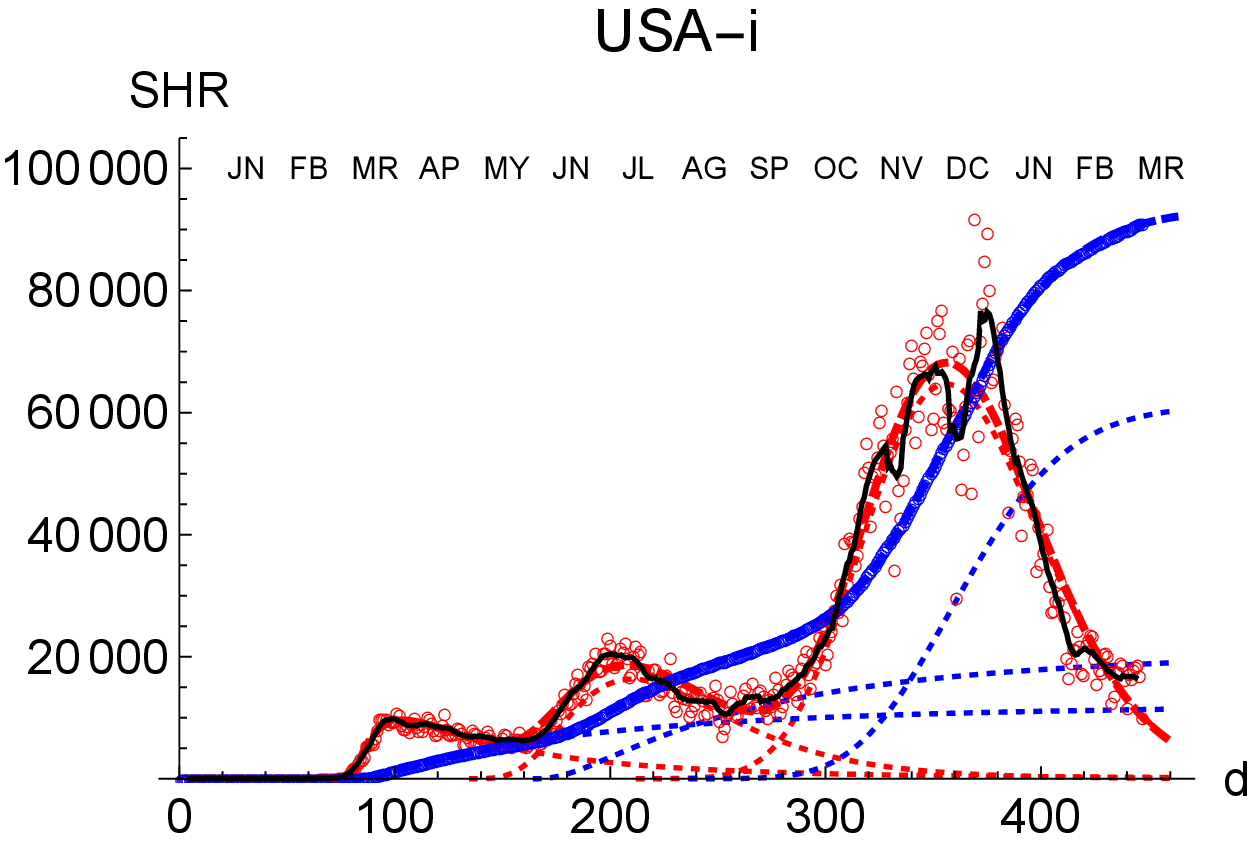} \\
\smallskip \caption{
{\bf SHR/Other countries (I).}
Symbols as in Fig.~\ref{fgr:shrESP}.
Changes in methodology took place 
in United Kingdom (GRB) on May 20th and July 3rd, and
in France (FRA) on May 28th.
}
\label{fgr:shrS1}
\end{figure}

\begin{figure}[!t]
\includegraphics[width=0.46\columnwidth]{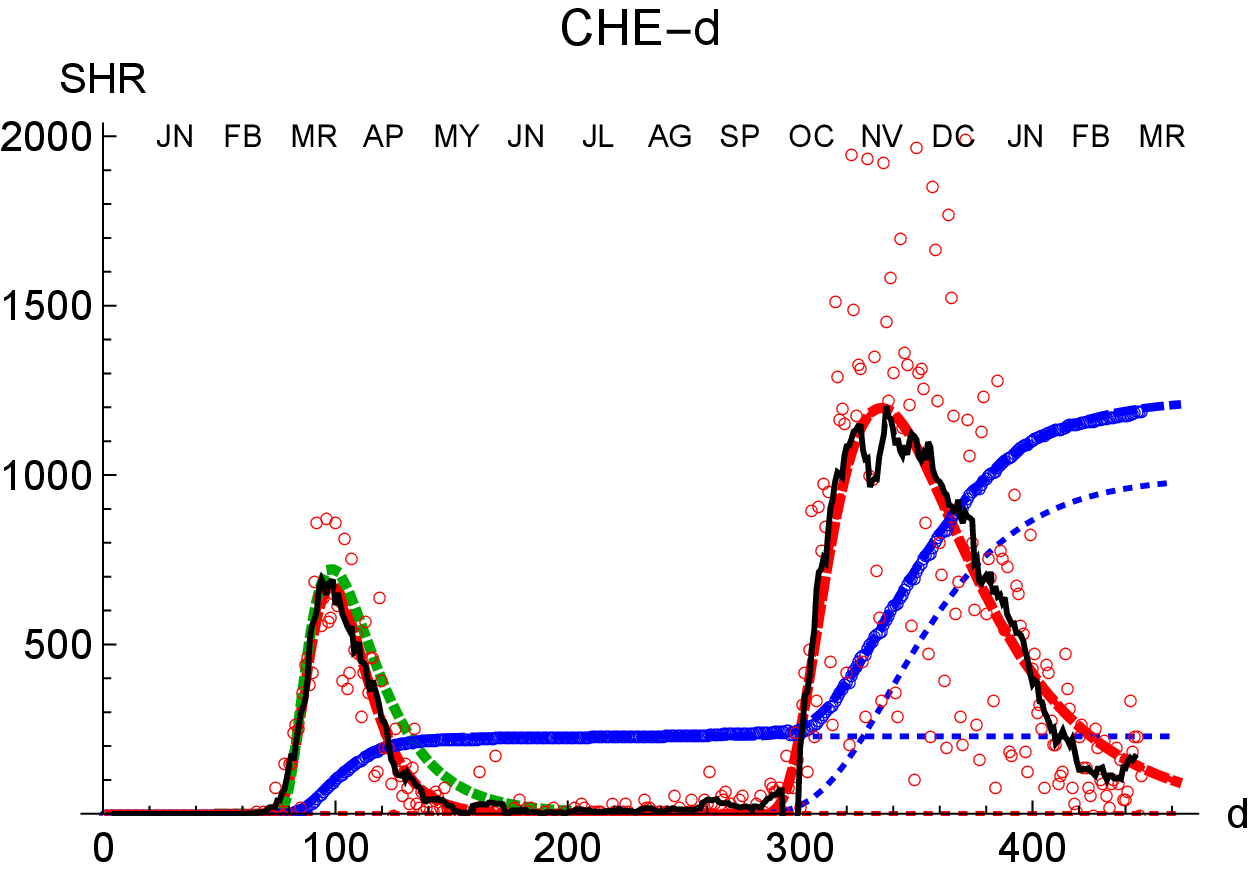}
\includegraphics[width=0.46\columnwidth]{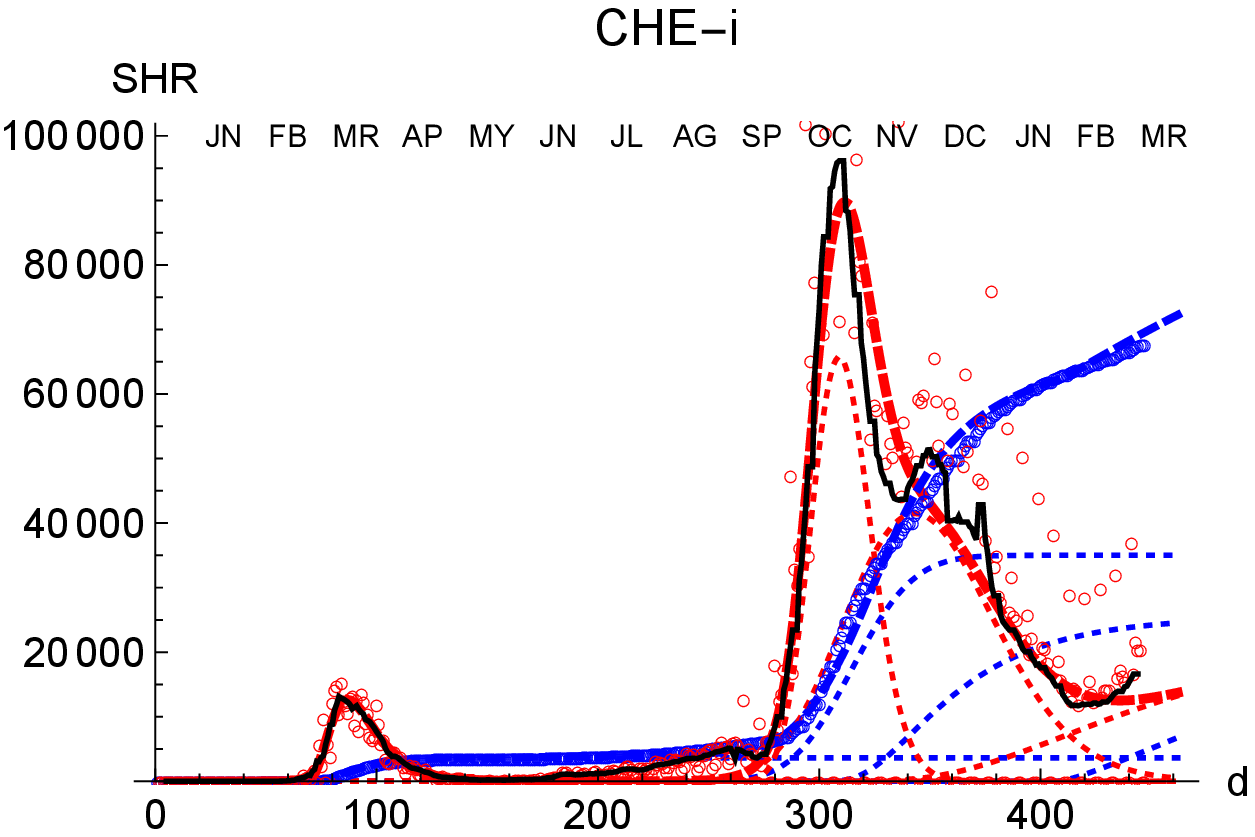} \\
\includegraphics[width=0.46\columnwidth]{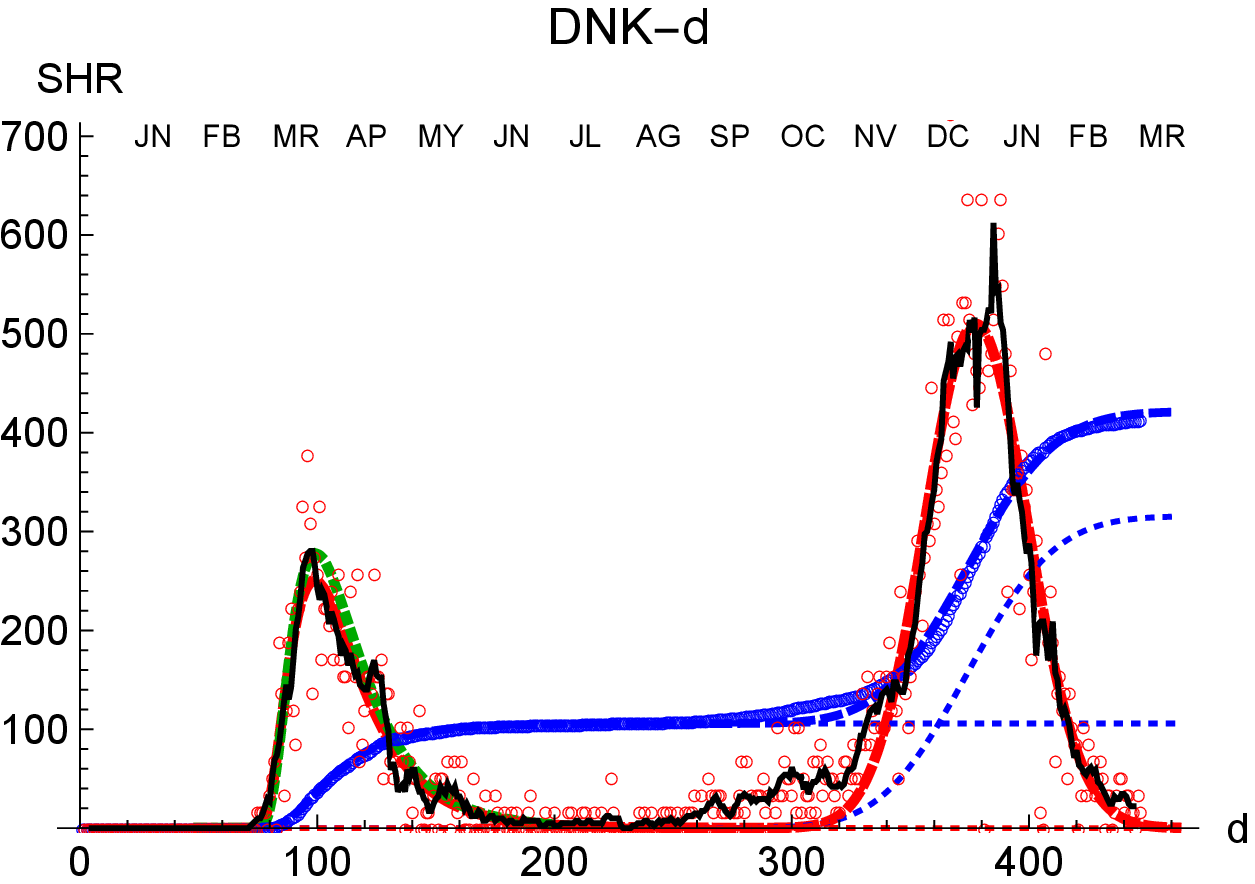}
\includegraphics[width=0.46\columnwidth]{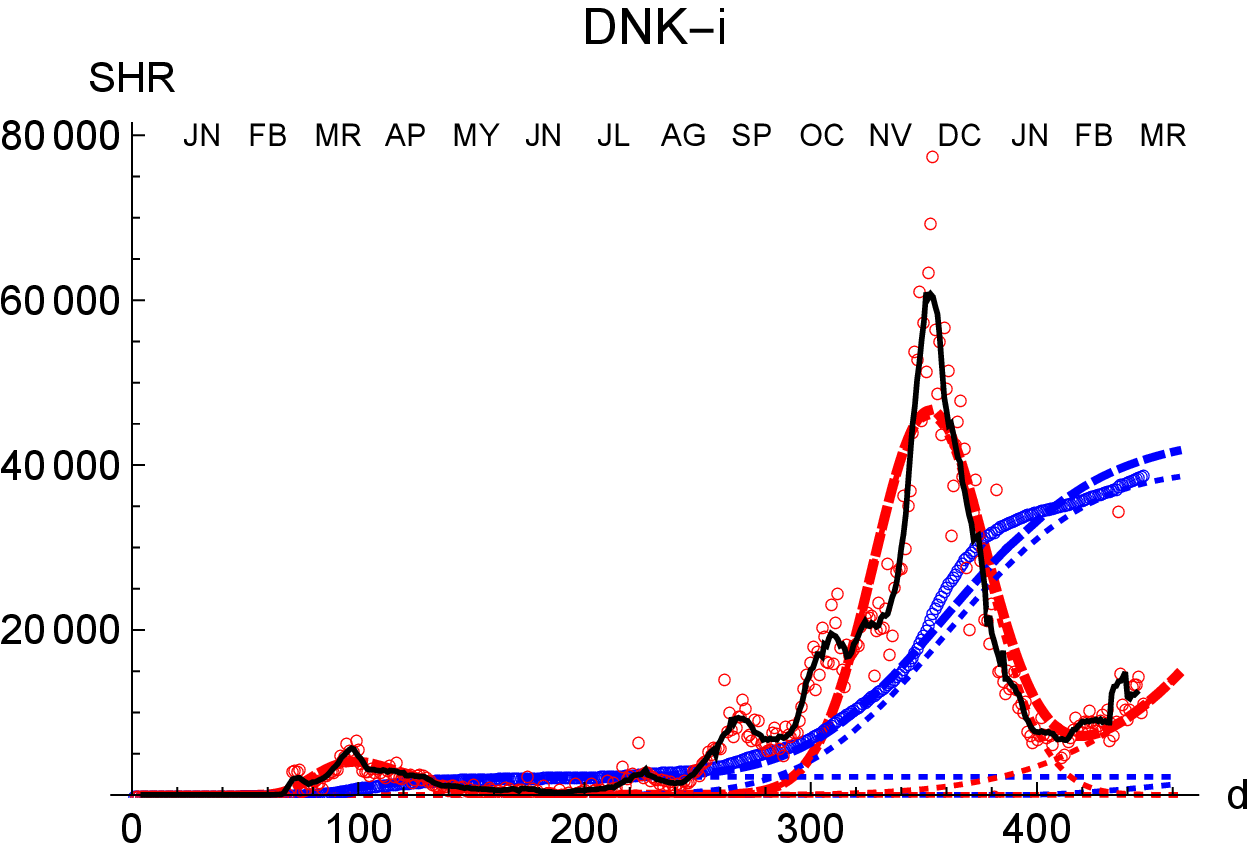} \\
\includegraphics[width=0.46\columnwidth]{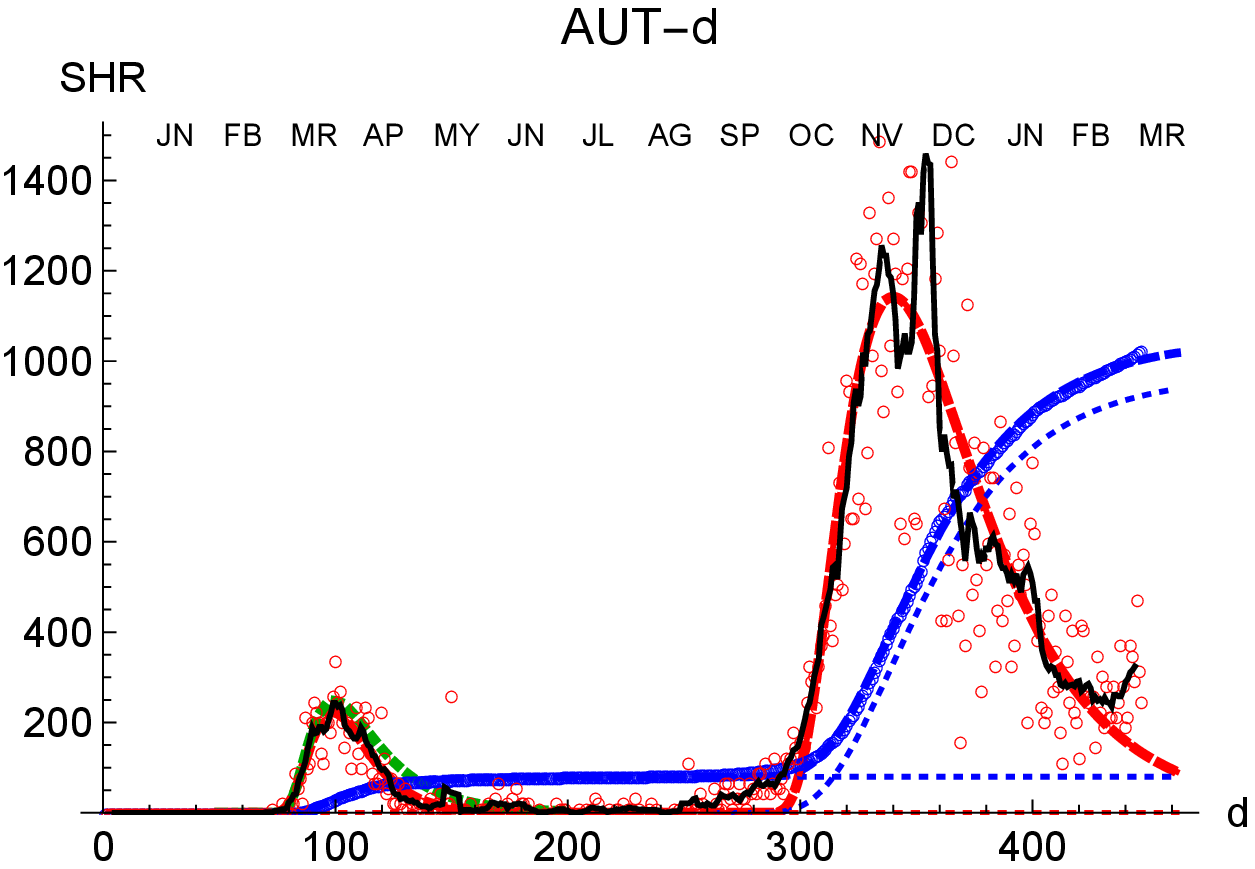}
\includegraphics[width=0.46\columnwidth]{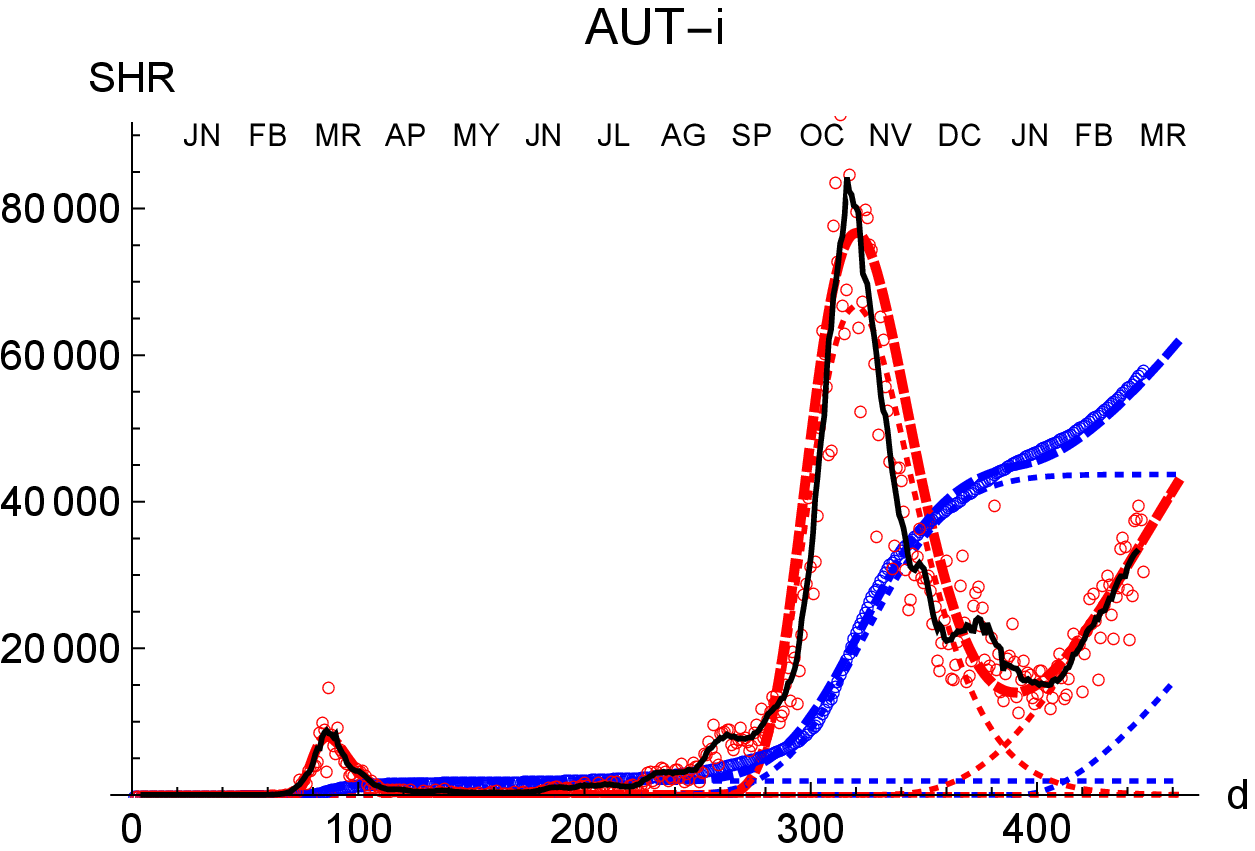} \\
\includegraphics[width=0.46\columnwidth]{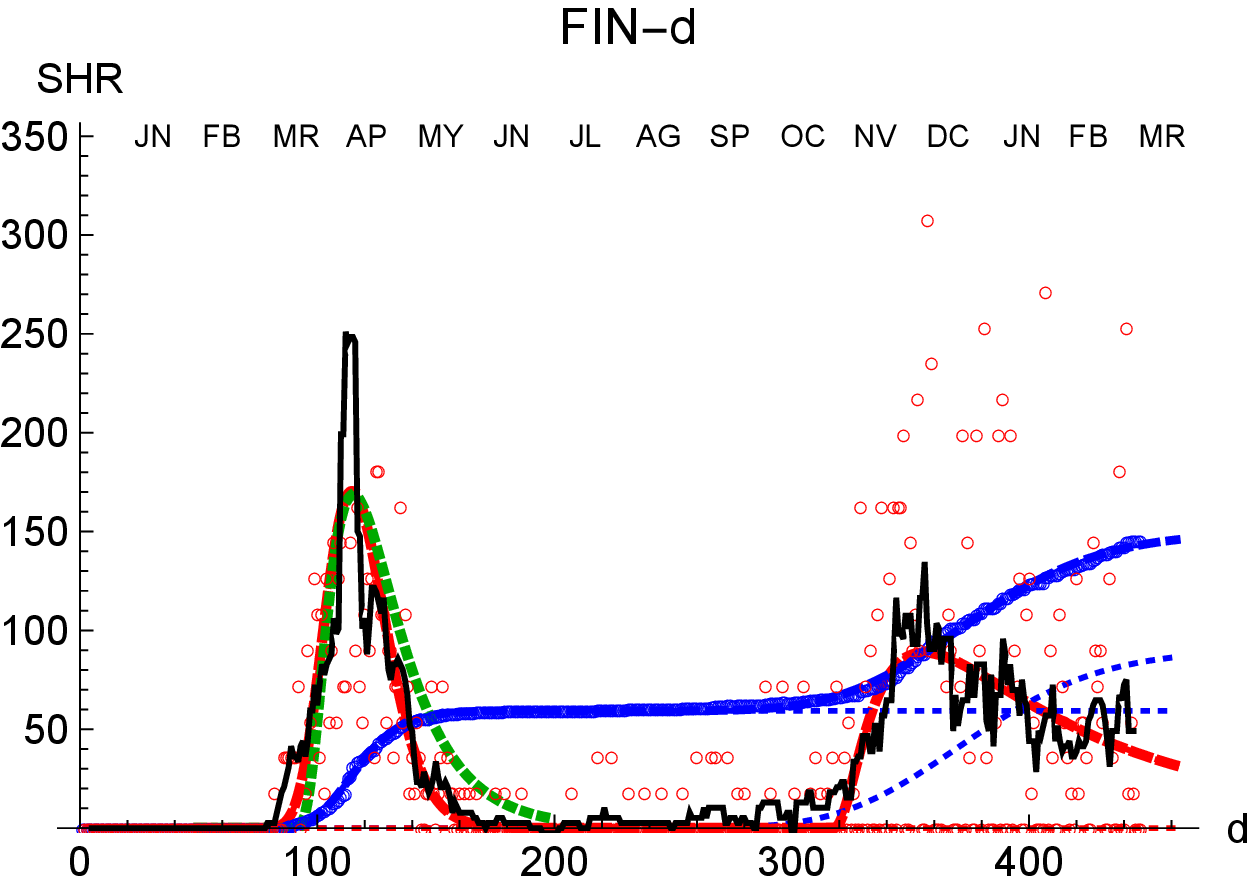}
\includegraphics[width=0.46\columnwidth]{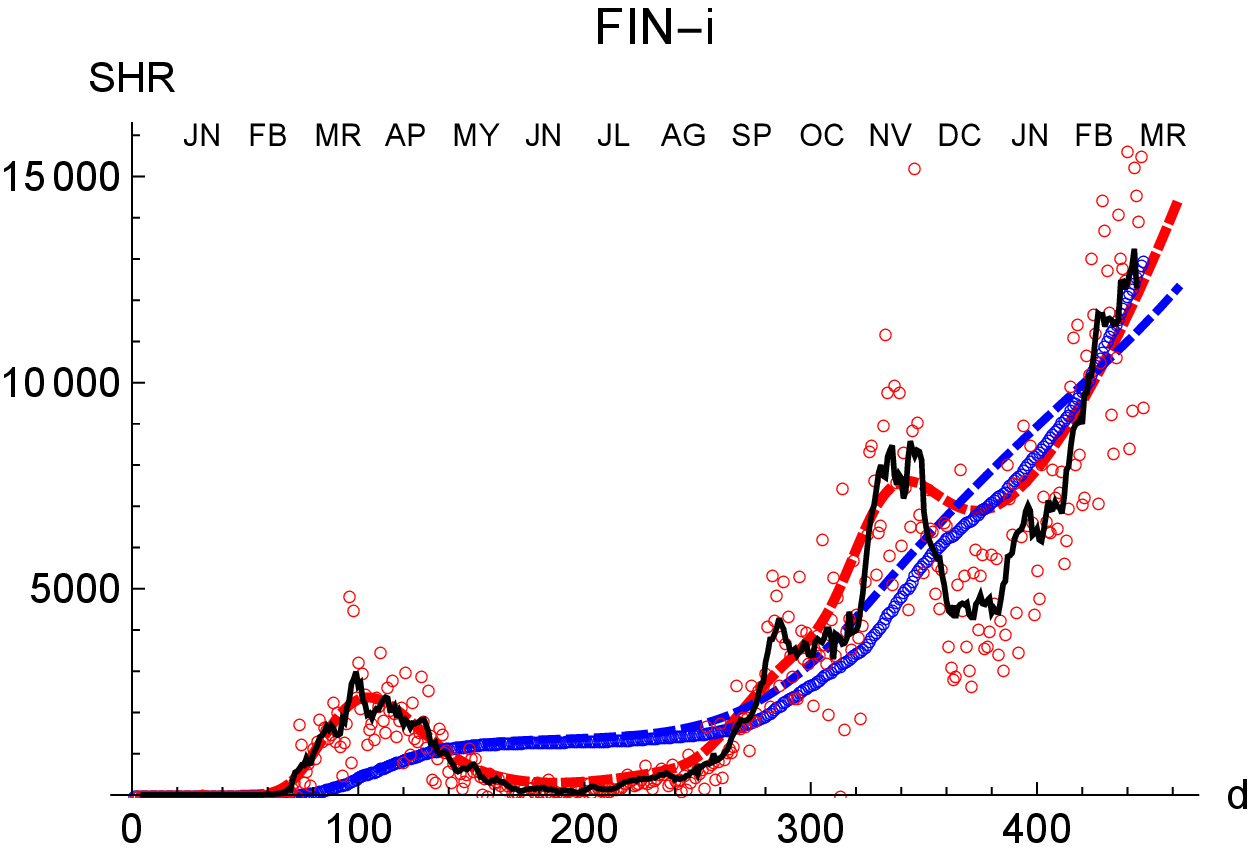} \\
\smallskip \caption{
{\bf SHR/Other countries (II).}
Symbols as in Fig.~\ref{fgr:shrESP}.
}
\label{fgr:shrS2}
\end{figure}

We also prove the capabilities and versatility of the SHR ansatz to reproduce the observed data 
by applying the same methodology to a pool
of western countries: Great Britain (GBR), 
Italy (ITA), United States (USA), France (FRA),
Switzerland (CHE), Denmark (DNK),
Austria (AUT) and Finland (FIN), cf. Figs.~\ref{fgr:shrS1} and~\ref{fgr:shrS2}.
In general, the agreement is quite good, both for deaths and infections. 
Among other advantages, 
this procedure allows a quick and simple monitoring of the evolution of the disease in the
different countries. In particular, it is a useful tool to identify and forecast the appearance of a second wave.
At the moment of writing, only the USA has fully developed the
maximum associated with the second
wave and, from the combined behaviour of deaths and infections, it could be argued
that the country is clearly heading towards a third wave.
Since this is the only case so far, it is not possible to characterize
well such a second wave by a proper average of different countries,
although it seems fair to say that it is represented by a wider 
distribution of daily deaths (e.g. the second component represented by 
the dotted red curve corresponds to 
having $\mu=5.5$ and $\sigma=1.2$)
and a lower value at the peak by about a factor $2$.

\subsubsection{Regions: NYC vs Madrid.}
\begin{figure}[!t]
\includegraphics[width=0.95\columnwidth]{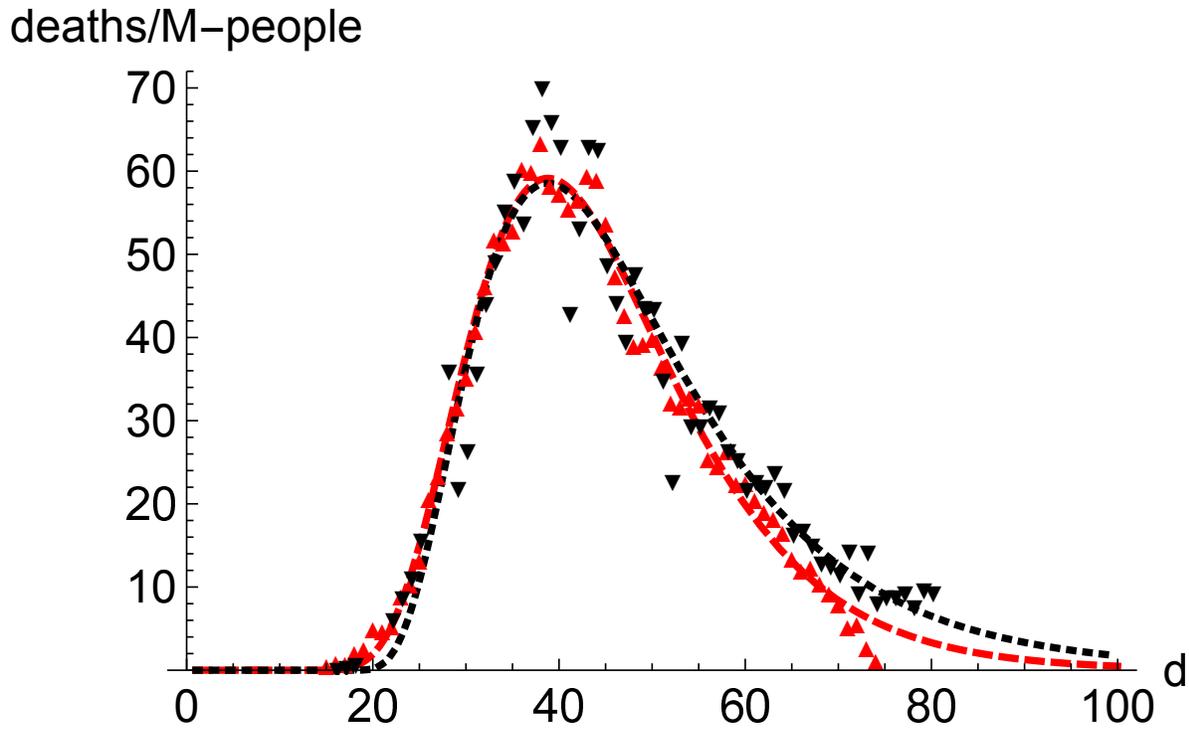}
\smallskip \caption{
{\bf Comparison of the evolution of number of deaths in NYC and the region of Madrid (CAM) during the first wave.} 
NYC (9.1 M people, red, down pointing triangles, $\triangledown$, dashed line) and Madrid (6.7 M people, black, upwards pointing triangles, $\triangle$, dotted line) during the first wave.
The data and SHR fits for both locations were juxtaposed matching the day with the maximum number of deaths, aiming to highlight the similarities. The values of the CAM were also scaled to the ratio of population between the two regions ($\frac{9.1}{7.6}$) to enable a better comparison.}
\label{fgr:D4}
\end{figure}

Prominent places where the infection spreads quickly are densely populated regions,
which constitute the core of the propagation of the disease.
Therefore, it is interesting to compare the distribution of cases
in those regions. 
We have juxtaposed the performance of New York City 
(9.1 M-people, NYC) and the Community of Madrid (6.7 M-people, CAM) in
the first wave (Fig.~\ref{fgr:D4}).
To highlight the similarities rather than the differences they are superimposed in such a way that the position (day) of the maximum coincides. Furthermore, CAM has been scaled by the ratio of respective populations, which makes the value at the maximum very similar for both regions. Despite all the differences between these regions, it is clear that a typical pattern emerges, which leads us to investigate the advantages of working with averages.


\subsubsection{Averaged profile.}
\label{sct:averaged}

\begin{figure}[!t]
\includegraphics[width=0.49\columnwidth]{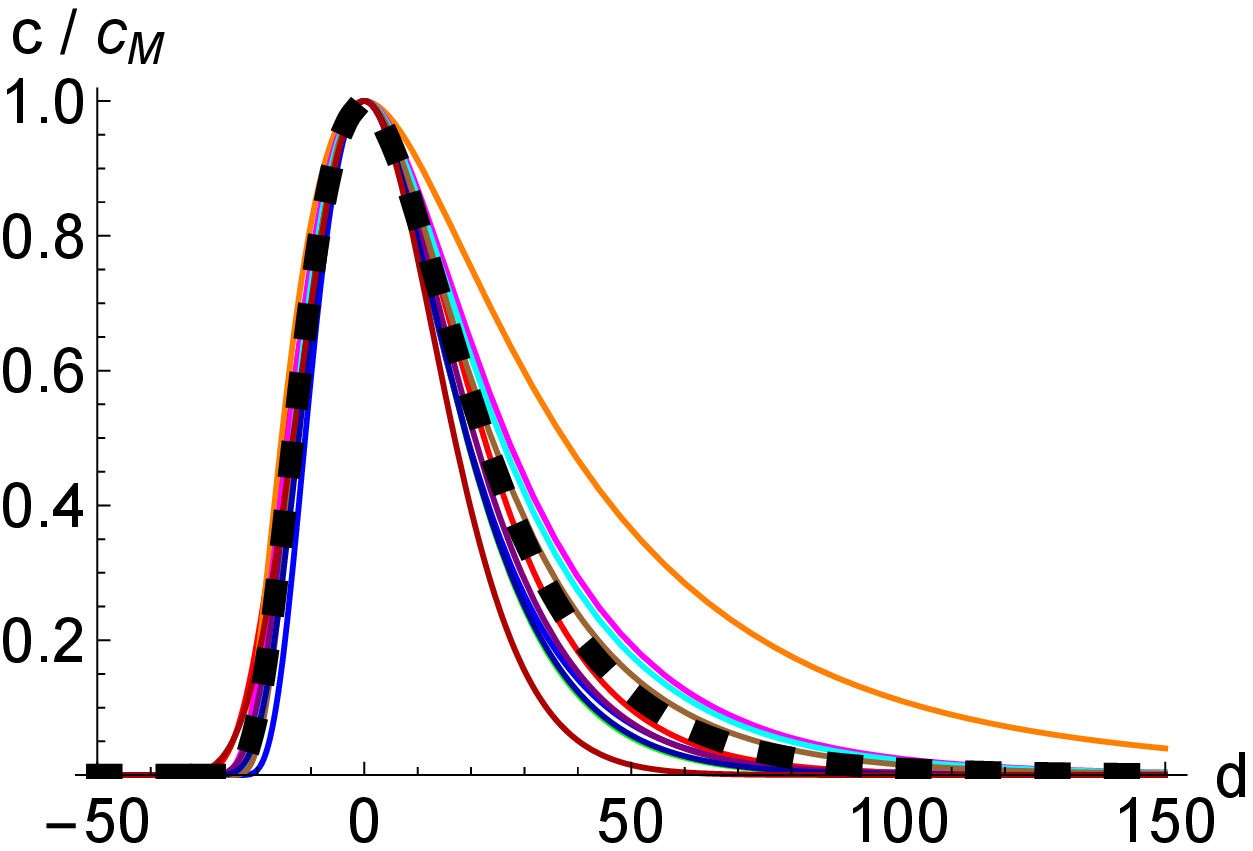}
\includegraphics[width=0.49\columnwidth]{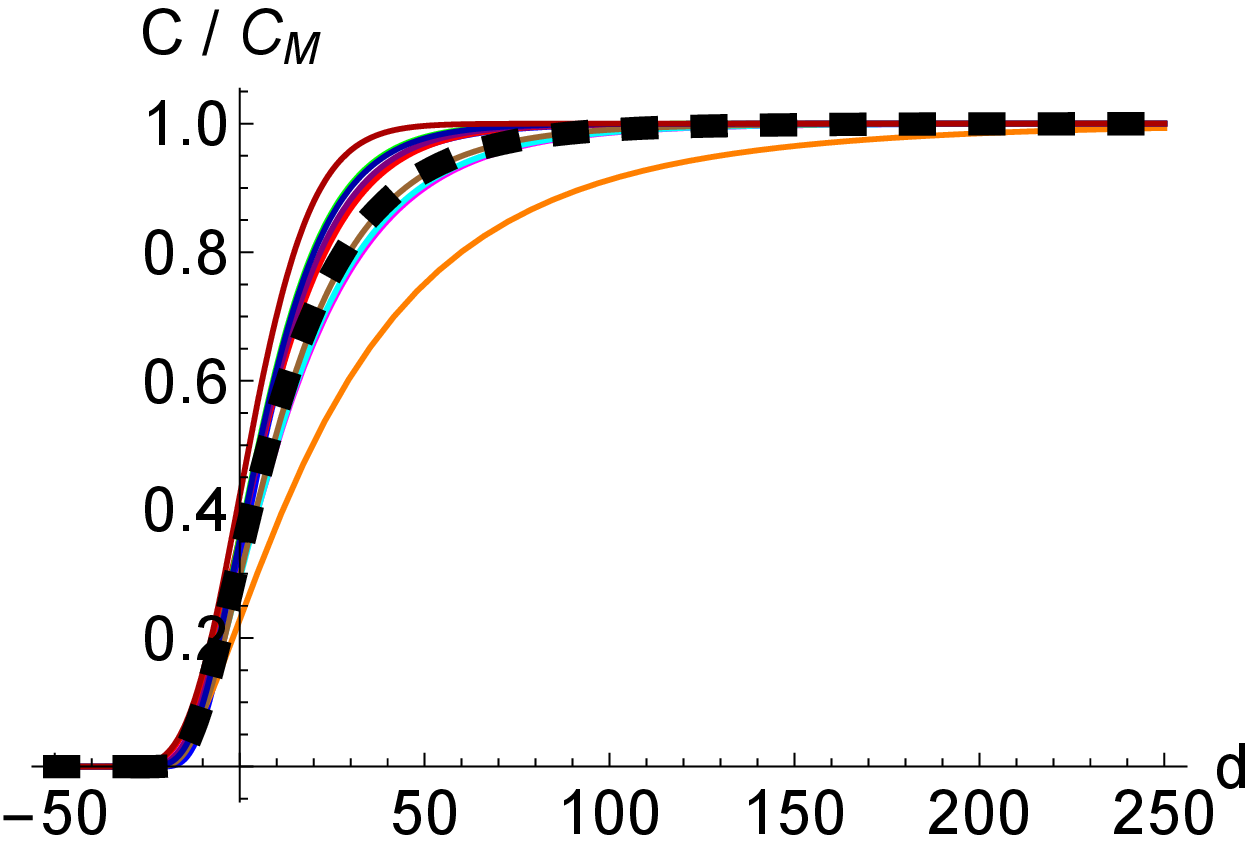}
\smallskip \caption{{\bf Averaged profile.}
Daily $c(t)$ (left panel) and accumulated $C(t)=r(t)$ cases (right panel)
for ten different
countries, normalized to its maximum value
and displaced rigidly in time
so $C''(t)=c'(t)=0$ the same day. 
Color codes are:
(1) Spain (blue), (2) Germany (red), (3) France (green),
(4) USA (orange), (5) Italy (magenta), (6) Great Britain (cyan),
(7) Switzerland (purple), (8) Denmark (brown), (9) Austria (darker blue),
(10) Finland (darker red).
The black thick dashed line gives the average over the ten countries,
with $\mu=3.53$, $\sigma=0.56$.
}
\label{fgr:AVR}
\end{figure}

Normalizing and superimposing the curves for COVID-19 deceases on different countries such that the maximum in $c(t)$ is in a common position, ($t_M$) allows us to focus on similarities. Despite slight differences, nine out of the ten arbitrarily-chosen western countries are all well represented by a normalized average function, $<c(t)>$, (Fig~\ref{fgr:AVR}).
USA shows as an outlier; a warning about the quite different boundary conditions from the other European countries. Since second waves are not fully developed (except in USA) it is not possible yet to ascertain if such a universal average could represent faithfully second waves, even if maybe with different effective values
of $\mu$ and $\sigma$ owing to the different boundary conditions that may apply.
We have not tried the same procedure with the infected because of the greater temporal and spatial variability of procedures used to define that variable. However, results for deaths in the first waves make us believe that such a representative average could also be applied to a properly defined observable for infections.
Excluding USA, the maximum average error made by substituting the actual data by the average function ($\epsilon=c(t)-<c(t)>$) is $\approx 0.03$ in units of $c_M$, which happens near the inflections points where the function $c(t)$ has decayed to $\approx 0.4$ (see inset in left panel in Fig~\ref{fgr:AVR}). 
Therefore, the averaged curve yields an answer with a fractional error of $\approx \pm 5\%$, which is an excellent initial guess taking into account that it
only depends on a single parameter, $c_M$.
Such parameter $c_M$ can 
be easily obtained from a single point: the maximum value in the daily distribution of cases for each wave, which we derive from a moving average of a few days (seven days makes appropriate averages that account for regular weekly routines and removes most of the noise for all the cases we have analyzed).

\subsubsection{Accuracy of SHR.}
\label{sct:accuracySHR}
To be able to confidently use a least-squares statistical regression to a given data set $\{ C_i \}$ ($i= 1, n$) the main question is how many data points, $n$, are needed to yield a reasonable estimation of the evolution of the epidemics based solely on the extrapolation of the fitted functions. Such question is relevant considering how unreliable extrapolations usually are~\cite{NumericalRecipes86}. Indeed, any simple algorithm to forecast the evolution of an epidemics can only be valuable if reasonable error bars can be assigned to predictions. 

\begin{figure}[!t]
\includegraphics[width=0.99\columnwidth]{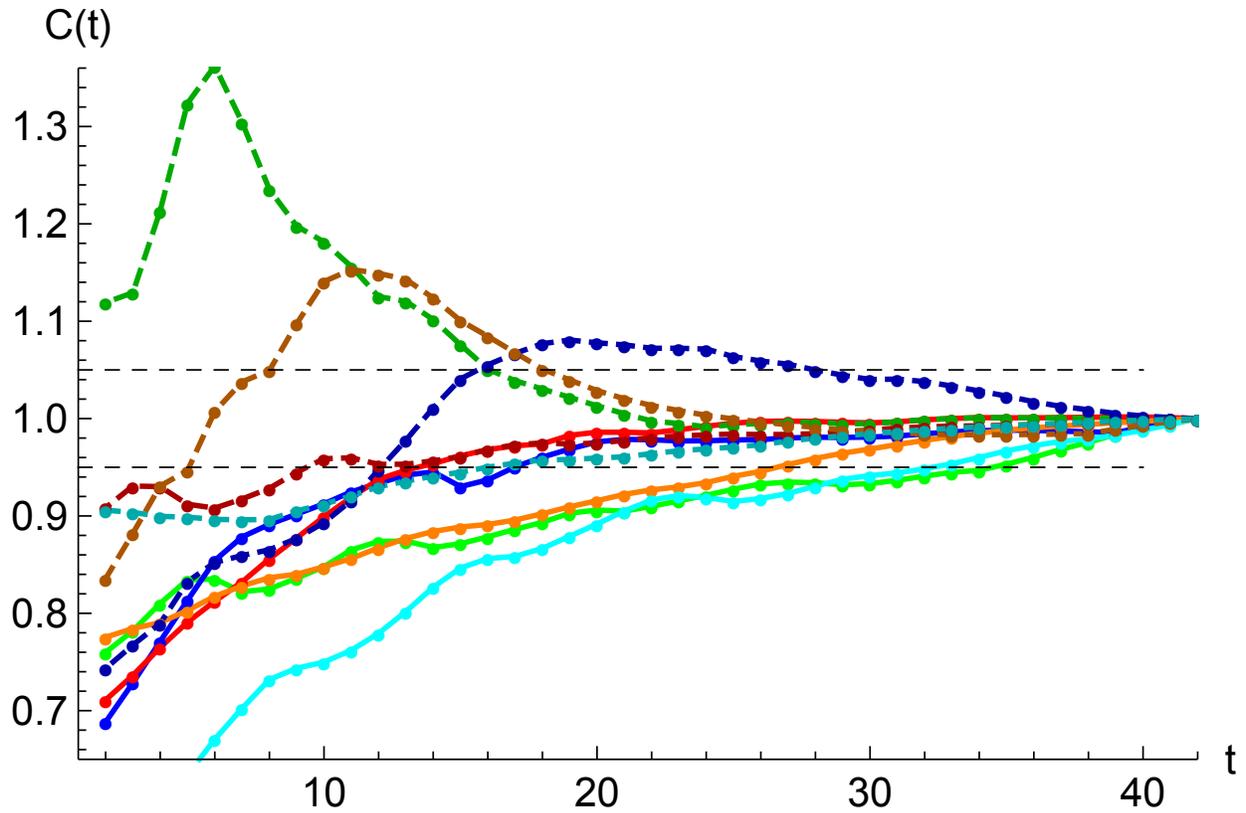}
\smallskip \caption{
{\bf Accuracy of SHR best fits.}
Starting at the second inflection point, $t_2$, 
fractional error in the evolution of the predicted accumulated number of cases 
$C(t)$ for:
(1) GBR (green), (1) ESP (blue), (2) ITA (red), (3) GBR (green), (4) FRA (orange),
(5) USA (cyan), (6) CHE (dashed darker green), (7) DNK (dashed darker blue), (8) DEU (dashed darker red),
(9) AUT (dashed darker orange), (10) FIN (dashed darker cyan).
The region $\pm 5\%$ is delimited by black dashed lines. 
A common normalizacion has been used by making
$C(t_2+40)=1$ for all cases.
}
\label{fgr:K}
\end{figure}

A simple target to quantify the error is to study 
the behavior of the expected total number of cases, 
$C(\infty)$, as a function of $n$. Fig.~\ref{fgr:K} shows the variation of the predicted asymptotic value as a function of the available amount of data after the second inflexion point. In most cases, a fractional accuracy of $\pm 15\%$ is achieved a fortnight after the second inflexion point, which is further decreased to $\pm 5\%$ in another fortnight.

\subsection{Susceptible-Infected-Removed (SIR)}

So far, we have shown that SHR qualifies as a quick and straightforward way to describe the evolution of an infectious disease. If adequately used, i.e. attached with appropriate error bars, it can be extrapolated to make predictions in the near future, since the functional forms associated with Equations~\ref{eqn:PDF} and~\ref{eqn:CDF} adapt so well to the observed data. 

However, 
a better understanding of the dynamics of the epidemics 
can be obtained from a set of 
differential equations which describe its time evolution. 
The simplest model for the evolution of a contagious disease is to postulate that the rate of new infections is proportional to the number of infected people itself, 
$\frac{\D I(t)}{\D t} = \frac{I(t)}{\tau_0}$, which results in an 
unbound exponential growth, $I(t) \propto e^{\frac{t}{\tau_0}}$, 
and makes a characteristic mark for the onset of a pandemic.

Such a simple model does not take into account how the rate of infections decreases as the number of infections approaches the
total population. Therefore,
a refined version is to divide a given population of size $N$ into three classes ($\mathcal{S},\mathcal{I},\mathcal{R}$):
(i) {\it susceptible} entities who can catch the disease, $S(t)$, (ii) {\it infected} ones who have the disease and transmit it, $I(t)$, and (iii) {\it removed} ones who have been isolated, died, or recovered and become immune, and are therefore not able to propagate the disease, $R(t)$. 
In this model, individuals pass from the susceptible class $S$ to the infective class $I$ and finally to the removed class $R$ with rates determined by a set of 
ordinary differential equations(ODEs)~\cite{kermak1927,ANDERSON19913,Hethcote,weiss2013}. 
The ODEs derive from the interactions of the entities in the different classes, which can be represented as 
$$\mathcal{S}+\mathcal{I} \to 2\mathcal{I},\quad \mathcal{I} \to \mathcal{R},$$
where we assume generalized mass-action kinetics \cite{GMAL} (with slightly different scaling with respect to $N$).
First, it is assumed that the number of susceptible individuals decreases 
at a rate proportional to 
the density of infected, $i(t)=\frac{I(t)}{N}$ times
the number of susceptible individuals, $S(t)$, 

\begin{equation}
\frac{\D S(t)}{\D t} = -\frac{\left( S(t) \right)^{n}}{\tau_0} ~ i(t)
\label{eqn:sir2}
\end{equation}
\noindent
where $\tau_0$ is an adjustable parameter that represents a typical time
to transmit the disease, and $n$ is a parameter that influences the  
ability of the disease to infect susceptible individuals in a non-linear way
(e.g. it might represent the effect of the viral load). 
Its main effect is to alter
the temporal scale of the epidemics, which 
in some circumstances facilitates the fitting of the
model to real data. 
The standard SIR model is recovered with $n=1$.

Removed entities originate from infected; therefore, its
variation is assumed to be proportional to the number of infected,

\begin{equation}
\frac{\D R}{\D t}= \frac{\left( I(t) \right)}{\tau_1}
\label{eqn:sirrr}
\end{equation}
\noindent
where $\tau_1$ is an adjustable parameter that represents a typical time
to recover from the disease. This equation merely helps to count the total
number of removed from the beginning of the infection up to a given day $t$,
\begin{equation}
R(t) = \int_{0}^{t} \frac{\left( I(u) \right)}{\tau_1} ~\D u
\label{eqn:sirr}
\end{equation}

Lastly, the infected vary according to the inflow of susceptible individuals who become infected minus the outflow of infected that have been removed,
\begin{equation}
\frac{\D I}{\D t}= \frac{\left( S(t) \right)^{n}}{\tau_0}~i(t) - \frac{I(t)}{\tau_1}
\label{eqn:sir1}
\end{equation}
\noindent
The derivative $\frac{\D I}{\D t}$ moves from positive to negative depending on the balance
between both terms in the equation and it determines
a single peak in $I(t)$ 
(for $n=1$, $I'(t)=0$ for $\frac{S(t)}{N}=\frac{\tau_0}{\tau1}$).

The task at hand for a given population of
$N$ elements is to determine the parameters, $\tau_0$, $\tau_1$ and $n$, 
that best reproduce the behavior of the epidemics by solving
the coupled system of differential equations~\ref{eqn:sir2} and~\ref{eqn:sir1},
subject to some initial conditions, e.g.
$S(0)=N-1$, $I(0)=1$. 
Good agreement with data can be used to lend an interpretative value to 
$\tau_0$ and $\tau_1$
(unlike parameters $\mu$ and $\sigma$ which only have
a statistical meaning).
The ratio $\Re_0 = \frac{\tau_1}{\tau_0}$ is called the effective
reproductive number; values $\Re_0>>1$ characterize a virulent
disease where $R(\infty)=S(0)$.

First, we focus on the task of simulating 
a population where $s(0)=\frac{S(0)}{N}=r(\infty)=\frac{R(\infty)}{N}=1$.
For this particular case, $\Re_0>>1$ 
and the entire susceptible population is removed at the end.
The proposed algorithm goes as follows,

\begin{enumerate}
\item We use the daily number of deaths to identify the
position and maximum value in the infections/deaths data: $t_M^{*}$ and $i_M^{*}$. 
\item $\tau_0$ is the main parameter that determines the position of the peak in $i(t)$. 
We estimate a value for $\tau_0$ that brings the maximum in $i(t)$ near $t_M^{*}$. 
\item We get an approximate value for $\tau_1$ from the expression 
$i_M^{*}=1-\frac{1}{\Re_0}\left(1+\ln{\Re_0}\right)$~\cite{weiss2013}.
\item The value $i_M^{*}$ yields $N$ in the particular case of $R(\infty)=N=c_M$.
We adjust the value of $N$ to agree with $i_M^{*}$.
\item We minimize the root-mean square deviation,
$\frac{\chi}{N}=\sqrt{\frac{1}{n}\sum_{i=1}^{n} (C_i-R(t_i))^2}$,
between the number of accumulated cases predicted by the model, $R(t)$, and
the recorded data, $C_i$, to find optimal values for $\tau_0$, and $\tau_1$.
\end{enumerate}

\subsubsection{Spain.} 
\begin{figure}[!t]
\includegraphics[width=0.49\columnwidth]{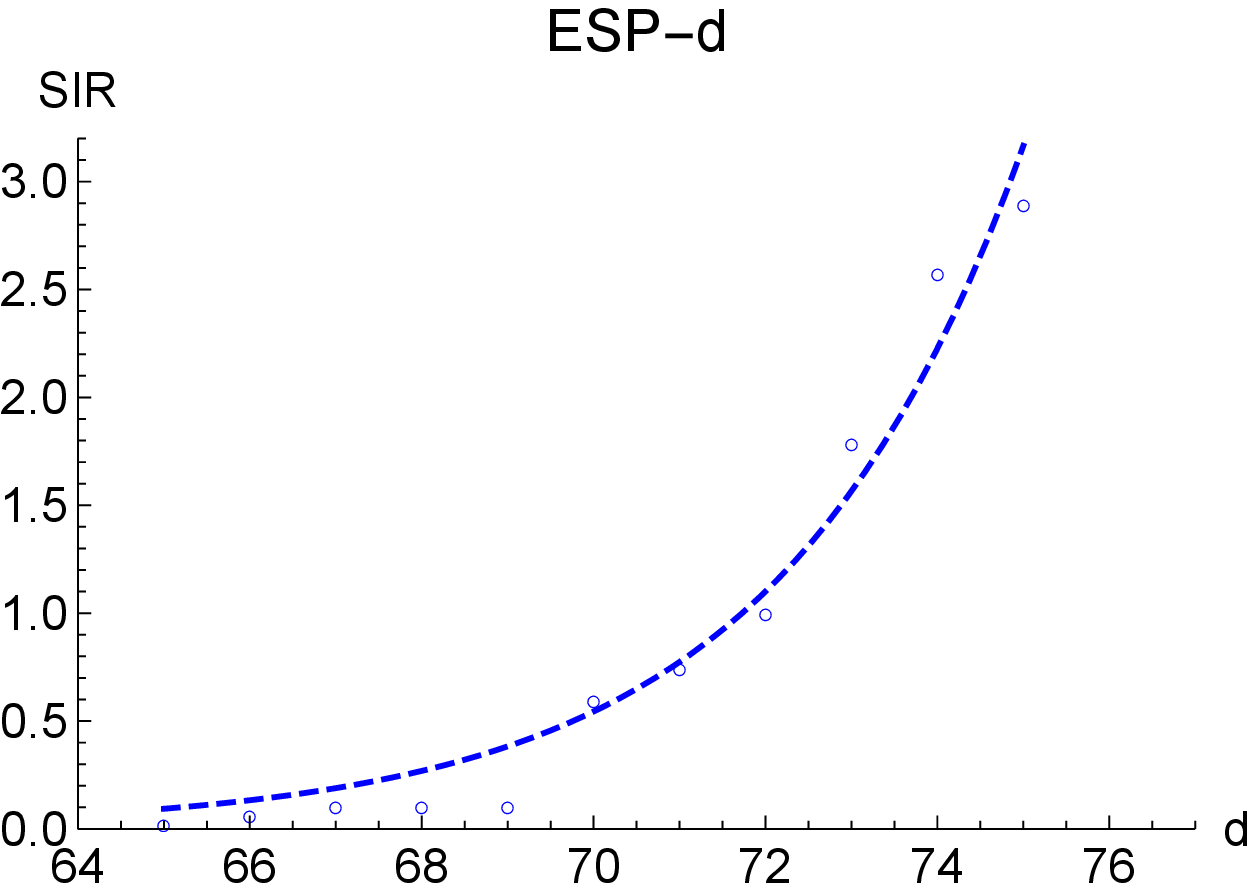}
\includegraphics[width=0.49\columnwidth]{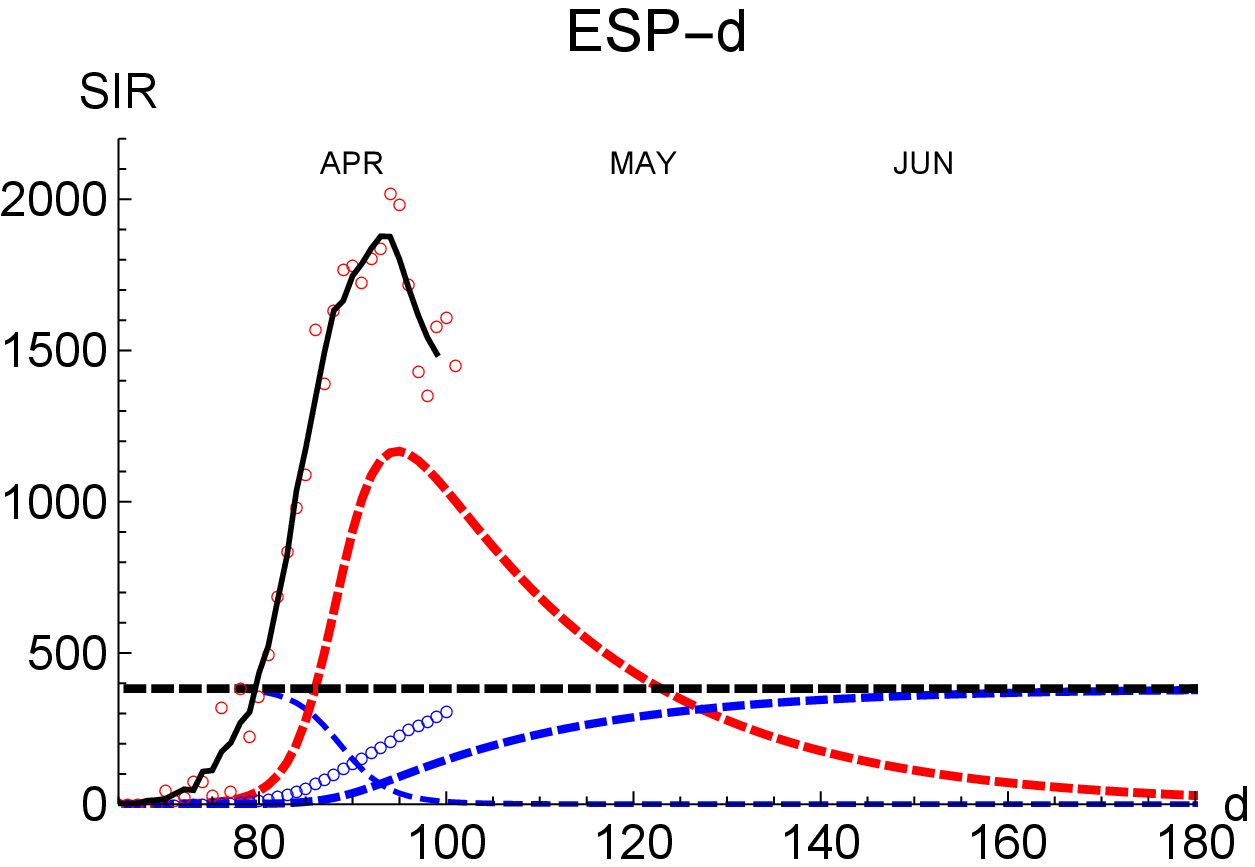} \\
\includegraphics[width=0.49\columnwidth]{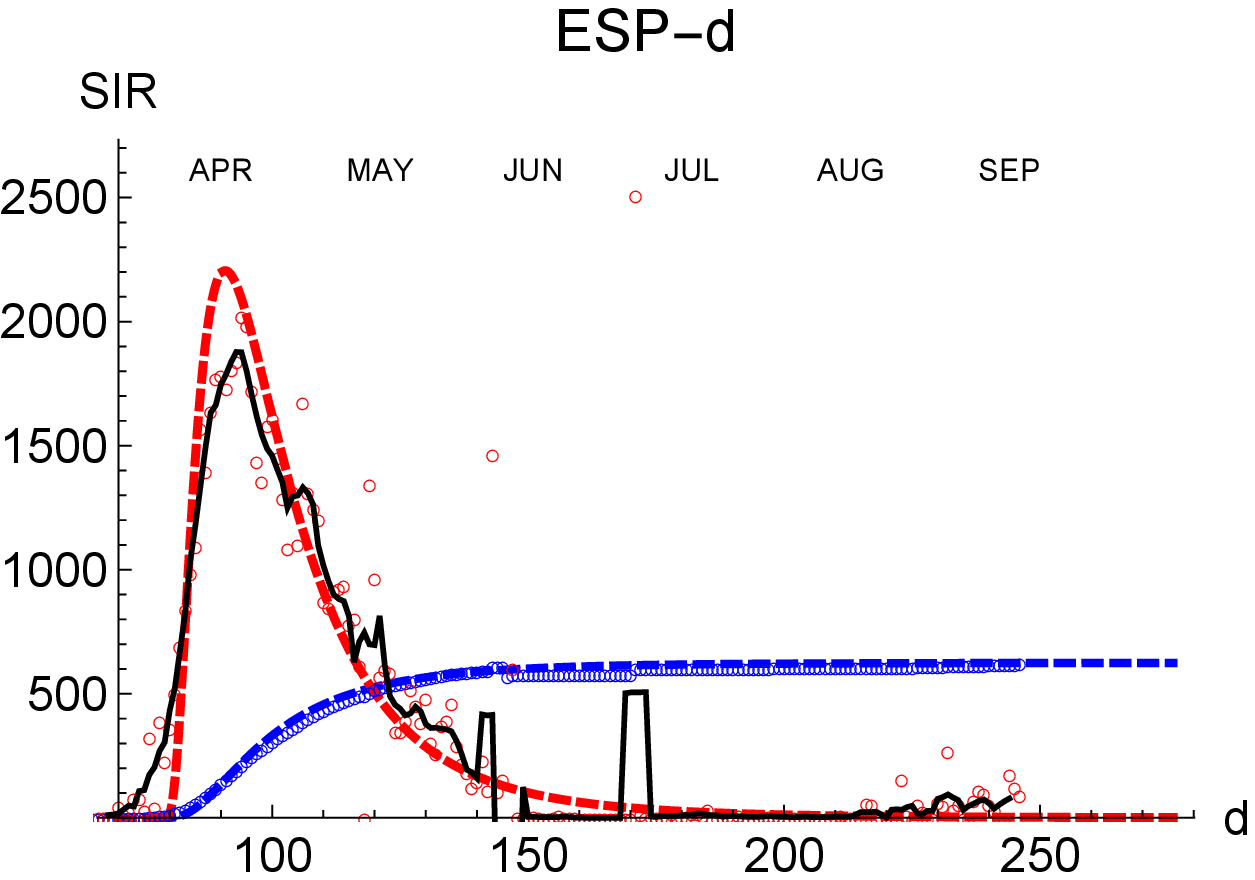}
\includegraphics[width=0.49\columnwidth]{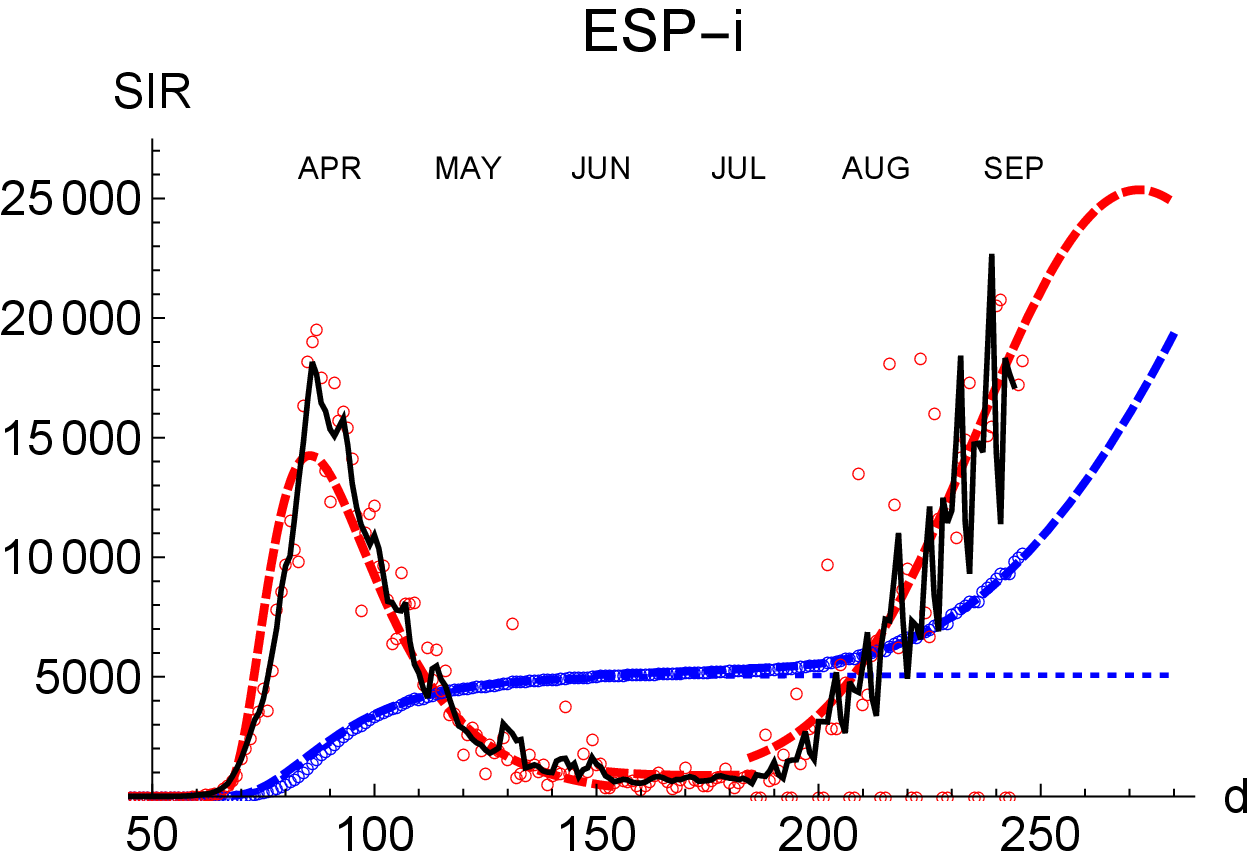} \\
\smallskip \caption{
{\bf SIR/Spain.}
Left upper panel: exponential fit near the onset.
Right upper panel: initial iteration for deaths (see text).
Left lower panel: final iterations for deaths.
Right lower panel: final iterations for infections.
Blue: accumulated cases, $R(t)$ (per million people). 
Red: daily cases, $I(t)$
($\times 100$ to increase visibility in the same scale as $R$).
Black is a 7-day moving average of data to help the eye.
}
\label{fgr:sirESP}
\end{figure}

Similarly to the SHR analysis we have presented above,
we illustrate the performance of the SIR model by first looking at
the distribution of deaths and infections in Spain
(Fig.~\ref{fgr:sirESP}).
The lower left panel shows how
numerical solutions to SIR equations match very well the 
temporal behavior of the epidemics under the condition
$s(0)=r(\infty)=1$ for optimized values of $\tau_0$ and $\tau_1$
(Table~\ref{tbl:SIRd}).
Dispersion of data in the daily reported cases is usually
smaller before the peak is reached (the quasi-exponential region) and
fluctuations grow in importance after the maximum is reached--
which is a general observation holding for most of the countries
we have studied. We assign it to the balance between different {\it currents} transferring
individuals between the three classes, the phenomenon responsible
for the appearance of a single maximum in daily cases for a given
wave in the pandemics.

The proposed procedure works for the deaths subset as follows.
First, the curve of daily cases is followed up to 
the appearance of its
maximum, which to circumvent the noise is identified
from a smoothed curve
obtained by a five-day moving average, $I_{M}^{*}(t_{M}^{*}=96)=17.1$ per million people
(the single daily maximum value is $I_M(t_M=94)=20.2$).
The SHR model for accumulated deaths using only data 
up to six days past the maximum yields a prediction of total deaths
of $N=383$, which is off the final mark by about 40\%.

Once the maximum is identified, the quasi-exponential behavior near the origin
is used to estimate an initial value for $\tau_0$ (supplementary material,
Eq.~S6).
For Spain, the first case happens at $t=65$, 
and the first inflexion point is at $t_1=82$. 
Therefore, the first 10 points 
(about halfway to $t_1$) are used to get an exponential fit 
to the accumulated number of cases that 
yields $\tau_0 \approx 2.8 \pm 0.3$. 
Such a value, combined with an initial guess $\Re_0=10$
produces a maximum in the curve of daily deaths at $t_M=103$. 
Accordingly, $\tau_0$ is decreased until we locate the
maximum closer to the right position.  
For $\tau_0=2.2$ we get $t_M=95$ and $I_M=11.7$ (per million inhabitant). 
Therefore, we update the value of N using the ratio
$\frac{17.1}{11.7}$ and start an efficient local 
Levenberg-Marquardt minimization of the root-mean-squared deviation
between the actual data and the computed values. This is done to simultaneously optimize
$N$, $\tau_0$ and $\tau_1$ (Fig.~\ref{fgr:sirESP}, left-lower panel). 
Taking into account that only data up to six days past the maximum have been used,
it is remarkable that
this self-consistent procedure reduces the fractional error between
the prediction of the SIR model and the data from $40$\% to $\pm 3$\%,
being the root-mean-squared deviation (RMSD) between the accumulated data and the predicted function
$\frac{\chi}{N}=0.6$\%.
Such a low RMSD value matches the good visual agreement observed.
We believe that the logic behind the steps proposed above amounts to more than a recipe to get a best
fit, yielding meaning to the values obtained
and their interpretation. 

Next, we explore how the SIR model represents the evolution of the number of infections.
The number of infections is a magnitude that carries larger error bars, but it can provide timely information on the evolution of the epidemics (Fig.~\ref{fgr:sirESP}, right-lower panel shows the case for Spain). As expected, infections start earlier than the deaths($t=32$
vs $t=65$), but need more time to attain its maximum value ($t=54$ vs $t=25$ after the first case). 

A prominent feature is the existence of the second wave of infections separated from the first one by a region of {\it sustained constant} number of cases, as we have discussed in Fig.~\ref{fgr:shrESP}.
To fit the data, we superpose the two waves, each with its own defining parameters. However, the constant region between waves cannot be easily accommodated in these models and it is a clear indication of a different stage in the epidemics with low but sustained transmission of the disease at a pace similar to the one at which individuals are removed (while in the SIR model usually it is assumed that $\tau_1 > \tau_0$). We shall come back to this point in the context of Monte-Carlo simulation.
Finally, we notice that this second wave of infections has finally overlapped with
a third one, as it is noticeable in Fig.~\ref{fgr:shrESP}.

\subsubsection{Germany.}
\begin{figure}[!b]
\includegraphics[width=0.49\columnwidth]{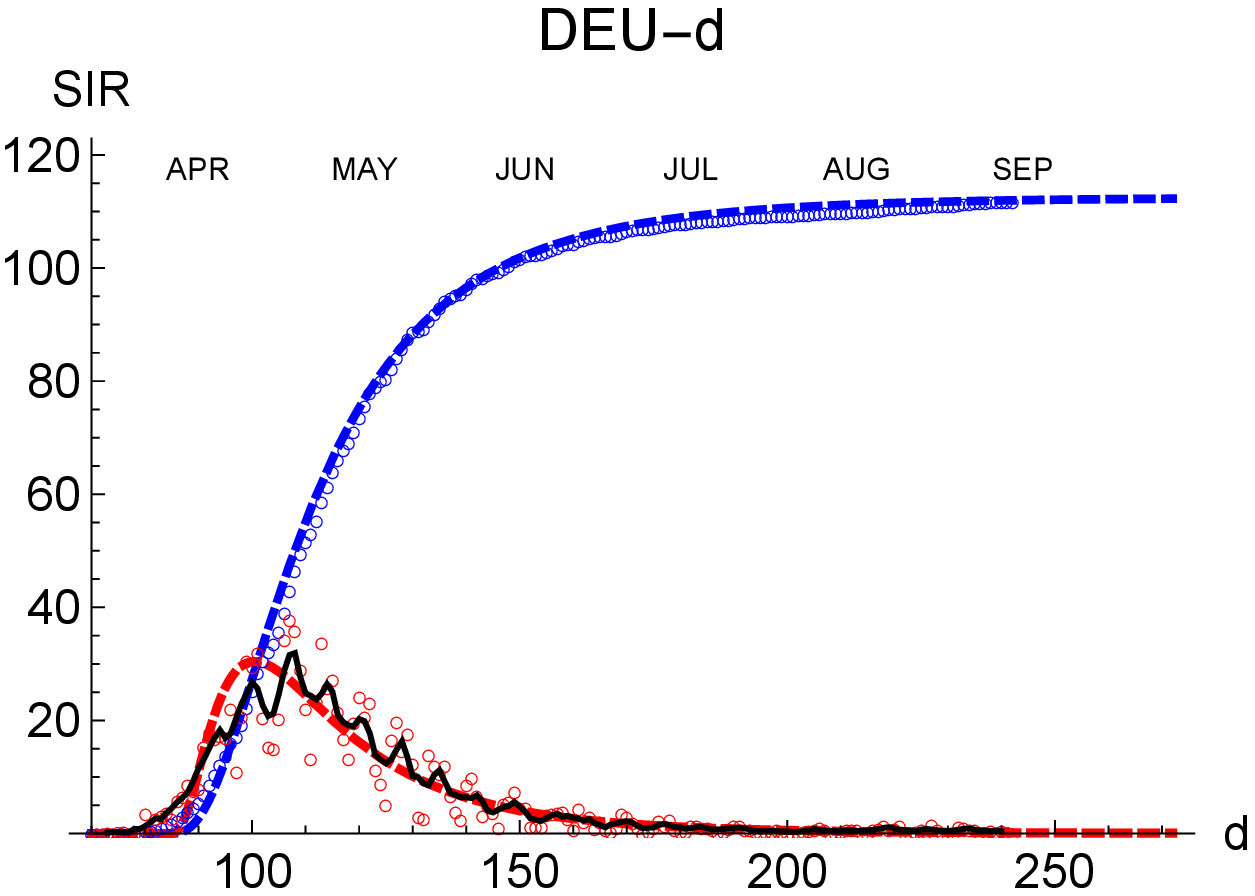}
\includegraphics[width=0.49\columnwidth]{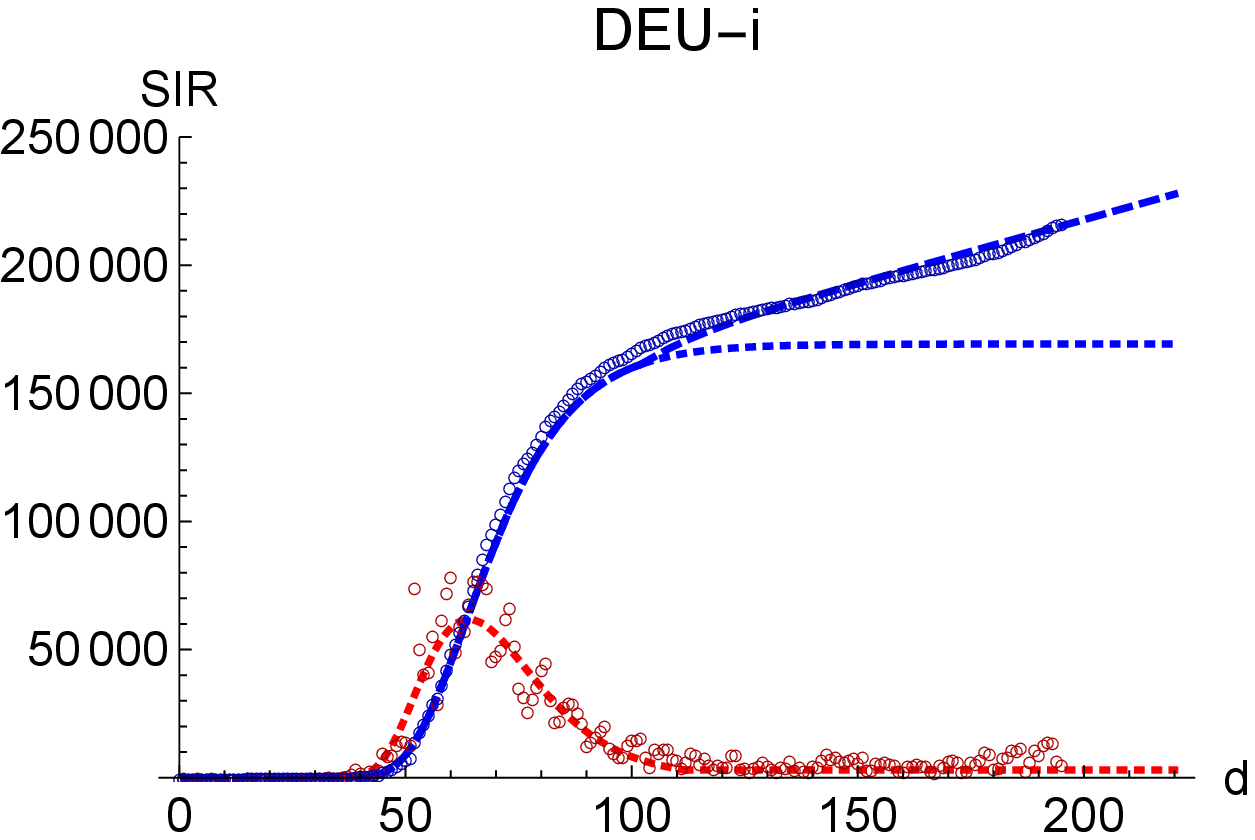}
\smallskip \caption{
{\bf SIR/Germany.}
Left/Right panels: Deaths/Infections.
Blue: total cases, $r(t)$. Red: daily cases, $i(t)$
($\times 10$).
Other symbols as in Fig.~\ref{fgr:sirESP}.}
\label{fgr:sirDEU}
\end{figure}

We have applied the same procedure to Germany, a country which had in the first wave about four times less
casualties per million inhabitant than Spain. 
The left panel of Fig.~\ref{fgr:sirDEU} shows the final
iteration for the daily and accumulated number of deaths, which again predicts the total number due to the first wave with accuracy $\approx \pm 3\%$ of the final true value, even if we
have only used data up to six days past the maximum.
The RMSD between the accumulated data and the predicted function is $\frac{\chi}{N}=0.5$\%,
which reflects the good visual agreement observed too.

Regarding the infections, it is interesting that the region of sustained infection is also observed, although a second wave is only weakly apparent up to the present day (t=200).  Again, infections start earlier ($t=28$) w.r.t deaths
($t=70$). Furthermore, maximum values are attained after a longer amount of time ($t=63$ for infections ($t=63$) than for deaths ($t=30$), counted after the first case, following a similar procedure to the one for Spain.

\subsubsection{Other countries.}
\begin{figure}[!t]
\includegraphics[width=0.49\columnwidth]{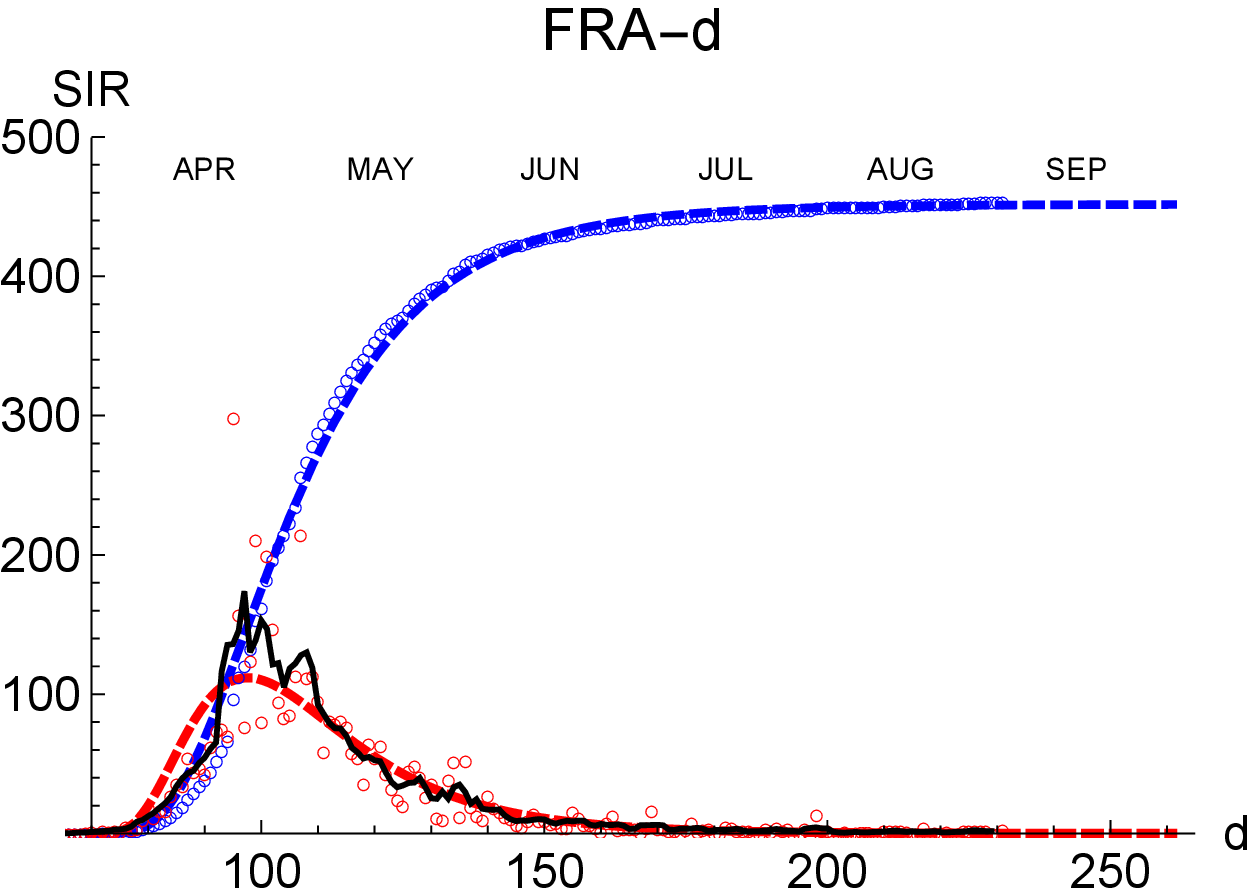} 
\includegraphics[width=0.49\columnwidth]{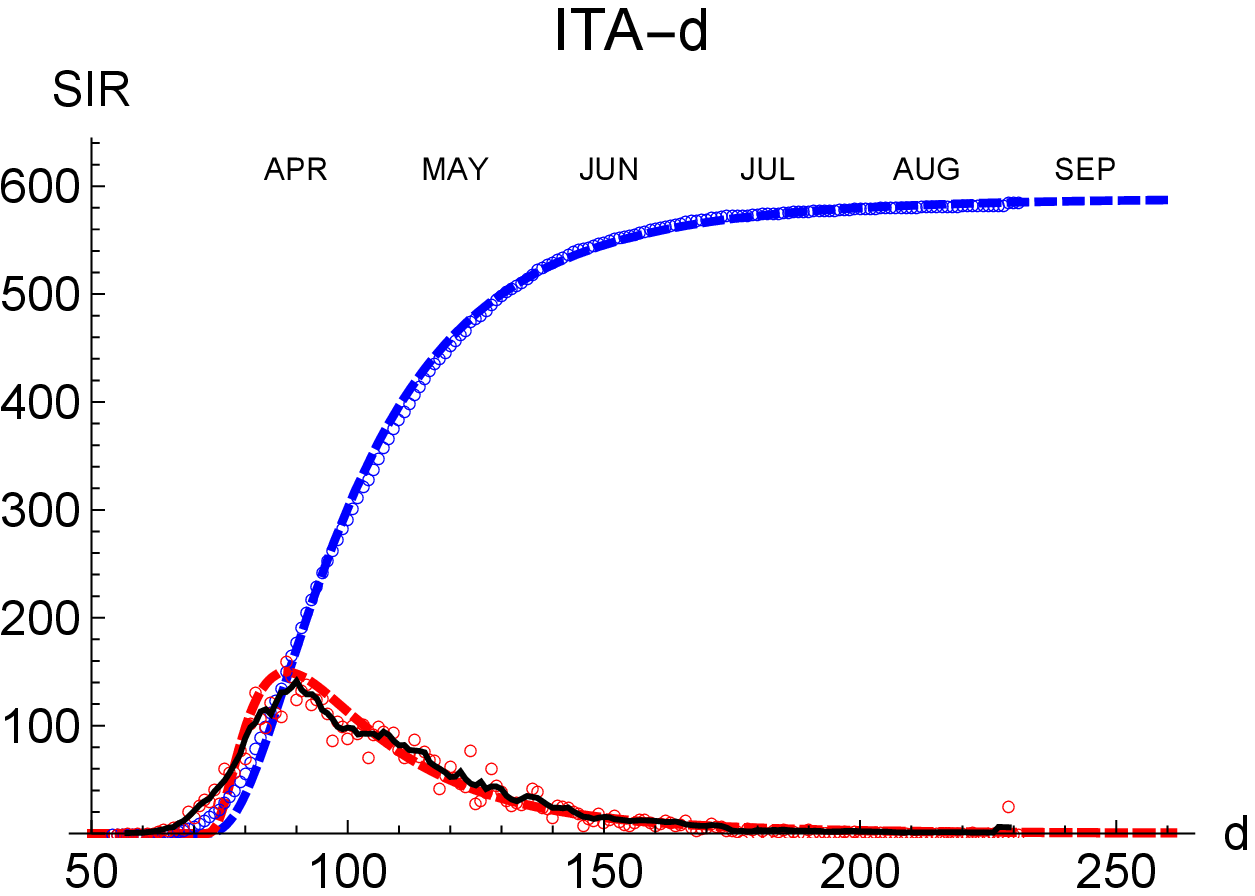} \\
\includegraphics[width=0.49\columnwidth]{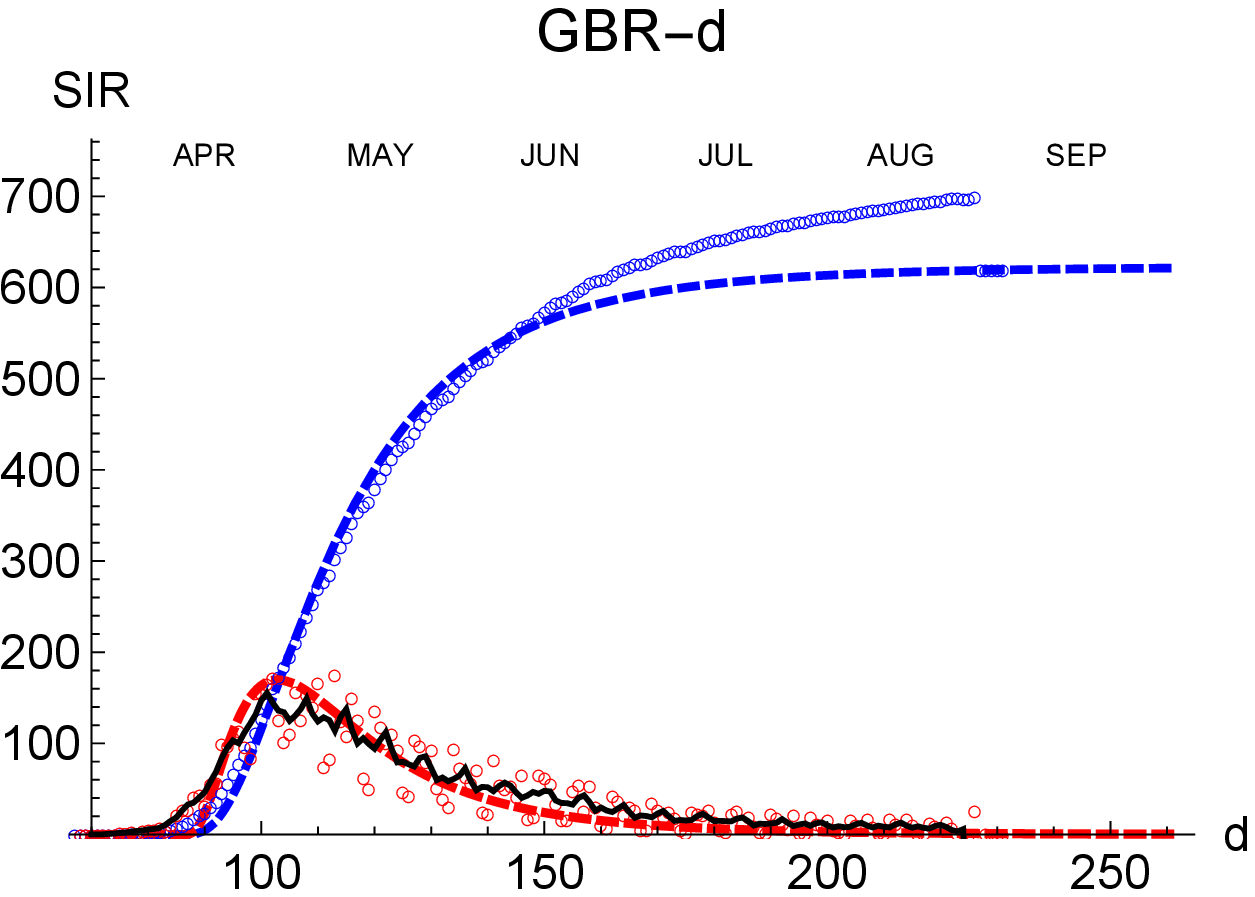}
\includegraphics[width=0.49\columnwidth]{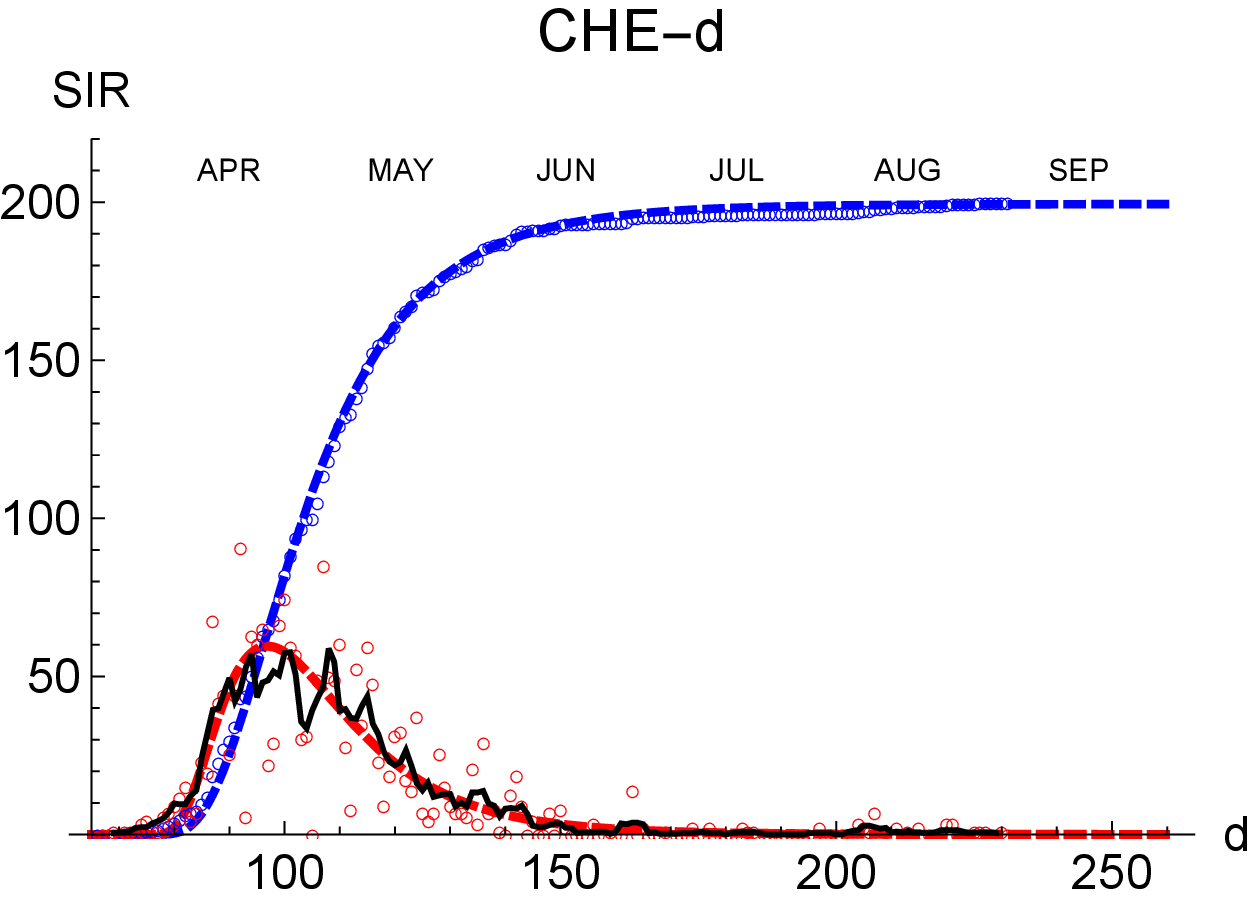}  \\
\smallskip \caption{
{\bf SIR (deaths)/FRA, ITA, GBR and CHE.}
Respectively left to right and top to bottom.
Symbols and lines as in Fig.~\ref{fgr:sirESP}.
}
\label{fgr:sir4C}
\end{figure}

Finally, similar results have been observed in four more countries: 
France, Italy, Great Britain and Switzerland (Fig.~\ref{fgr:sir4C} 
and Table~\ref{tbl:SIRd}).

\subsection{Spatial individual-based model.}
To gain further insight into the spatio-temporal evolution of COVID-19, we consider next a stochastic discrete-time individual-based model in which the spread propagates on a two-dimensional
$N\times N$ lattice, where each node represents an 
individual.
The dynamics are Markovian, and we will use Monte Carlo (MC) to sample from its distributions in time,
which is a technique known to handle well difficult collective effects in
many-body systems, like e.g. the magnetic phase transition in the 2D Ising model\cite{peliti2011}.
The $N^2$ individuals can be in any of the three states of the SIR 
model, making transitions between them with two probabilities: (i) for someone susceptible to be infected $\mathcal{S} \to \mathcal{I}$, $p_i$ and, (ii)
for someone infected to recover and be removed $\mathcal{I} \to \mathcal{R}$ , $p_r$. At each time-step, individuals make transitions between classes according to the corresponding probabilities. 
We consider various scenarios of uniform and spatially dependent Markov dynamics.

\begin{figure}[!b]
\includegraphics[width=0.32\columnwidth]{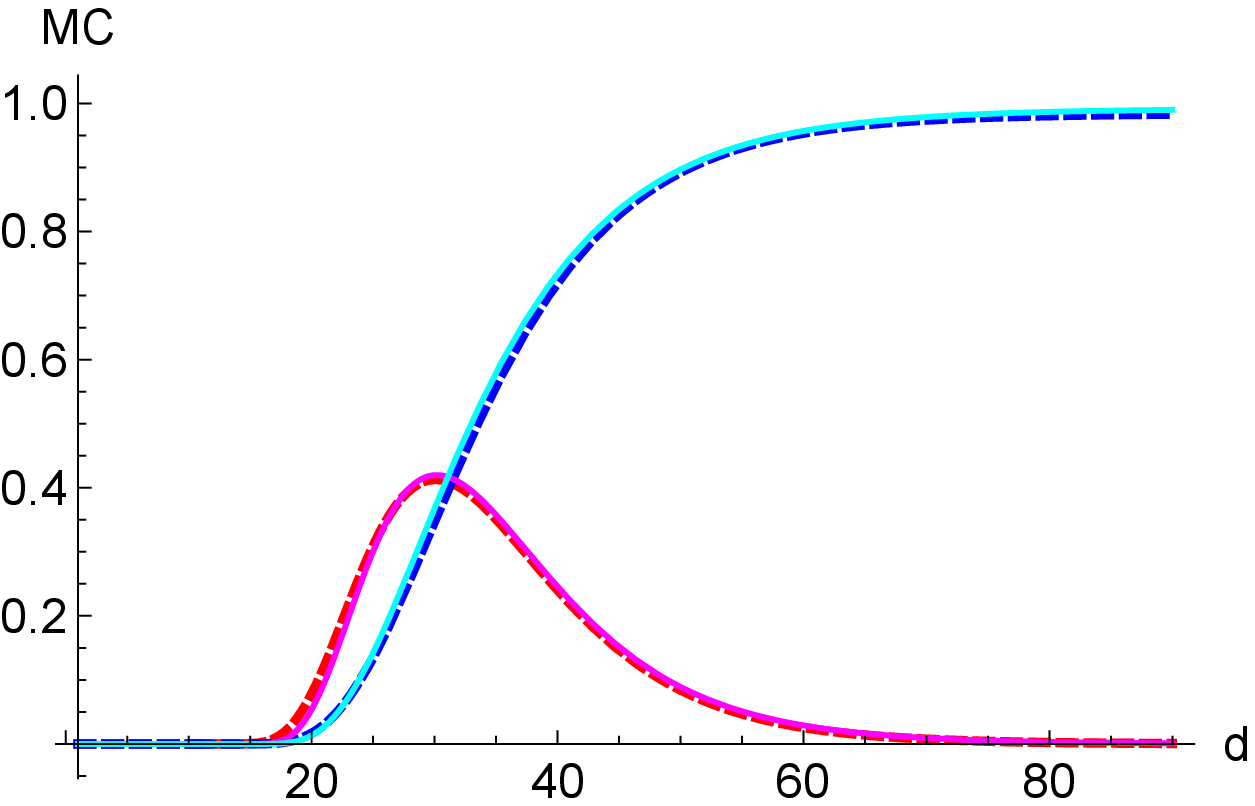}
\includegraphics[width=0.32\columnwidth]{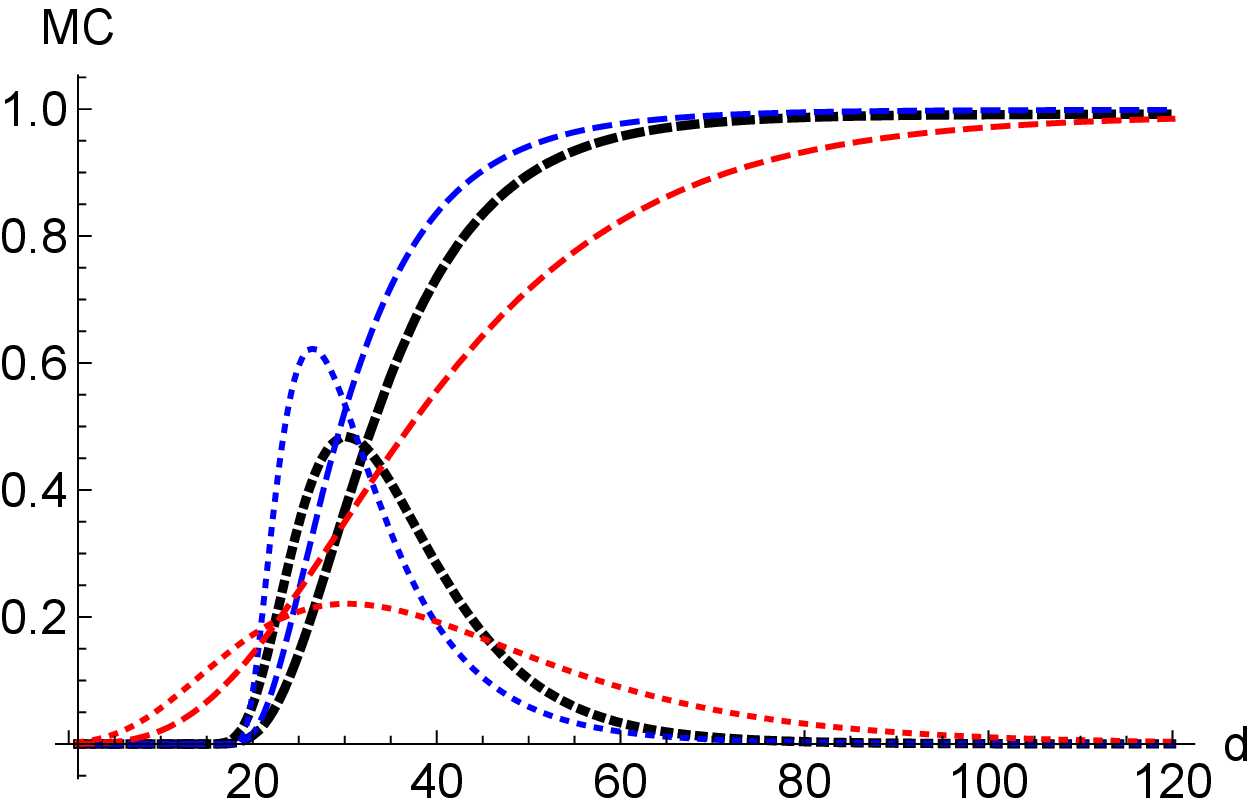}
\includegraphics[width=0.32\columnwidth]{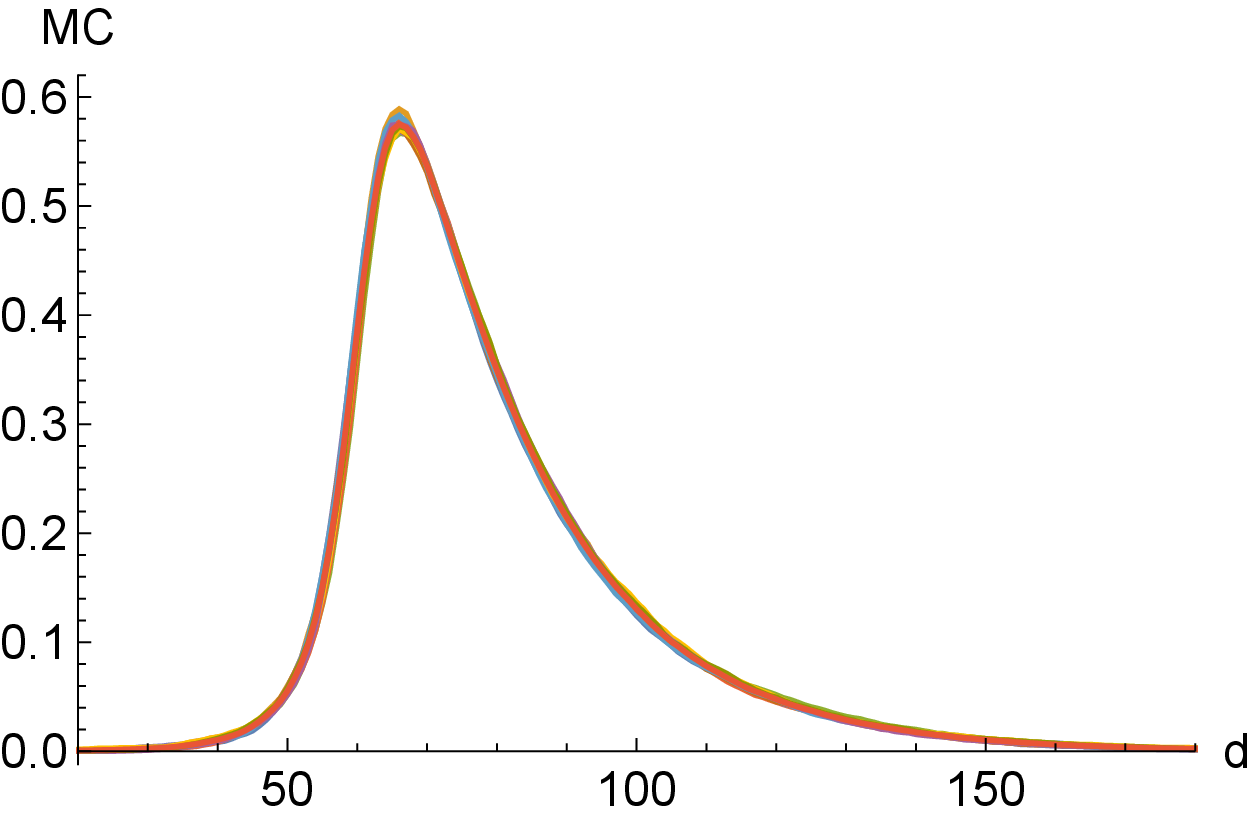}
\smallskip \caption{{\bf Monte-Carlo simulations}.
Left panel: MC (continous) vs SIR (dashed). $N=10000$. 
Red/Magenta: infected.
Blue/Cyan: removed. Parameters used in MC are:
$p_{i}=0.5*i(t)$,
$p_{r}=0.1$.
Parameters used in SIR are:
$\tau_0=2$,
$\tau_1=8$.
Middle panel:
various MC scenarios for infected (dotted) and removed (dashed).
Thick black: probabilities as in left panel.
Blue: nearest-next neighbors increased probability of infection increases to
$p_{i}=0.75~i(t)$.
Red: both probabilities for infection and removal are kept constant values,
$p_{i}= 0.1$ and $p_{r}=0.05$.
Right panel:
average and standard deviation (inset) of 10 random realizations for
$p_{i}=0.2*i(t)$ (double for nearest-neighbors infections) and,
$p_{r}=0.05$.
}
\label{fgr:mc0}
\end{figure}

First, we start with a single isolated case of
infection per $10^{4}$ individuals, and we use 
$p_i(t)=\frac{i(t)}{\tau_0}$ for $\mathcal{S} \to \mathcal{I}$
in close analogy to the SIR model, while we assign 
a constant value $p_r=\frac{1}{\tau_1}$ to the second transition probability,
$\mathcal{I} \to \mathcal{R}$.
Comparing MC simulations for $N=100$
with $p_i(t)=\frac{1}{2}~i(t)$ and $p_r=\frac{1}{10}$ to the deterministic SIR with
$\tau_0=2.1$ and $\tau_1=9$ yields an excellent agreement
between both approaches for 
same initial values, which
confirms the adequacy of Monte-Carlo techniques (left panel in 
Fig.~\ref{fgr:mc0}, where both results cannot be distinguished).
By way of example, we modify the model to increase the
probability of infection of individuals in next-neighbors contact
with members already infected to $p_i=\frac{3}{4}~i(t)$.
As expected, infections grow faster near the onset, the
daily maximum happens earlier and results in a larger and narrower peak (while keeping the
final total number the same, Fig.~\ref{fgr:mc0} blue dotted line
compared to thick dotted line).

\begin{figure}[!t]
\includegraphics[width=0.19\columnwidth]{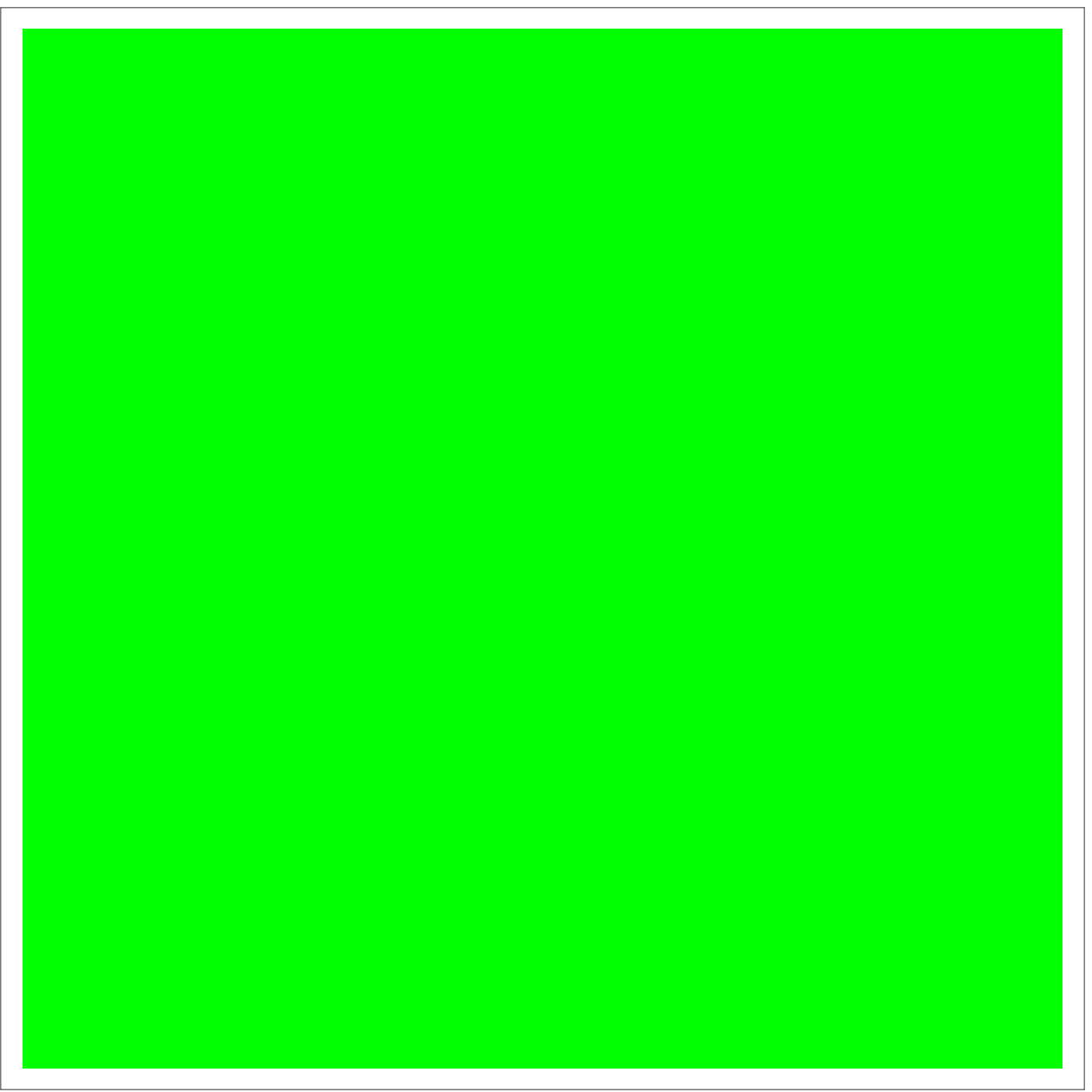} 
\includegraphics[width=0.19\columnwidth]{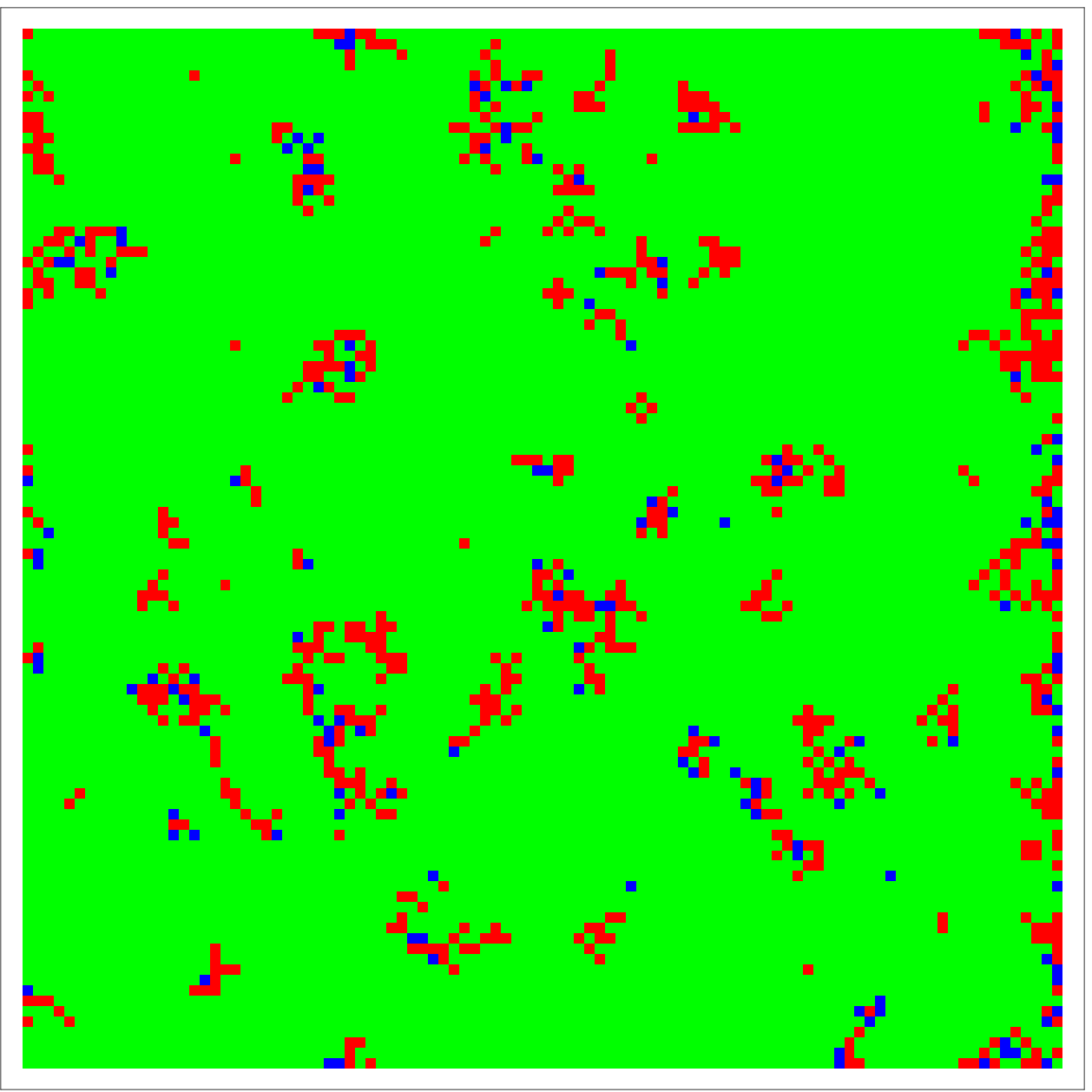}
\includegraphics[width=0.19\columnwidth]{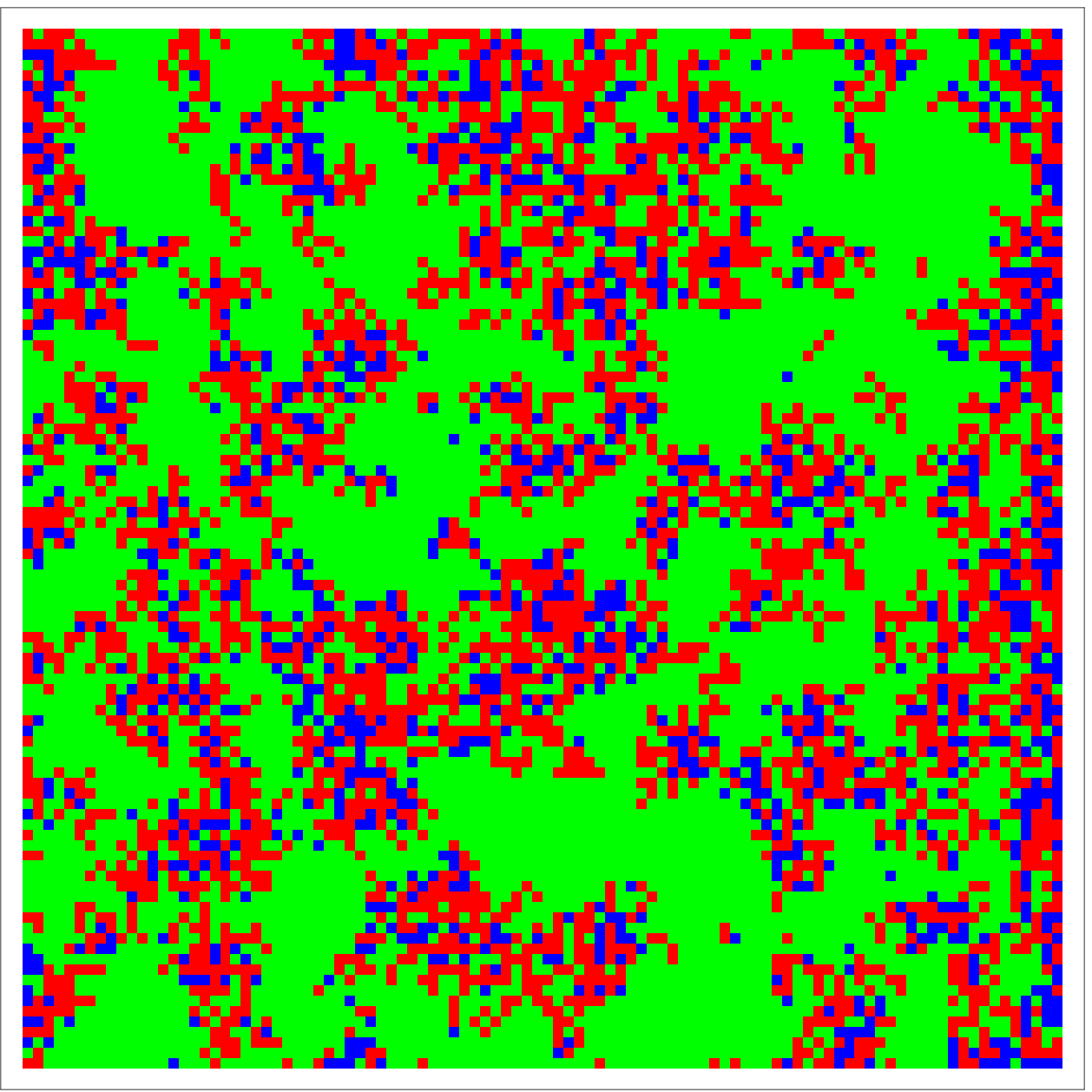}
\includegraphics[width=0.19\columnwidth]{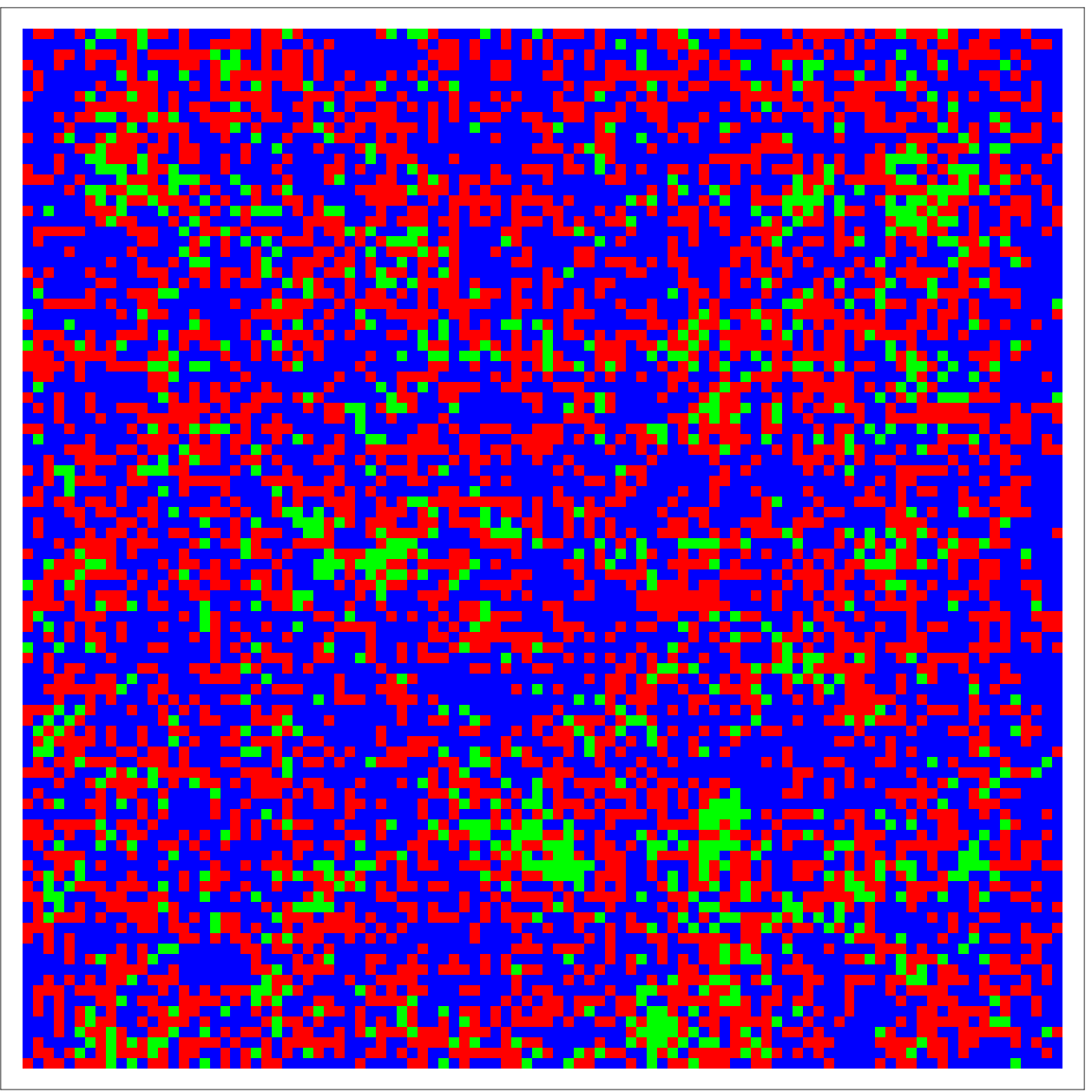}
\includegraphics[width=0.19\columnwidth]{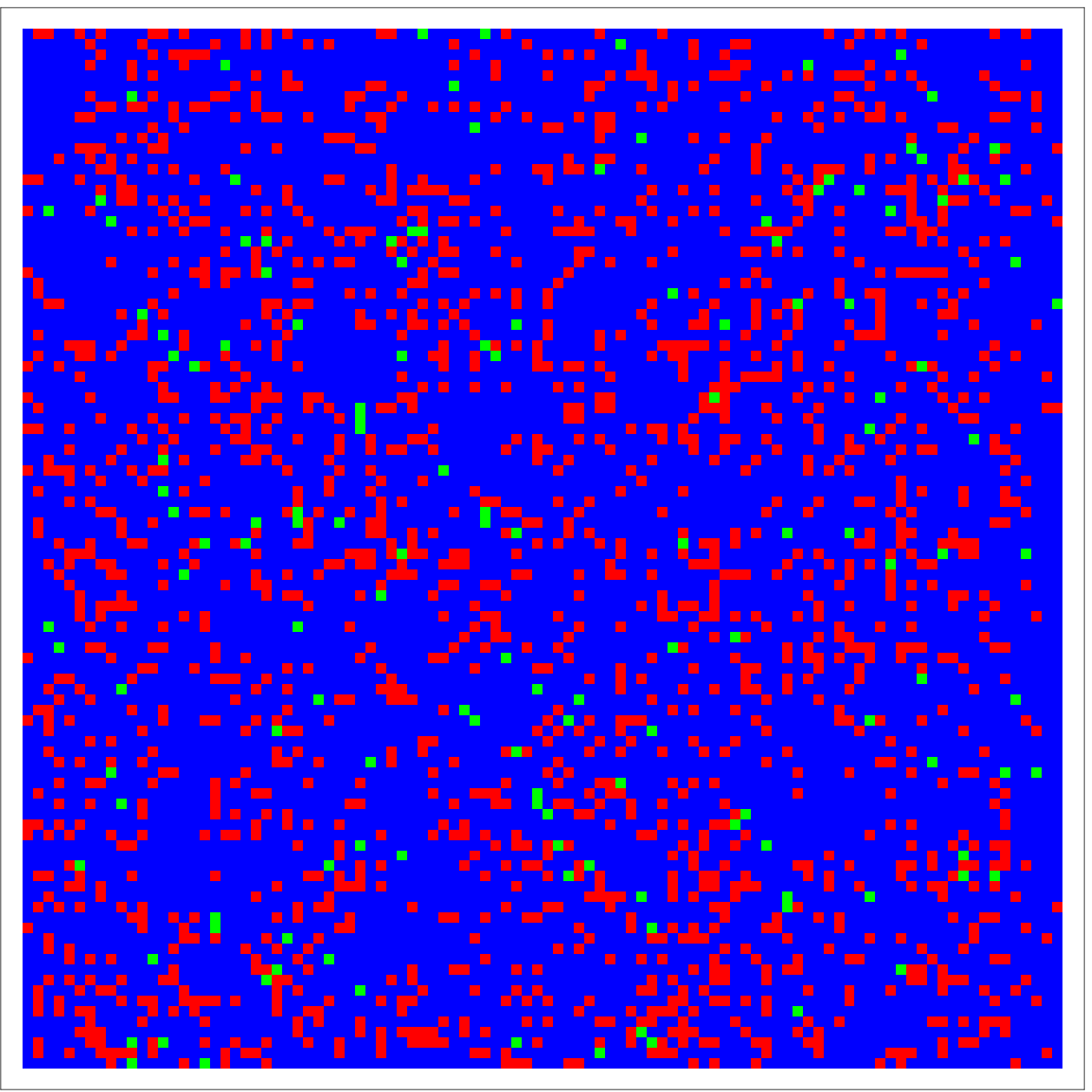} \\
\includegraphics[width=0.19\columnwidth]{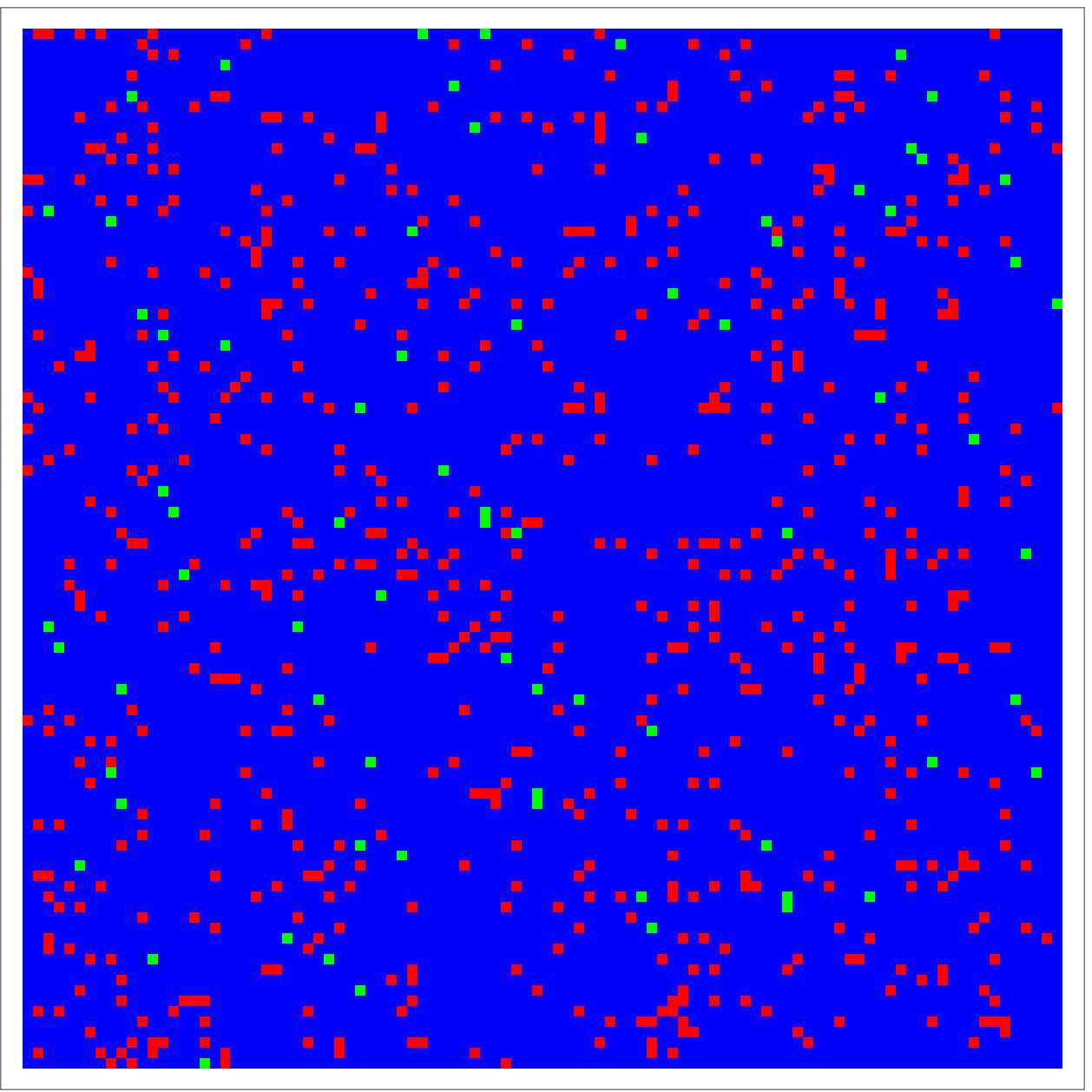} 
\includegraphics[width=0.19\columnwidth]{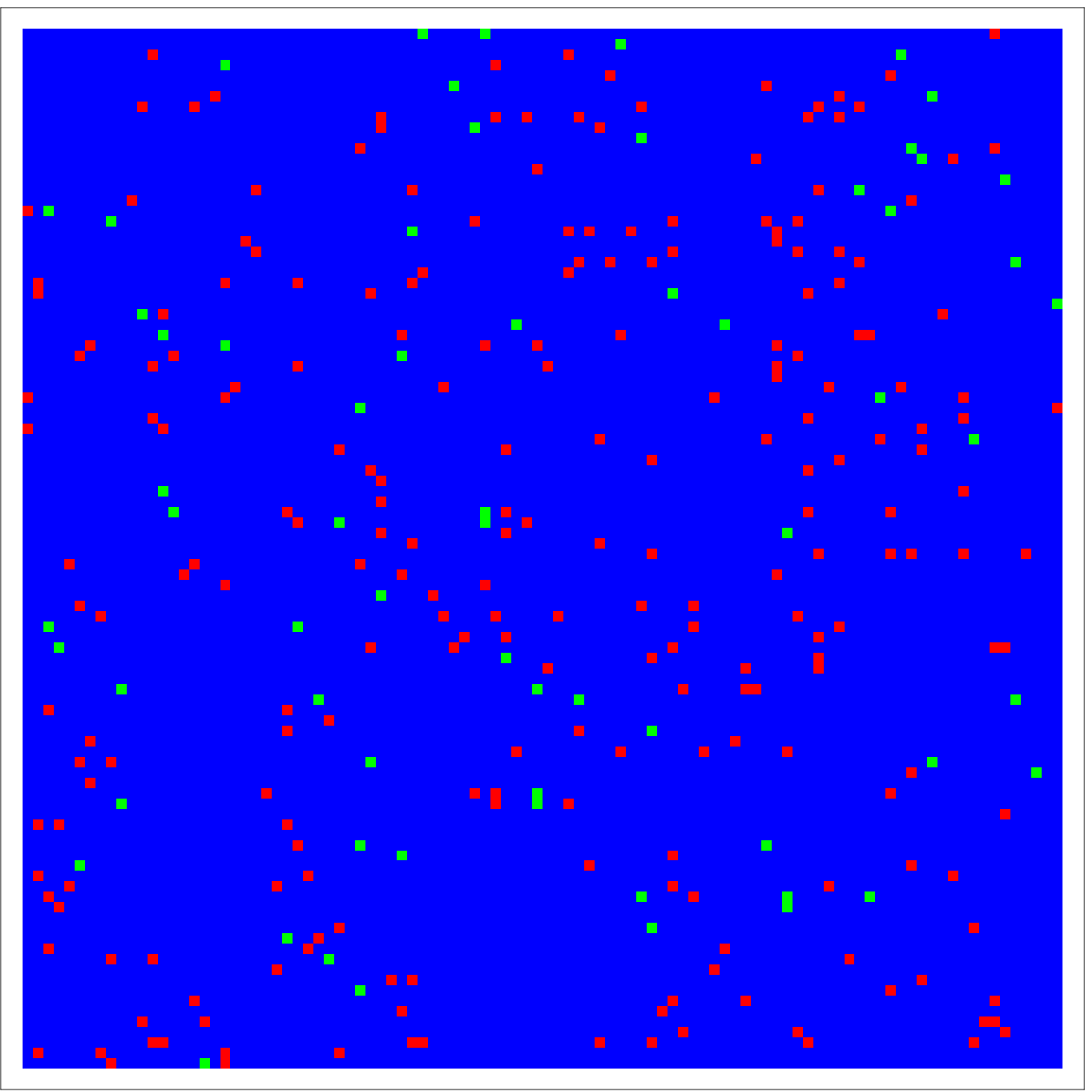}
\includegraphics[width=0.19\columnwidth]{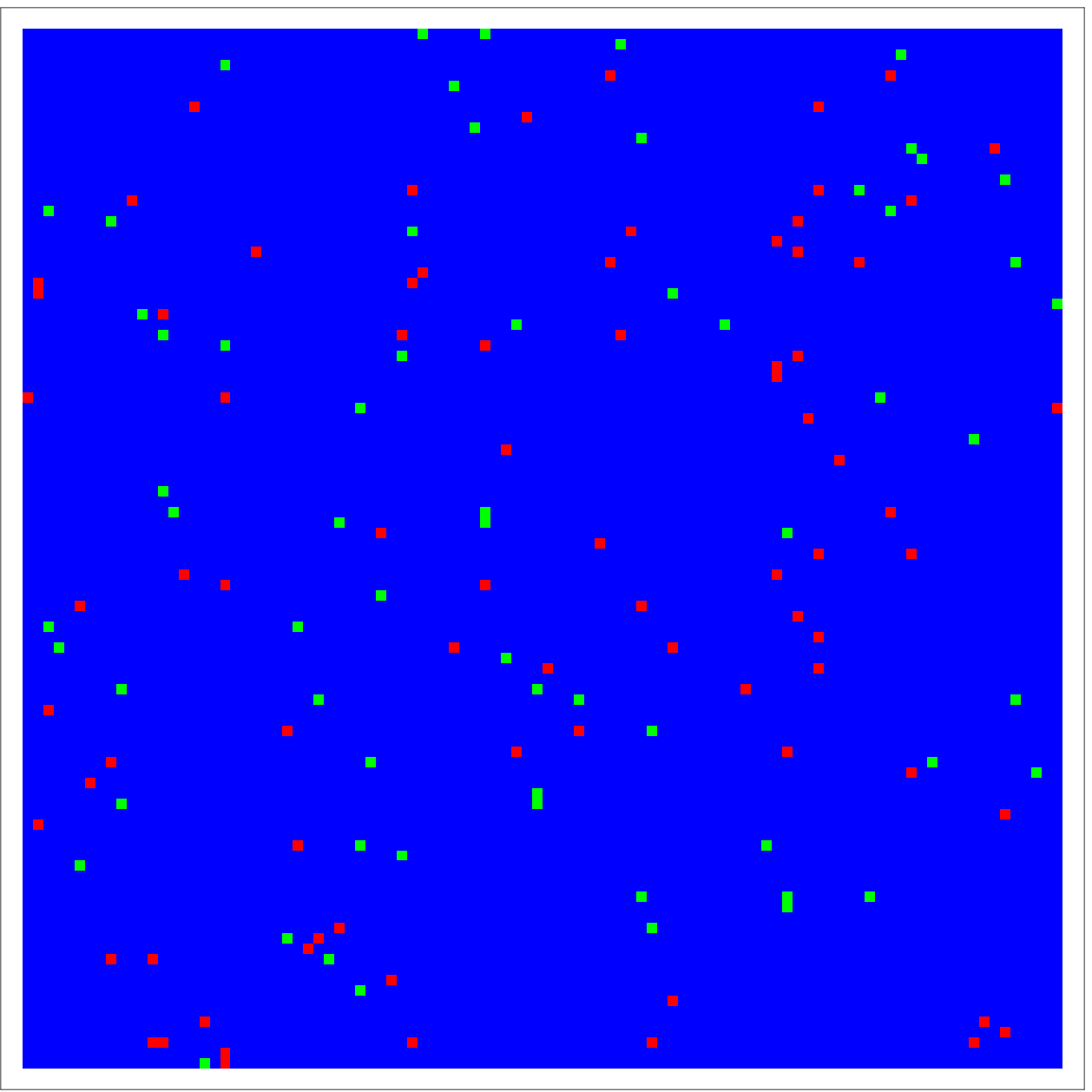}
\includegraphics[width=0.19\columnwidth]{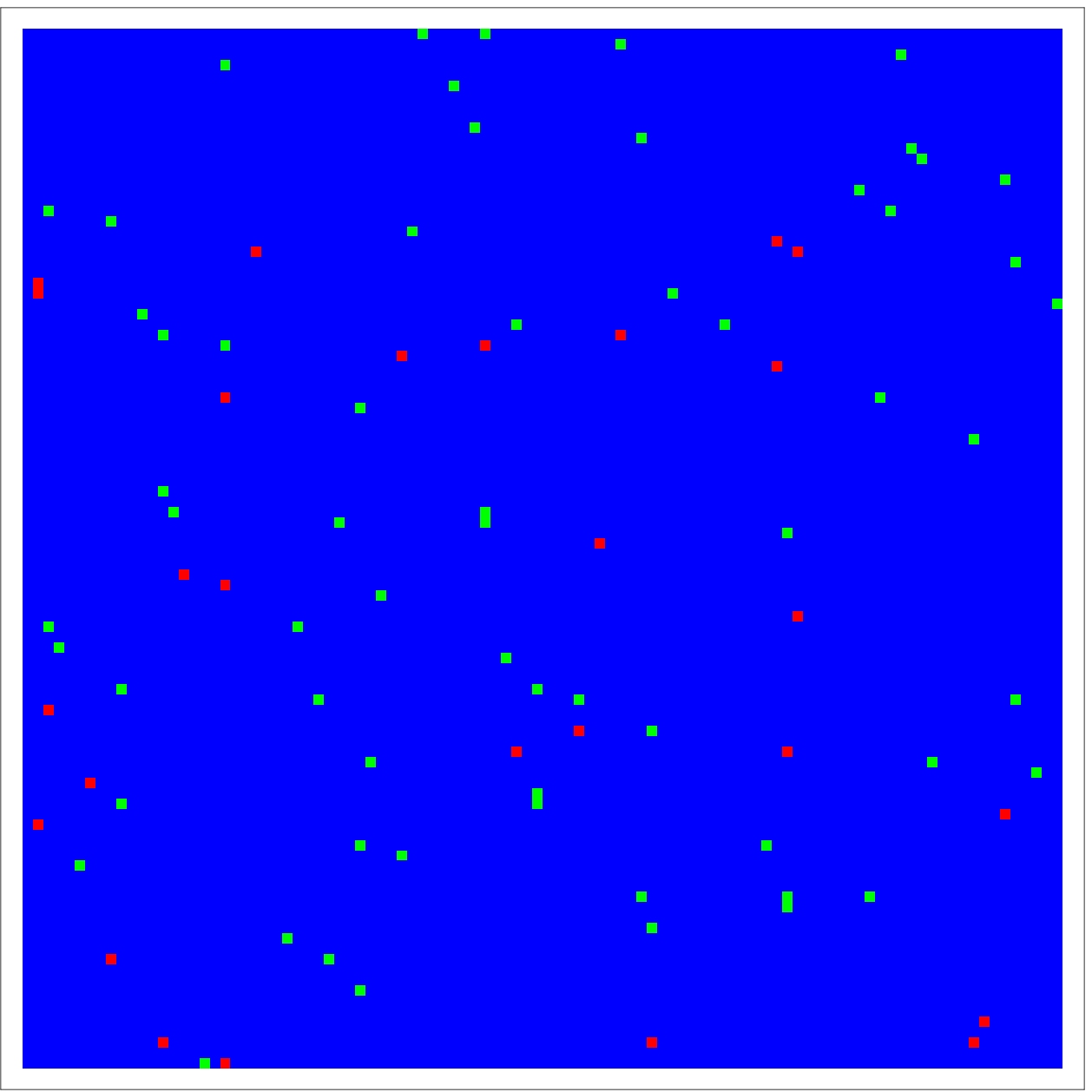}
\includegraphics[width=0.19\columnwidth]{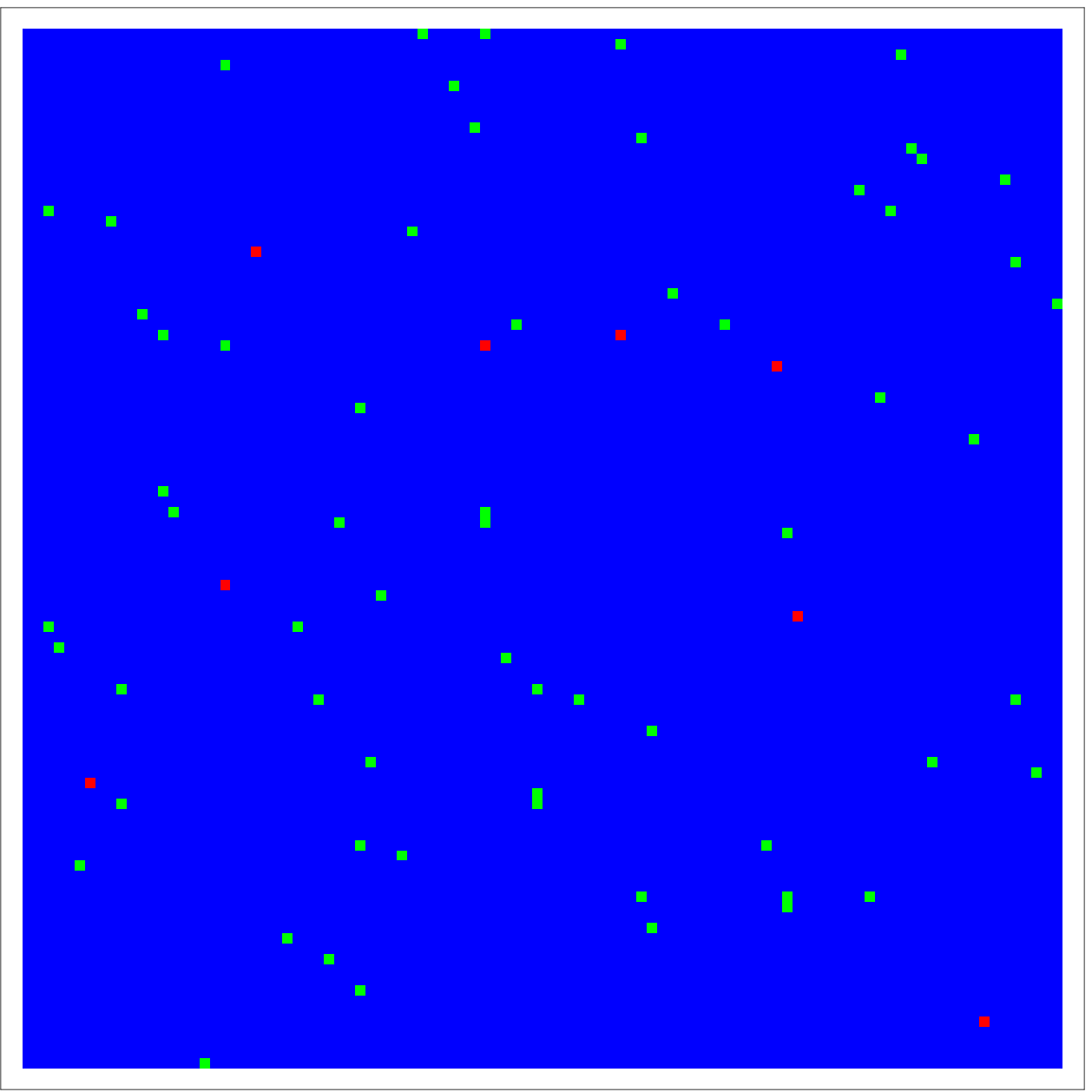} 
\smallskip \caption{{\bf MC/Spatial.} Typical evolution of individuals (pixels)
with MC steps (10 steps between frames). 
Green, red and blue correspond to susceptible, infected and
recovered. Other parameters are:
$N=100$,
$p_{i_0}=0.001$,
$p_{i_1}=0.1$,
$p_{r}=0.05$.
}
\label{fgr:mcM}
\end{figure}

On the other hand, a scenario where the infection probability is kept constant
($p_i=0.1$, $p_r=0.05$) results in a wider and smaller maximum
(the infection and recover constant probabilities have been adjusted
to yield the peak near the same MC steps on the previous cases, Fig.~\ref{fgr:mc0} red dotted line
compared to thick dotted line).
For these conditions, a typical temporal evolution of individuals (pixels) is shown in Fig.~\ref{fgr:mcM}. A weak tendency to clustering is observed, although the
system is seen to reach a quasi-homogeneous state fast. 

Unlike SIR, this model can sustain in a natural way a constant background of infections if at some point in the epidemics $p_{i}$ becomes very similar to $p_{r}$, establishing a transient regime which we categorize as qualitatively different from the region where the daily distribution derived from SHR or SIR is simply too low. This is a feature that can be observed in real data (Fig.~\ref{fgr:shrESP} inset in the right-hand side).

Finally,  
we checked how statistical properties of the model perform and scale under different lattice sizes and parameters via simulation. The distributions over time for $N=100$ and $N=1000$
are virtually indistinguishable as long as the initial infectious
individuals are kept in the same ratio. In order to further visualize the stochasticity under the chosen scale, we show in the right panel of Fig.~\ref{fgr:mc0} ten randomly chosen realizations out of one hundred runs with random initial positions of infectious in the lattice. As the starting day where the infection expands is random, we have rigidly displaced the time of the samples such that they peak on the same day. Then, the
ten different realizations and their averaged value lie nicely on the same
curve and the standard deviation displayed in the inset is seen
to be acceptably small.







\section*{Conclusions and future venues}
We have analysed and compared three mathematical approaches of increasing complexity to investigate the dynamics of COVID-19. 
A take-home message is that all three approaches have enough flexibility to embody the pandemics' actual behaviour for ten arbitrarily chosen countries. However, they display different error bars and have different abilities to be extrapolated into the future to produce valuable predictions

We have proved that a least-squares SHR-model based on the lognormal distribution is well suited to describe the epidemic's evolution using only two free parameters per infection wave. 
Confronted against real data up to the second inflexion point, the values determined for these parameters are well converged and stable, guaranteeing fractional error bars of $\pm 5\%$. Therefore, the SHR-model is suitable to extrapolate tendencies to the next one or two weeks, even in the presence of noisy data. 
A simpler averaged version depending only on a single free parameter per wave has been shown to be adequate to be used as a first approximation, albeit with larger associated incertitudes.
We have also considered a generalised deterministic SIR dynamics to analyse the temporal evolution of the disease. In this case, the corresponding two free parameters are
well converged and stable once the maximum in the daily
distribution of cases is passed, i.e. about a fortnight before the SHR
reaches a similar accuracy. 
Besides the two deterministic models, we have considered stochastic individual-based dynamics reflecting the daily changes in individuals' classes.
We examined both the case of uniform and neighbour-dependent transitions via a Monte-Carlo simulation, which has an excellent correspondence with the analogue SIR model's temporal evolution.

While such simple dynamics ignore individual, spatial or further inhomogeneities (e.g., genetic, socioeconomic, or other differences) we have proven that 
they can reproduce, predict and forecast relevant features of the actual COVID-19 dynamics.
In particular, they provide reasonably robust ways to monitor and forecast the actual
temporal evolution of contagious diseases in different environments, while only requiring basic mathematical tools.

The analysis of ten different countries makes us conclude that the SHR model can be extrapolated into the future with at most a 5\% fractional error after a fortnight passed the second inflexion point. On the other hand, the SIR model, which includes two free parameters only too, seems more stable and can be used with a similar accuracy about one week passed the maximum. Finally, the MC model is helpful to study the interactions between separated regions developing the epidemics. 

By comparing SHR and SIR we find an excellent correlation between functions $c_{\sigma,\mu}(t)$ and $i_{\tau_{0},\tau_{1}}(t)$, and their respective cumulative distributions $C(t)$ and $r(t)$, which suggest that an analytical parametrized solution for SIR might be possible by trying a variational-like approach:

$$
i_{\tau_{0},\tau_{1}}(t) := c_{\sigma,\mu}(t) + \delta(t)
$$

\noindent
Our results strongly suggest that useful bounds can be found for $\delta(t)$. Such promising venue will be explored in the future. 

On the other hand, the excellent agreement between SIR and MC (provided the transition probabilities are chosen in accordance with the hypothesis behind the SIR model) opens new prospects to whrite spatially resolved SIR-like models that might be solved applying Markov chains techniques. 

\newpage
\clearpage
\section*{Tables}

\begin{table}[!ht]
\centering
\caption{{\bf Parameters for SHR model (confirmed deaths, first wave).}
P: country's population (millions).
$\mu$ and $\sigma$: parameters in the lognormal distribution.
$C(\infty)$: asymptotic value for accumulated cases (per million person).
$R^2$ and $r^2$: R-squared correlation factors for $C(t)$ and $c(t)$, respectively.
$t_M$ and $t_2$: maximum and second inflection point (origin is the 1st of January, 2020).}
\begin{tabular}{l|ccccccccc}
COUNTRY    & P    & $\mu$          & $\sigma$       & $C(\infty)$ & $R^2$   & $d_1$ &  $t_M$ & $t_2$ \\ \hline
G. BRITAIN & 66.65& $3.61\pm 0.01$ & $0.76\pm 0.02$ & $688\pm 4$   & 0.9999 & 07/03 & 37 & 52 \\ 
SPAIN      & 46.94& $3.55\pm 0.06$ & $0.43\pm 0.03$ & $606\pm 2$   & 0.9997 & 05/03 & 28 & 40 \\
ITALY      & 60.36& $3.75\pm 0.02$ & $0.51\pm 0.01$ & $581\pm 1$   & 0.9999 & 23/02 & 37 & 52\\
USA        &327.20& $3.95\pm 0.01$ & $1.08\pm 0.04$ & $456\pm 5$   & 0.9997 & 01/03 & 47 & 64\\ 
FRANCE     & 67.00& $3.27\pm 0.03$ & $0.59\pm 0.02$ & $451\pm 1$   & 0.9998 & 15/02 & 53 & 63\\ \hline
SWITZERLAND& 10.23& $3.77\pm 0.03$ & $0.36\pm 0.02$ & $199\pm 1$   & 0.9999 & 06/03 & 32 & 45\\
GERMANY    & 83.02& $3.71\pm 0.02$ & $0.45\pm 0.01$ & $110\pm 0.3$ & 0.9999 & 09/03 & 36 & 50 \\ 
DENMARK    &  5.81& $3.60\pm 0.02$ & $0.53\pm 0.01$ & $106\pm 0.4$ & 0.9999 & 16/03 & 23 & 36 \\
AUSTRIA    &  8.86& $3.30\pm 0.04$ & $0.62\pm 0.03$ & $80.5\pm 0.2$& 0.9997 & 13/03 & 26 & 38 \\
FINLAND    &  5.52& $4.31\pm 0.14$ & $0.20\pm 0.03$ & $59.6\pm 0.3$& 0.9995 & 22/03 & 33 & 45 \\ \hline
\end{tabular}
\label{tbl:SHRd}
\end{table}

\begin{table}[!ht]
\centering
\caption{{\bf Parameters for SHR model (confirmed infections, first wave).}
P: country's population (millions).
$\mu$ and $\sigma$: parameters in the lognormal distribution.
$C(\infty)$: asymptotic value for accumulated cases (per million person).
$R^2$ and $r^2$: R-squared correlation factors for $C(t)$ and $c(t)$, respectively.
$t_M$ and$t_2$: maximum and second inflection point (origin is the 1st of January, 2020).}
\begin{tabular}{l|ccccccccc}
COUNTRY    & P    & $\mu$          & $\sigma$       & $C(\infty)$ & $R^2$    & $d_1$  & $t_M$ & $t_2$ \\ \hline
G. BRITAIN & 66.65& $4.09\pm 0.01$ & $0.47\pm 0.02$ & $4511\pm 11$  & 0.9999 & 31/01 & 78 & 99 \\ 
SPAIN      & 46.94& $3.55\pm 0.06$ & $0.43\pm 0.03$ & $5082\pm 14$  & 0.9996 &  1/02 & 56 & 67 \\
ITALY      & 60.36& $3.75\pm 0.02$ & $0.51\pm 0.01$ & $4083\pm 9$   & 0.9999 & 31/01 & 56 & 71 \\
FRANCE     & 67.00& $3.66\pm 0.03$ & $0.40\pm 0.02$ & $2743\pm 12$  & 0.9998 & 25/01 & 68 & 80 \\ 
USA        &327.20& $4.23\pm 0.04$ & $0.93\pm 0.03$ & $13300\pm 400$& 0.9391 & 21/01 & 49 & 107\\ 
SWITZERLAND& 10.23& $3.79\pm 0.03$ & $0.35\pm 0.02$ & $4028\pm 5$   & 0.9999 & 26/02 & 31 & 41 \\
GERMANY    & 83.02& $3.84\pm 0.02$ & $0.38\pm 0.01$ & $2038\pm 10$  & 0.9999 & 27/01 & 63 & 23 \\ 
DENMARK    &  5.81& $4.35\pm 0.12$ & $0.28\pm 0.04$ & $2056\pm 30 $ & 0.9994 & 27/02 & 42 & 61 \\
AUSTRIA    &  8.86& $2.92\pm 0.05$ & $0.57\pm 0.03$ & $2334\pm 7   $& 0.9996 & 26/02 & 29 & 37 \\
FINLAND    &  5.52& $4.41\pm 0.06$ & $0.29\pm 0.02$ & $1341\pm 12  $& 0.9998 & 30/01 & 74 & 96 \\ \hline
\end{tabular}
\label{tbl:SHRi}
\end{table}

\begin{table}[!ht]
\centering
\caption{{\bf Parameters for SIR model (first wave).}
$N$ (number of individuals), $\tau_0$ and $\tau_1$ (given in days).
Upper: deaths per million people.
Lower: infections per million people.
}
\begin{tabular}{l|ccc}
COUNTRY    &$N$    & $\tau_0$  &$\tau_1$  \\ \hline
SPAIN      &622    & 1.81      & 18.23    \\
G. BRITAIN &621    & 2.57      & 22.47    \\
ITALY      &586    & 2.47      & 25.59    \\
FRANCE     &454    & 3.68      & 19.41    \\
SWITZERLAND&200    & 2.97      & 16.70    \\ 
GERMANY    &112    & 2.48      & 23.15    \\ \hline \hline
COUNTRY    &$N$    & $\tau_0$  &$\tau_1$  \\ \hline
SPAIN      &5082   & 3.11      & 18.05    \\
GERMANY    &2111   & 3.49      & 12.50    \\ \hline
\end{tabular}
\label{tbl:SIRd}
\end{table}


\clearpage
\newpage


%

\appendix

\section*{Supplementary Material: Main public-health actions suggested by the models.}
\label{ACT1}

In this appendix, we summarize
a few known public-health actions suggested by the
SIR model~\cite{weiss2013} which are useful to follow
the lines of reasoning developped in the paper.

\begin{enumerate}
\item Since the maximum value for $R(\infty)$ is the entire population, $N<\infty$, the disease always dies out, $I(t>t_0)=0$. 
Otherwise, if for some initial conditions we could have $I(\infty) \ne 0$, Eq.~\ref{eqn:sir2} would imply that
$R(t)$ could grow without limit, which proves the fact by {\it reductio ad absurdum}.
\item For $n=1$ the ratio $\Re_0=\frac{\tau_1}{\tau_0}$ determines whether the disease
grows or dies. 
Since $S(t)$ can only decrease, we have from the second Eq. in~\ref{eqn:sir1} that 
\begin{equation}
\frac{\D I}{\D t}= 
\frac{ S(t)~i(t)}{\tau_0} - \frac{I(t)}{\tau_1} \le 
\frac{1}{\tau_1}~ \left( \frac{\tau_1}{\tau_0} s(0) - 1 \right) I(t) =
\frac{1}{\tau_1} \left( \Re_0 - 1 \right) I(t) =\frac{I(t)}{\tau}
\label{eqn:expO}
\end{equation}
\noindent
at the onset, for $\Re_0>>1$, an initial estimation for $\tau_0$
can be obtained from $\tau \approx \tau_0$.
\footnote{$s(0)\approx 1$ for large N.}
Therefore,
if $\Re_0<1$, $I(t)$ is a monotonically decreasing function and the
infection dies quickly.
On the other hand, 
if $\Re_0>1$, $I(t)$ increases in the region near $t=0$,
it reaches a maximum value, $I_M(t_M)$, and then it goes to zero, as proved
in the point above. $\Re_0$ is called the {\bf basic reproductive number}
and it sets up a threshold for the expansion of the disease which is not obvious
without a mathematical analysis of the underlying differential equations. 
\item The maximum number of infected people can be obtained by dividing the
two equations~\ref{eqn:sir2} and~\ref{eqn:sir1}, 
\begin{equation}
\frac{\D S}{\D I}= -\frac{ S(t)~i(t) }{ s(t)~I(t) - \frac{I(t)}{\tau_1}}
\end{equation}
\noindent
which can be integrated to yield for $S(0)\approx N$ and $I(0)\approx 1$,
\begin{equation}
i_M  = 1 - \frac{1}{\Re_0} \left( 1 + \ln{\Re_0} \right)
\end{equation}
\item Similarly, dividing the equation~\ref{eqn:sir1} by~\ref{eqn:sirrr},
we get
\begin{equation}
\frac{\D S}{\D R}= -\Re_0 S
\end{equation}
\noindent
i.e., $S(t)=S(0) e^{-\Re_0 R(t)}$, assuming $R(0)=0$.
Notice that for $\Re_0>>1$, $S(\infty)$ might take the value zero,
which corresponds to a very virulent epidemics where everybody dies.
\end{enumerate}

\begin{figure}[!b]
\includegraphics[width=0.32\columnwidth]{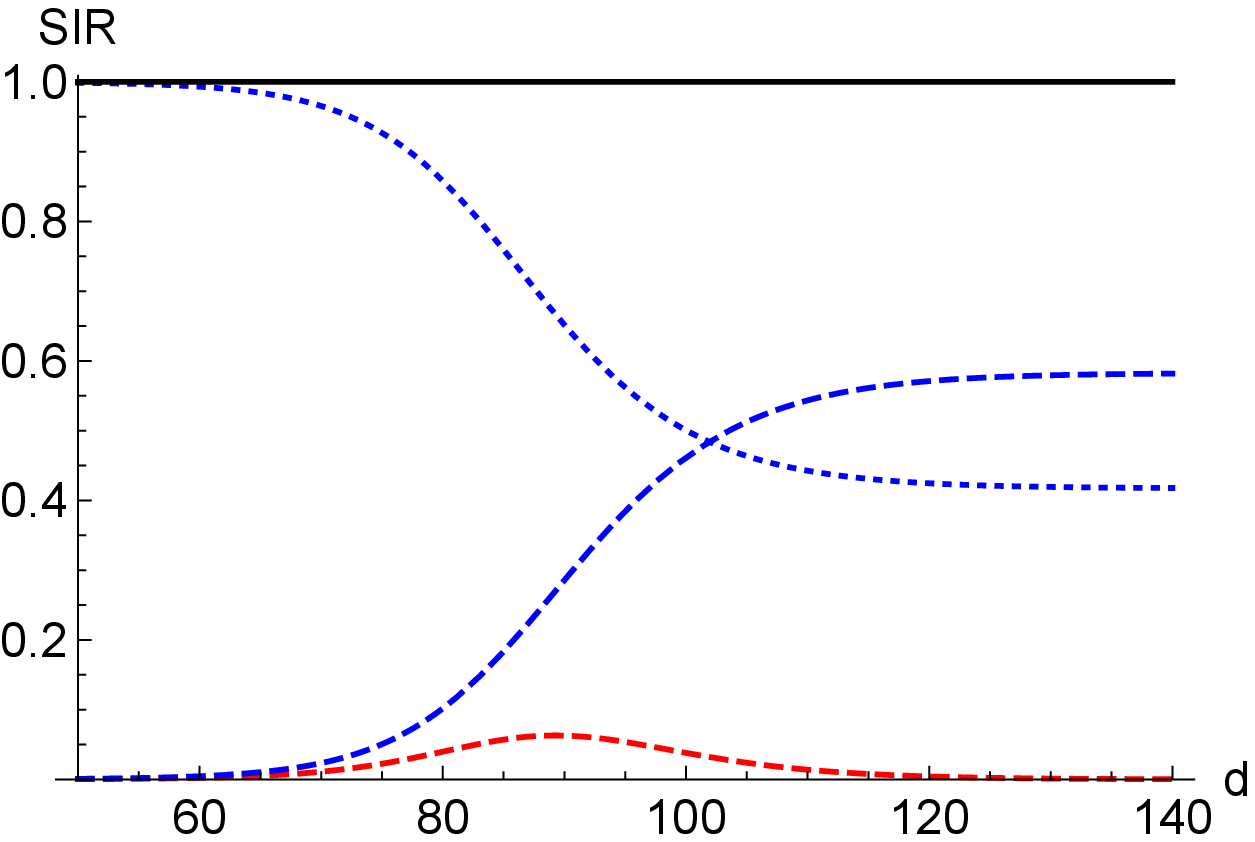}
\includegraphics[width=0.32\columnwidth]{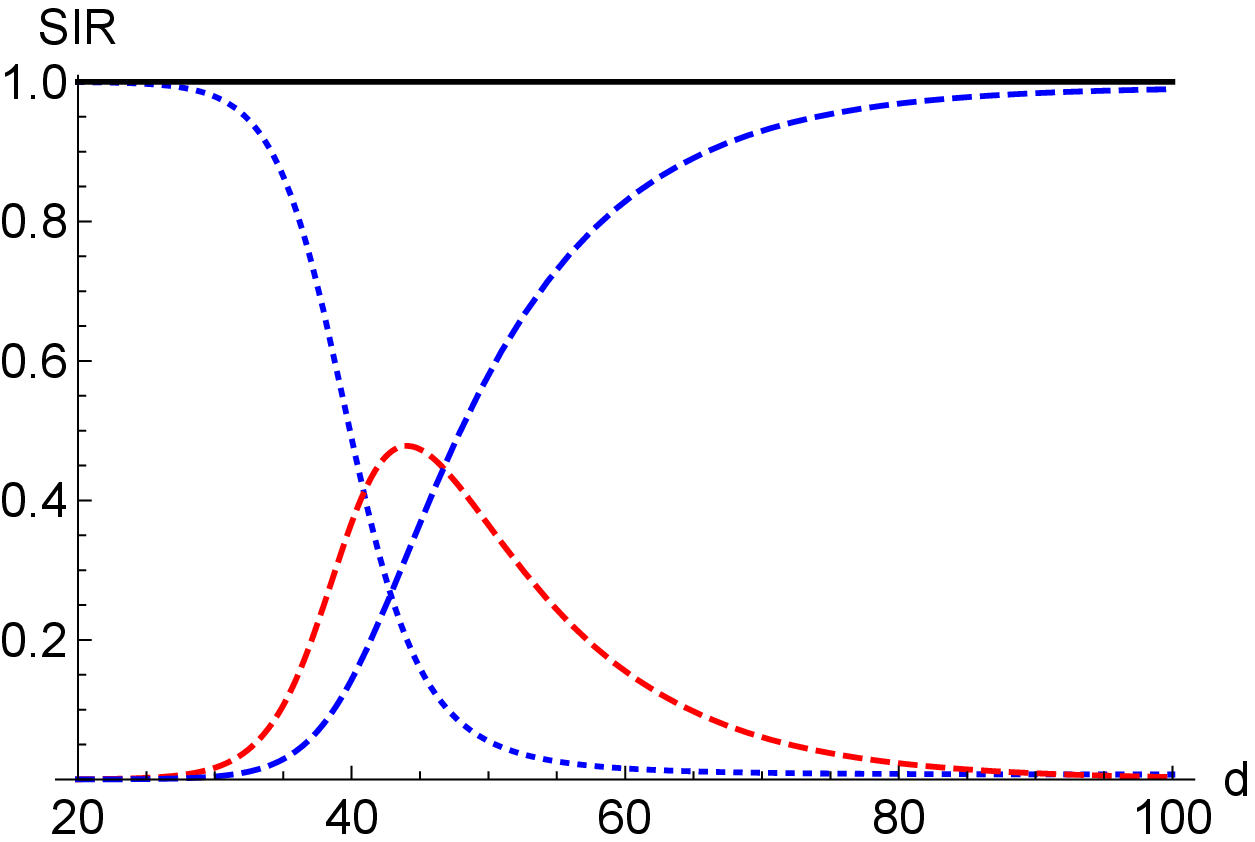}
\includegraphics[width=0.32\columnwidth]{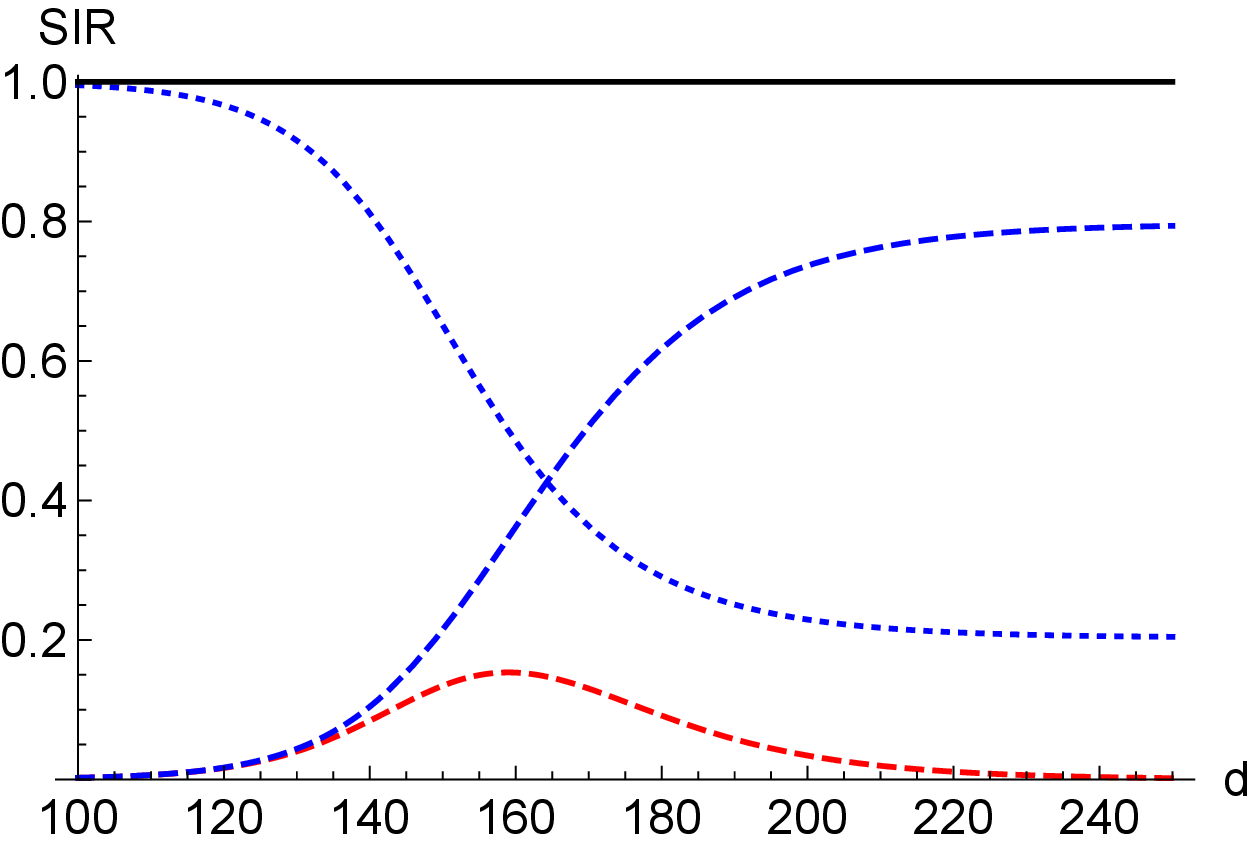}
\smallskip \caption{
{\bf Representative behaviour of the SIR model ($n=1$) depending on the parameters
($\Re_0 \ge 1.5$).}
Blue dotted: $S(t)$. Red dotted: $I(t)$. Blue dashed: $R(t)$. 
Initial conditions
$N=10000000=S(0)+1$, $I(0)=1$.
Left to right:
(I)
$\tau_1=2$, $\tau_0=3$, $t_M=90$, $r(\infty)=0.58$ ;
(II)
$\tau_1=2$, $\tau_0=10$, $t_M=44$, $r(\infty)=0.99$;
(III)
$\tau_1=5$, $\tau_0=10$, $t_M=159$, $r(\infty)=0.80$.
}
\label{fgr:sir1}
\end{figure}
\begin{figure}[!b]
\includegraphics[width=0.32\columnwidth]{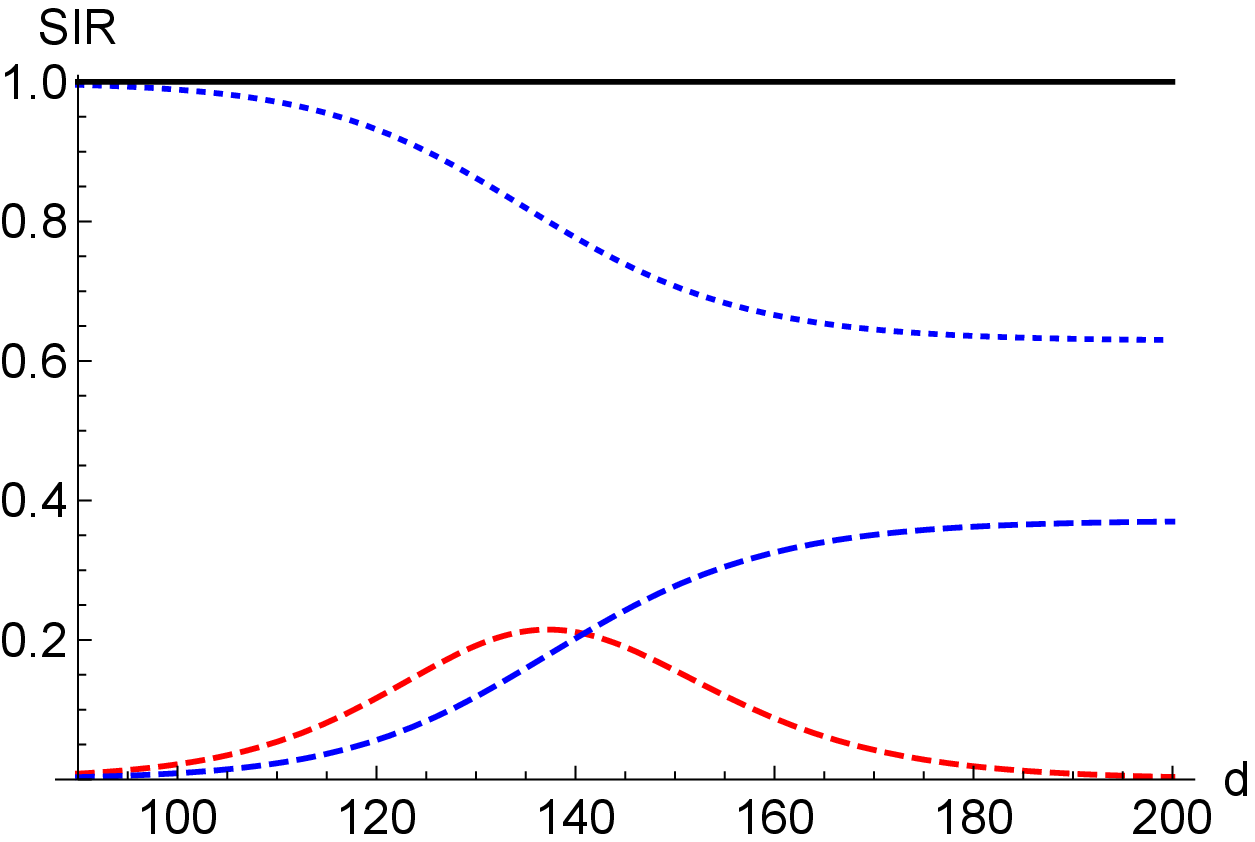}
\includegraphics[width=0.32\columnwidth]{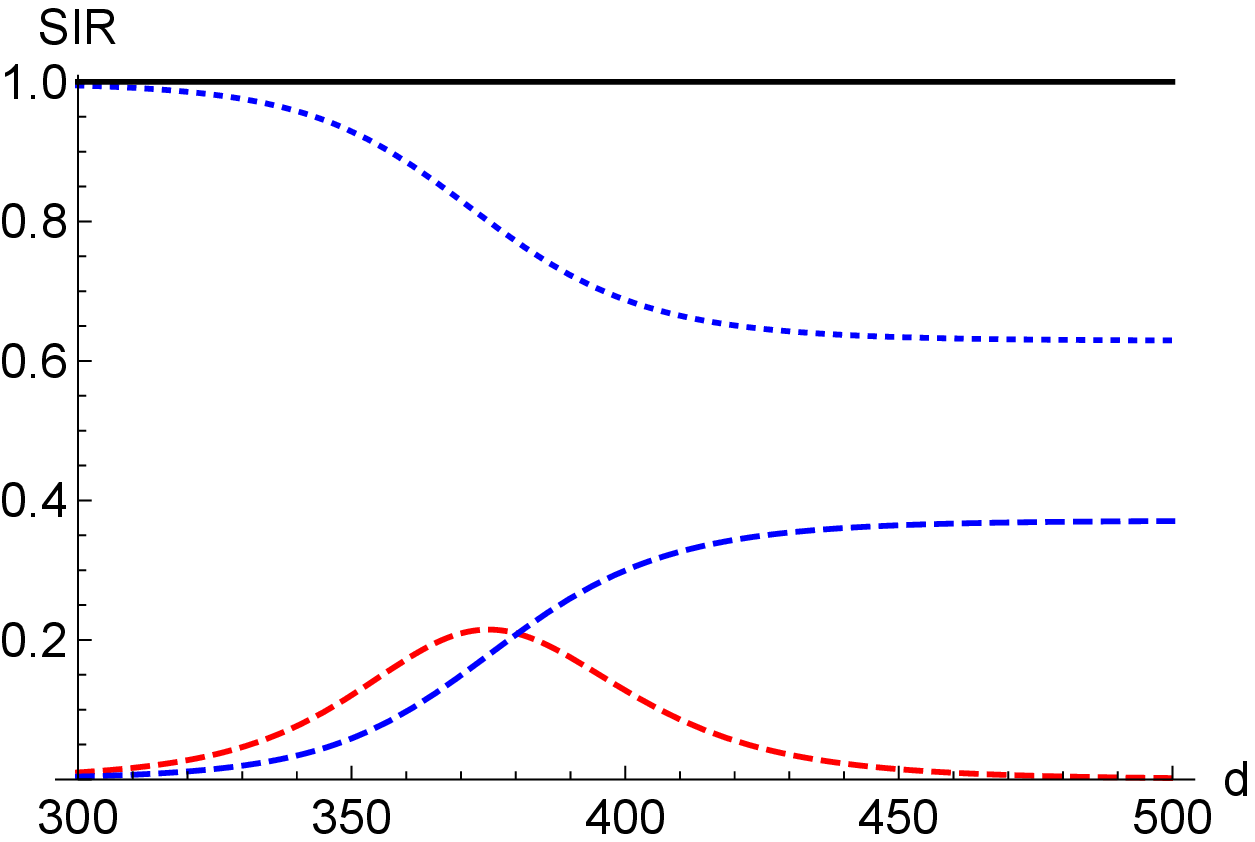}
\includegraphics[width=0.32\columnwidth]{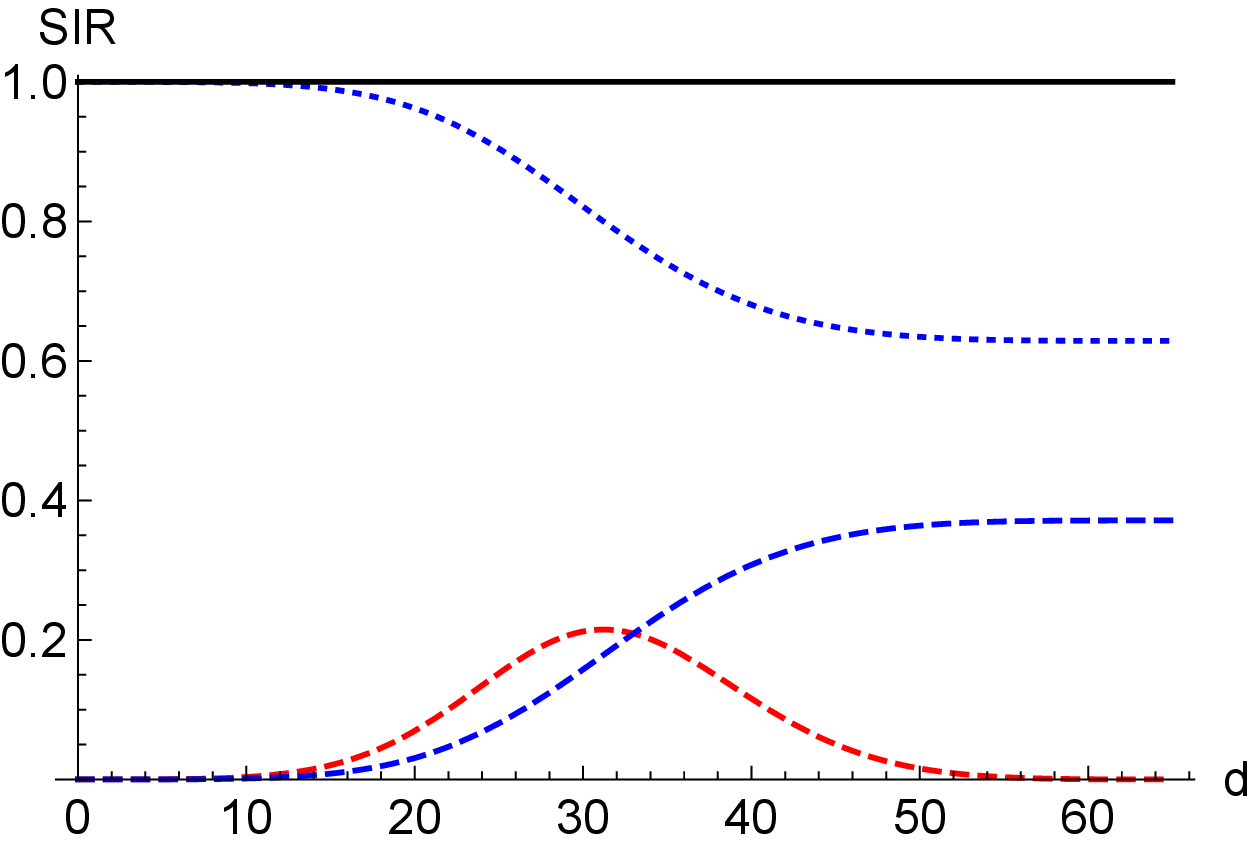}
\smallskip \caption{
{\bf Representative behaviour of the SIR model ($n \ne 1$) depending on the parameters
($\Re_0 \le 1.5$).}
Blue dotted: $S(t)$. Red dotted: $I(t)$. Blue dashed: $R(t)$. 
Initial conditions:
$N=10000000=S(0)+1$, $I(0)=1$.
Parameters:
$\tau_1=2.5$, $\tau_0=2$; $\Re_0=1.25$, $r(\infty)=0.37$.
Left to right:
(I)
$n=1$, $t_M=137$;
(II)
$n=1.1$, $t_M=375$;
(III)
$n=0.818$, $t_M=31$.
}
\label{fgr:sir2}
\end{figure}

The above considerations yield to the following public-health actions while dealing with
an infectious disease (Figs.~\ref{fgr:sir1} and~\ref{fgr:sir2}):

\begin{enumerate}
\item Reduce the contact rate or transmissibility, $\frac{1}{\tau_0}$, by 
isolating infectious nodes, encouraging frequent hand washing and
the use of face masks. 
{\bf Increasing values of $\tau_0$ displace $t_M$ towards larger times and decrease the value of $I_M$}.

\item Decrease $\tau_1$ to reduce the duration of infection. {\bf Increasing values
of $\tau_1$ displace $t_M$ towards the origin and increase the 
value of $I_M$}.

\item Reduce $N=S(0)$ by vaccination or any other kind of immunity.
{\bf Increasing $S(0)$ displaces $t_M$ towards larger values,
decreases $i_M$ and $r_M$.}.

\item Decrease $n$. {\bf Decreasing $n<1$ moves $t_M$ towards larger values and, it decreases $i_M$ and $r_M$.}
\end{enumerate}

\section*{Conflict of Interest Statement}

Since July 2020, the co-author LA-B has been employed by Frontiers Media SA. L-AB declared her affiliation with Frontiers, and the handling Editor states that the process nevertheless met the standards of a fair and objective review.

The remaining authors declare that the research was conducted in the absence of any commercial or financial relationships that could be construed as a potential conflict of interest.

\section*{Author Contributions}

All the authors contributed equally to the research and the manuscript.


\section*{Funding}
L. Hoessly is supported by the Swiss National Science Foundations Early Postdoc.Mobility grant (P2FRP2\_188023). 
This work has been financed by the Spanish MINECO (MAT2017-85089-C2-1-R) and the European Research Council under contract (ERC-2013-SYG-610256 NANOCOSMOS). Computing resources have been provided by CTI-CSIC. 
Open access is partly funded by CSIC.
The funders had no role in study design, data collection and analysis, decision to publish, or preparation of the manuscript.
The authors have declared that no competing interests exist.

\section*{Acknowledgments}
The authors are grateful to Profs. F. Flores and R. Ramirez for useful suggestions and comments on this manuscript.

\clearpage
\newpage

\end{document}